\documentclass[seceq]{ptptex}

\usepackage{graphicx}

\def\lsim{\raise0.3ex\hbox{$<$\kern-0.75em\raise-1.1ex\hbox{$\sim$}}}
\def\gsim{\raise0.3ex\hbox{$>$\kern-0.75em\raise-1.1ex\hbox{$\sim$}}}
\newcommand{\nc}[1]{\newcommand{#1}}
\nc{\R}{{\cal R}}
\nc{\Ca}{{\cal C}}
\nc{\Om}{\hat\Omega_{\rm M}}
\nc{\Oe}{\hat\Omega_{\rm E}}
\nc{\Omo}{\hat\Omega_{\rm M-}}
\nc{\Oee}{\hat\Omega_{\rm E+}}
\nc{\Ome}{\hat\Omega_{\rm M+}}
\nc{\Oeo}{\hat\Omega_{\rm E-}}
\nc{\Mmei}{m_{_{\rm M+}}}
\nc{\Meoi}{m_{_{\rm E-}}}

\title{
Ab initio Study of QCD Thermodynamics on the Lattice\\ at Zero and Finite Densities
}

\author{
Shinji \textsc{Ejiri},$^{1,}$\footnote{ejiri@muse.sc.niigata-u.ac.jp}
Kazuyuki \textsc{Kanaya}$^{2,}$\footnote{kanaya@ccs.tsukuba.ac.jp}
and
Takashi \textsc{Umeda}$^{3,}$\footnote{tumeda@hiroshima-u.ac.jp}\\
for WHOT-QCD Collaboration\footnote{
Sinya Aoki, 
Shinji Ejiri, 
Tetsuo Hatsuda, 
Noriyoshi Ishii, 
Kazuyuki Kanaya,  
Yoshiyuki Nakagawa,
Yu Maezawa, 
Hiroshi Ohno, 
Hana Saito, 
Naoya Ukita, 
and
Takashi Umeda.
}
}

\inst{
$^1$Graduate School of Science and Technology, Niigata University, \\Niigata 950-2181, Japan 
\\
$^2$Faculty of Pure and Applied Sciences, University of Tsukuba, \\Tsukuba 305-8571, Japan
\\
$^3$Graduate School of Education, Hiroshima University, \\Hiroshima 739-8524, Japan
}

\abst{
The WHOT-QCD Collaboration is pushing forward a series of lattice studies of QCD at finite temperatures and densities using improved Wilson quarks. 
Because Wilson-type quarks require more computational resources than the more widely adopted staggered-type quarks, various theoretical and computational techniques have to be developed and applied. 
In this paper, we introduce the fixed-scale approach armed with the $T$-integration method, the Gaussian method based on the cumulant expansion, and the histogram method combined with the reweighting technique.
Adopting these methods, we have carried out the first study of finite-density QCD with Wilson-type quarks and the first calculation of the equation of state with $2+1$ flavors of Wilson-type quarks.
We present results of these studies and discuss perspectives towards a clarification of the properties of $2+1$ flavor QCD at the physical point, at finite temperatures and densities.
}

\begin{document}

\maketitle

\section{Introduction}
\label{sec:intro}

\begin{figure}[b,t]
\centerline{
    \includegraphics[width=100mm]{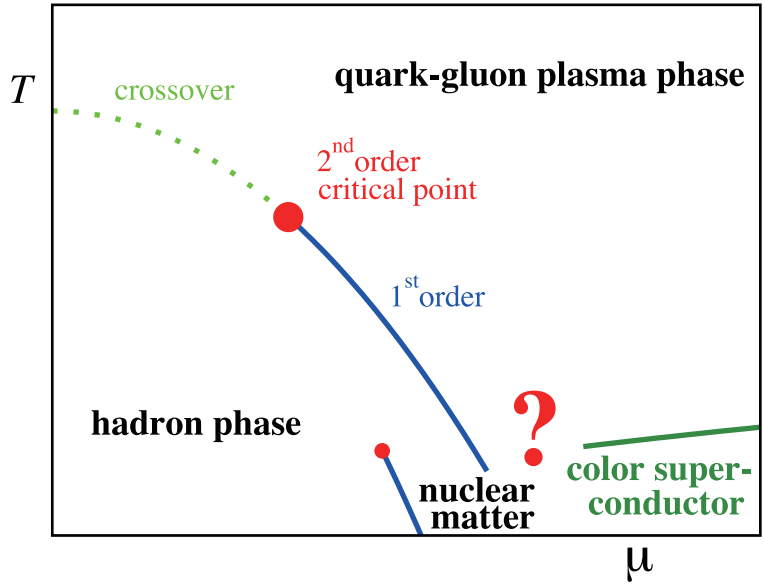}
}
\caption{Prospected phase diagram of QCD at finite temperatures and densities. 
}
   \label{fig:QCDpd}
   \end{figure}

Temperature $T$ and density are controllable parameters of the system. 
At sufficiently high $T$, we expect that the confinement is violated and the chiral symmetry is recovered
because the effective coupling at the thermal energy scale becomes small due to the asymptotic freedom of QCD.
Then, the systems with quarks and gluons will form a colored plasma state called ``quark-gluon plasma'' (QGP).
Although the humankind has never experienced the QGP, 
QGP is expected to play an important role in the creation of matter during the early development of the Universe.
Furthermore, QGP is considered to be observed by relativistic heavy-ion collision experiments at RHIC and LHC \cite{experiment}.
Similar to the case of high temperatures, we expect deconfinement at sufficiently high density too because the average distance between quarks becomes small there and the property of the system will be dominated by the asymptotic freedom.
The density is controlled by the chemical potential $\mu$.
At very large $\mu$ and low $T$, we expect a BCS-like state called ``color superconductor'' due to an attractive interaction among quarks.
At lower densities around the nuclear density, we expect a nuclear fluid state, which may appear around the core of neutron stars.
We thus expect a rich phase structure in QCD as a function of $T$ and $\mu$.
See, e.g., Ref.~\citen{FukushimaHatsuda} for a recent review.
A prospected phase diagram is shown in Fig.~\ref{fig:QCDpd}.

\begin{figure}[b,t]
\centerline{
    \includegraphics[width=100mm]{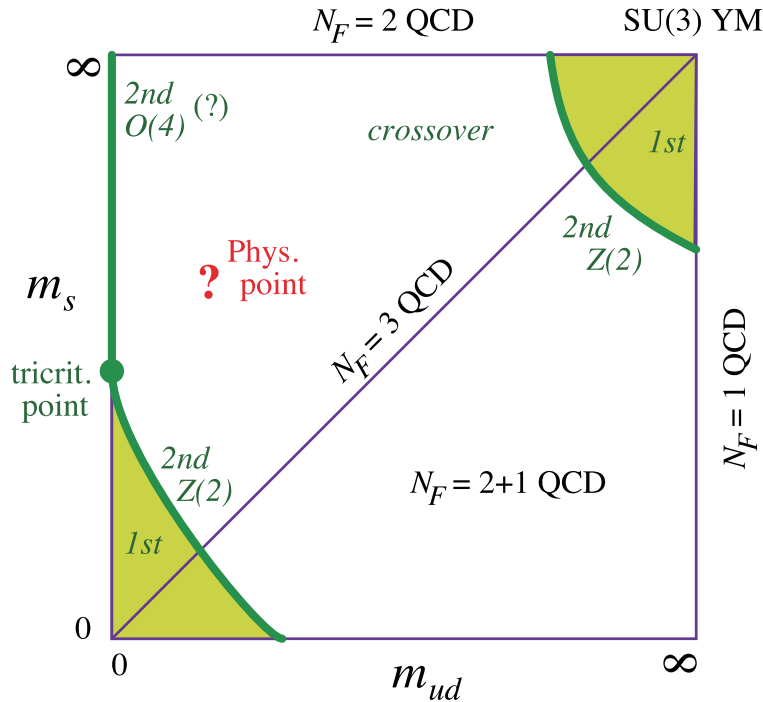}
}
\caption{Prospected order of the finite-temperature QCD transition at $\mu=0$ as a function of the light quark mass $m_{ud} \stackrel{\rm def.}{=} m_u = m_d$ and the strange quark mass $m_s$.  
The top-right corner corresponds to the quenched limit of QCD.
Lattice simulations with improved staggered quarks suggest that the physical point is located in the crossover region. 
The phase diagram is not fully established yet.
See Ref.~\citen{reviews1} for discussions and caveats.
}
\label{fig:QCDpd2}
\end{figure}

When we vary the quark masses off the physical point, the nature of the quark matter may be different depending on them.
The usual expectation for the order of the finite-temperature QCD transition at $\mu=0$, based on effective model studies and lattice simulations, is summarized in Fig.~\ref{fig:QCDpd2}. 
Details of the phase diagram as well as the nature of the quark matter at finite $T$ and $\mu$ are, however, not well clarified yet.
Because the issue is essentially non-perturbative, numerical studies on the lattice is the only systematic way to investigate the phase structure directly from the first principles of QCD\cite{reviews1,reviews2}.

Most lattice studies of hot/dense QCD have been done with computationally less demanding staggered-type lattice quarks\cite{reviews1,reviews2}. 
In particular, in the study of the equation of state (EOS), extrapolations to the physical point and to the continuum limit have been achieved only with staggered-type quarks.
However, the theoretical basis such as locality and universality are not well established with them\cite{reviews1}. 
Therefore, to evaluate the effects of lattice artifacts and thus to obtain reliable predictions to be compared with experiment, it is important to perform simulations using theoretically sound lattice quarks, such as the Wilson-type quarks. 
Here we note that, until recently, the O(4) scaling property expected around the chiral transition of two-flavor QCD\cite{PisarskyWilczek} has been observed only with Wilson-type quarks\cite{O4scaling,CPPACS00}.
Quite recently, an O($N$) scaling behavior\footnote{Presumably the O(2) scaling, as expected from the symmetry of staggered-type quarks.} was observed with an improved staggered quarks by letting the light quark mass much lighter than the physical $u$ and $d$ quark masses\cite{RBCBi09}.
Therefore, some of the chiral properties around the transition temperature may be easier to extract with Wilson-type quarks.

A reason that Wilson-type quarks have not been widely adopted in the study of hot/density QCD is that the computational cost is larger than that for staggered-type quarks, in particular at small quark masses. 
Previous studies with Wilson-type quarks were limited to large quark masses and the case of two-flavor QCD at vanishing chemical potentials \cite{CPPACS00,CPPACS01}.
The WHOT-QCD Collaboration is pushing forward studies of lattice QCD at finite $T$ and $\mu$
adopting improved Wilson quarks\cite{SW} coupled to RG-improved Iwasaki glues\cite{Iwasaki}.
We want to extend the studies to more realistic $2+1$ flavor QCD at finite chemical potentials with physical light quarks.
Towards this goal, we made a series of simulations by implementing and developing efficient methods for Wilson-type quarks.
We developed the $T$-integration method to make the fixed-scale approach applicable\cite{WHOT09}, tested the Gaussian method to tame the sign problem\cite{WHOT10dense}, and extended the histogram method by combining with the reweighting technique to investigate the phase structure\cite{Saito1,Saito2,Nakagawa}. 
With these techniques, we have studied thermodynamic properties of the quark matter 
through the equation of state\cite{Umeda:2010ye,WEOS12}, 
heavy-quark free energies and screening masses\cite{WHOT07,WHOT10screening,Maezawa2011}, 
spectral functions\cite{OhnoLat08,WHOT11-SPF},   etc.

In Sect.~\ref{sec:FTLQCD}, we review techniques adopted in our investigation of finite-temperature QCD on the lattice.
We also introduce the $T$-integration method  to calculate the EOS in the fixed-scale approach.
We then report our calculation of EOS with 2+1 flavors of improved Wilson quarks in Sect.~\ref{sec:EOS}.
Sections~\ref{sec:FDLQCD}, \ref{sec:Nf2Mu} and \ref{sec:Histogram} are devoted to our study of finite-density QCD.
We first discuss in Sect.~\ref{sec:FDLQCD} major methods to perform simulations of QCD at finite densities.
We introduce our approach using the cumulant expansion in a hybrid method of Taylor and reweighting methods. 
We then present in Sect.~\ref{sec:Nf2Mu} our results of the pressure and the quark number susceptibility at finite densities in two-flavor QCD, adopting these methods.
In Sect.~\ref{sec:Histogram}, we introduce another method, a histogram method, to investigate the first order transition and its boundary. 
We apply the method to calculate the location of the critical point where the first-order deconfining transition of QCD in the heavy quark limit turns into a crossover as the quark masses are decreased.
We then study the critical point at non-zero chemical potentials.
We also present our on-going project to study finite density QCD with light dynamical quarks by combining the histogram method with phase-quenched simulations.
Finally, in Sect.~\ref{sec:HQFE}, our results of the heavy-quark free energy are summarized for zero and finite densities.
A short summary is given in Sect.~\ref{sec:conclusions}.

\section{Thermodynamics of QCD on the lattice}
\label{sec:FTLQCD}

On a lattice with a size $N_{\rm s}^3\times N_{\rm t} \equiv N_{\rm site}$, the temperature of the system is given by 
$ 
T = 1/ (N_{\rm t} a) ,
$ 
where $a$ is the lattice spacing.
To vary $T$, we may either vary $a$ at fixed $N_{\rm t}$, or vary $N_{\rm t}$ at a fixed $a$.
Let us call the former as the {\em fixed-$N_{\rm t}$ approach}, and the latter as the {\em fixed-scale approach}. 

In the fixed-$N_{\rm t}$ approach, we can vary $T$ continuously through a continuous variation of $a$.
This is a reason that the fixed-$N_{\rm t}$ approach has been widely adopted in many simulations. 

The value of $a$ is controlled by the coupling parameters, which we denote as $\vec{b}$. 
For QCD with Wilson-type quarks, $\vec{b} = (\beta, \kappa_{u}, \kappa_{d}, \kappa_{s}, \cdots)$. 
%
We first define the lines of constant physics (LCP's) in the coupling parameter space by the fixed dimension-less ratios of physical observables such as $m_\pi/m_\rho = m_\pi a / m_\rho a$.
Here, to remove additional dependence on $T$, these observables have to be measured at $T=0$.
A LCP represents a physical system at different values of $a$.
In two-flavor QCD with improved Wilson quarks, LCP's defined by $m_\pi/m_\rho$ are determined in Refs.~\citen{CPPACS01} and \citen{WHOT07}. 
In 2+1 flavor QCD, we have to fix one more dimension-less ratio such as $m_K/m_\rho$ or $m_{\eta_{ss}}/m_\phi$.
Our world corresponds to the LCP with $m_\pi/m_\rho \approx 135/770$, etc.

The beta function $a\, d \vec{b} / d a$ is defined as the variation of $\vec{b}$ along a LCP. 
In the fixed-$N_{\rm t}$ approach, we vary $T$ of a given physical system by varying $\vec{b}$ along a LCP on a lattice with a fixed value of $N_{\rm t}$.

The energy density $\epsilon$ and the pressure $p$ of the system are given by derivatives of the partition function $Z$ in terms of $T$ and the physical volume $V=(N_{\rm s} a)^3$:
\begin{eqnarray}
\epsilon = -  \left\langle \frac{1}{V} \frac{\partial \ln Z}{\partial T^{-1}}  \right\rangle_{\rm \! sub}, 
\hspace{5mm} 
 p       =  \left\langle T \frac{\partial \ln Z}{\partial V}  \right\rangle_{\rm \! sub} . 
\label{eqn:ep}
\end{eqnarray}
where $\langle \cdots \rangle_{\rm sub}$ is the thermal average with zero temperature contribution subtracted for renormalization.
To vary $T$ and $V$ independently, we need to introduce anisotropic lattices.
When $a_{\rm s}$ and $a_{\rm t}$ are the lattice spacings in spatial and temporal directions, 
$V$ and $T$ are given by $V = (N_{\rm s} a_{\rm s})^3$ and $T = 1 / (N_{\rm t} a_{\rm t})$. 
Then, in principle, (\ref{eqn:ep}) can be evaluated by independent variations of $a_{\rm s}$ and $a_{\rm t}$.
However, this requires a systematic study of a set of physical observables on anisotropic lattices with varying both $a_{\rm s}$ and $a_{\rm t}$, which is, however, quite demanding. 

Here, we note that the combination 
\begin{eqnarray}
\epsilon - 3p = - \frac{T}{V} \left\langle  \left( T^{-1} \frac{\partial }{\partial T^{-1}}  + 3 V \frac{\partial}{\partial V} \right) \ln Z \right\rangle_{\rm \! sub}
\end{eqnarray}
is given in terms of a uniform rescaling 
$ 
\displaystyle{ a_{\rm t} \frac{\partial }{\partial a_{\rm t}}  + a_{\rm s} \frac{\partial}{\partial a_{\rm s}}  }
$, which can be realized without introducing anisotropic lattices. 
We thus obtain on isotropic lattices 
\begin{eqnarray}
\epsilon - 3p = - \frac{T}{V} \left\langle a \frac{d \ln Z}{d a} \right\rangle_{\rm \! sub} \!
= - \frac{T}{V} \, a \frac{d \vec{b}}{da} \cdot \left\langle \frac{\partial \ln Z}{\partial \vec{b}} \right\rangle_{\rm \! sub} \!
= \frac{T}{V} \, a \frac{d \vec{b}}{da} \cdot \left\langle \frac{\partial S}{\partial \vec{b}} \right\rangle_{\rm \! sub}, 
\;\;\;\;
\label{eqn:e3p}
\end{eqnarray}
where $S$ is the lattice action. 
The coefficient $a\, d \vec{b} / d a$ is just the beta-function of the system, whose non-perturbative values can be determined by simulations on isotropic lattices. 
Eq.~(\ref{eqn:e3p}) enables us to study $\epsilon - 3p$ non-perturbatively without introducing anisotropic lattices.
The combination $\epsilon - 3p$ is nothing but the trace of the energy-momentum tensor, and called the trace anomaly.
$\epsilon - 3p$ vanishes for free gasses, but will have non-trivial values with interacting matters.

\subsection{Fixed-$N_{\rm t}$ approach and integration method}

In order to obtain $\epsilon$ and $p$ separately, we need one more independent input. 
The most widely adopted is the {\em integration method}\cite{Engels:1990vr}, with which we can determine the pressure $p$ non-perturbatively through an integration in the coupling parameter space:
\begin{eqnarray}
p = \frac{T}{V} \int^{\vec{b}}_{\vec{b}_0} \! d\vec{b} \cdot \left\langle \frac{1}{Z}
\frac{\partial Z}{\partial \vec{b} } \right\rangle_{\rm \! sub}
= -\frac{T}{V} \int^{\vec{b}}_{\vec{b}_0} \! d\vec{b} \cdot \left\langle
\frac{\partial S}{\partial \vec{b}} \right\rangle_{\rm \! sub} .
\label{eq:Integral}
\end{eqnarray}
This relation is obtained by differentiating and then integrating the thermodynamic relation $p = (T/V) \ln Z$ in the coupling parameter space. 
The integration path is free to choose as far as the initial point $\vec{b}_0$ is located in the low temperature phase such that $p(\vec{b}_0) \approx 0$.
See Appendix A of ref.~\citen{CPPACS01} for a concrete demonstration of the path-independence.

Several points to be kept in mind with the fixed-$N_{\rm t}$ approach are as follows:
(i) When we fix $N_{\rm s}$, the spatial volume $V=(N_{\rm s} a)^3$ is varied simultaneously as we vary $T$. 
In the high $T$ region, $V$ may be quite small with a fixed $N_{\rm s}$.
To keep $V$ around a fixed value, $N_{\rm s}$ has to be increased as we increase $T$.
(ii) At low $T$'s, the lattice may be coarse. 
To ensure asymptotic scaling around the QCD transition temperature, a large $N_{\rm t}$ together with improvement of the lattice action is mandatory. 
(iii) For the zero-temperature subtraction, we have to carry out zero-temperature simulations at all of the finite-temperature simulation points. 
Together with the systematic zero-temperature simulations to determine LCP in a wide range of the coupling parameter space, an indispensable fraction of the total computational cost is required to carry out the systematic zero-temperature simulations in the fixed-$N_{\rm t}$ approach.

\subsection{Fixed-scale approach and $T$-integration method}
\label{sec:Tintegration}

In the fixed-scale approach, $T$ is varied by varying $N_{\rm t}$ at a fixed $\vec{b}$ (and thus at a fixed $a$), i.e., the simulations are done at the same $\vec{b}$ point for all $T$'s.
Therefore, all the simulations are automatically on a LCP without fine-tuning. 
Furthermore, the $T=0$ subtractions can be done by a common $T=0$ lattice. 
We may even borrow high statistic configurations at $T=0$ on the International Lattice Data Grid (ILDG)\cite{Maynard:2010wi}.
Therefore, we can largely reduce the cost for the zero-temperature simulations with the fixed-scale approach\cite{WHOT09}. 

On the other hand, the conventional integral method to obtain $p$ by an integration in the coupling parameter space is inapplicable, because data are available at only one $\vec{b}$ point in the fixed-scale approach.
To overcome the problem, we have developed a new method, the {\em $T$-integration method}\cite{WHOT09}:
Using a thermodynamic relation at vanishing chemical potential, we obtain
\begin{eqnarray}
T \frac{\partial}{\partial T} \left( \frac{p}{T^4} \right) =
\frac{\epsilon-3p}{T^4}
\hspace{6mm}\Longrightarrow\hspace{6mm}
\frac{p}{T^4} = \int^{T}_{T_0} dT \, \frac{\epsilon - 3p}{T^5}
\label{eq:Tintegral}
\end{eqnarray}
with $p(T_0) \approx 0$.
When we vary $T$ by varying $\vec{b}$ along a LCP, the integral in (\ref{eq:Tintegral}) is equivalent to that in (\ref{eq:Integral}) with the integration path chosen to be on this LCP.
However, (\ref{eq:Tintegral}) allows us to integrate over $T$ without varying $\vec{b}$.

In the fixed-scale approach, various $T$'s are achieved by varying $N_{\rm t}$.
Because $N_{\rm t}$ is discrete, we have to interpolate the data with respect to $T$ to carry out the integration of (\ref{eq:Tintegral}).
We need to check the magnitude of systematic errors from the interpolation of the trace anomaly in $T$.
Application of the method to finite $\mu$ is straightforward when we reweight from $\mu=0$.

We find that the fixed-scale approach is complemental to the conventional fixed-$N_{\rm t}$ approach in many respects:
At very high temperatures, typically at $T$ \gsim\ $1/3a$, 
the fixed-scale approach suffers from lattice artifacts as discussed below, 
while the fixed-$N_{\rm t}$ approach can keep $N_{\rm t}$ finite and can keep the lattice artifact small by adopting a sufficiently large $N_{\rm t}$.
On the other hand, at small $T$, typically at $T$ \lsim\ $T_c$, the fixed-scale approach can keep $a$ small at the price of larger cost due to large $N_{\rm t}$, while the fixed-$N_{\rm t}$ approach suffers from lattice artifacts due to large $a$.

Another attractive point of the fixed-scale approach in a study with improved Wilson quarks is that, unlike the case of the fixed-$N_{\rm t}$ approach, we can keep the lattice spacing small at all temperatures and thus can avoid extrapolating the non-perturbative clover coefficient $c_{\rm SW}$ to coarse lattices on which the improvement program is not quite justified.

It should be kept in mind that the fixed-scale approach is not applicable at very high temperatures where, besides the artifacts due to a quite small value of $N_{\rm t}$, the lattice spacing $a$ may also be too coarse to resolve thermal fluctuations\cite{WHOT09,Maezawa2011}.
To get an idea about the latter effects, we estimate a typical length scale of thermal fluctuations by the thermal wave length $\lambda \sim 1/ E$ where $E$ is an average energy of massless particles at finite $T$. 
We find $E \sim 3 T \zeta(4)/\zeta(3) \sim 2.7 T$ for the Bose-Einstein distribution and $E \sim 3 T \zeta(4)/\zeta(3) \times 7/6 \sim 3.15 T$ for the Fermi-Dirac distribution.
We then obtain $\lambda \sim 1/3T$.
Thus, data at $T$ \gsim\ $1/3a$ for which $a$ \gsim\ $\lambda$ should be taken with care \cite{Maezawa2011}.

   \begin{figure}[b,t]
       \centerline{
    \includegraphics[width=67mm]{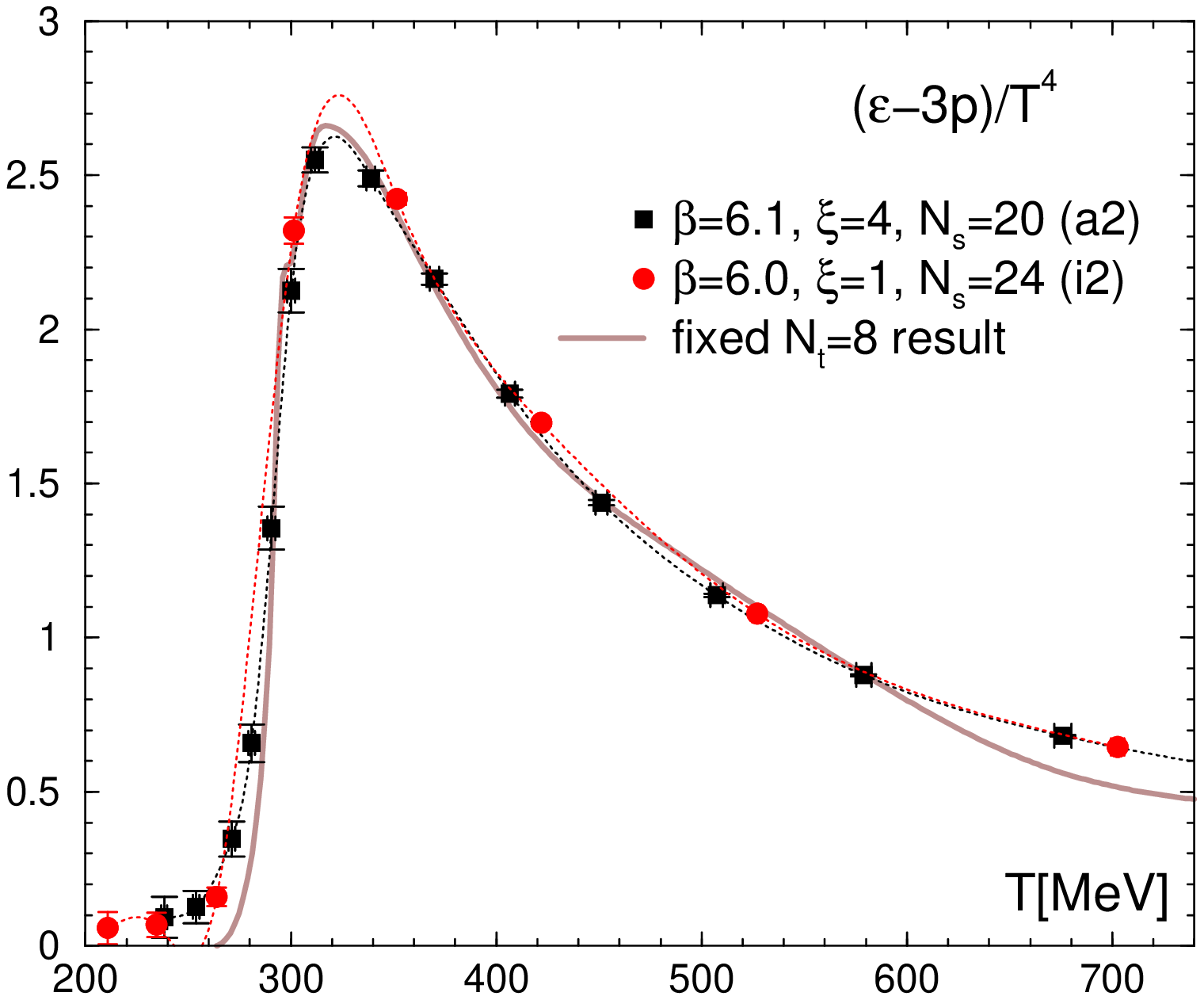}
    \hspace{3mm}
    \includegraphics[width=65mm]{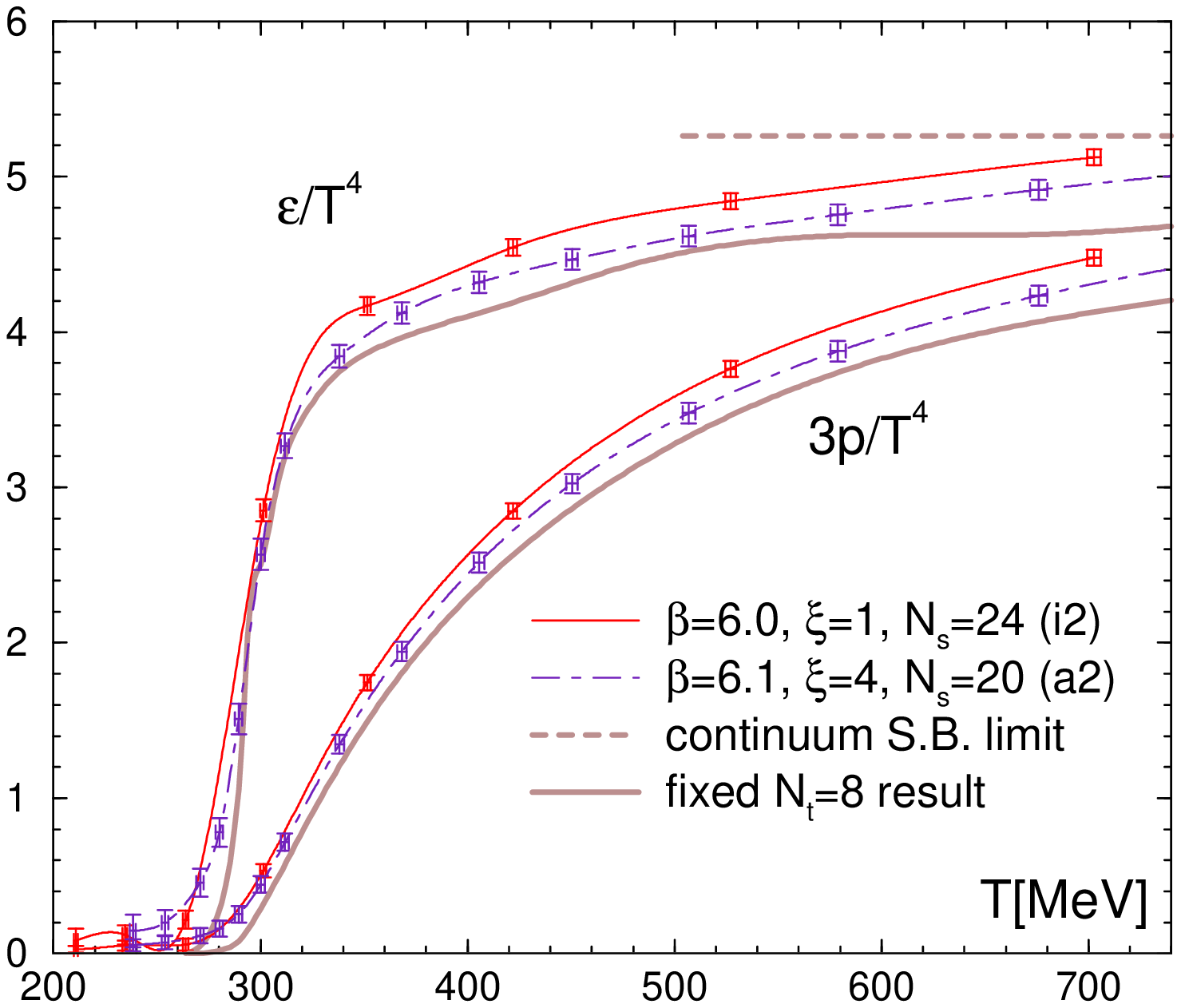}
       }
   \caption{Test of the fixed-scale approach armed with the $T$-integration method in quenched QCD\cite{WHOT09}. 
    {\em Left}: 
    trace anomaly on an anisotropic lattice (a2) compared with the isotropic lattice with similar spatial lattice spacing and volume (i2).
    {\em Right}: energy density and pressure by the T-integral method.
    The shaded curves represent the results of the conventional fixed-$N_{\rm t}$ method at $N_{\rm t}= 8$ \cite{Boyd:1996bx}.}
   \label{fig:fixedscale}
   \end{figure}

We have tested the fixed-scale approach and the $T$-integration method in quenched QCD\cite{WHOT09}. 
Main results are summarized in Fig.~\ref{fig:fixedscale}.
Comparing EOS' obtained on various lattices as well as the result from the fixed-$N_{\rm t}$ approach on large lattices, we find that the fixed-scale approach is reliable and powerful to calculate EOS, in particular at low and intermediate temperatures.
The systematic errors  due to the interpolation in $T$ is well under control in these studies.
The EOS from the fixed-scale approach was shown to be well consistent with that from the fixed-$N_{\rm t}$ approach with large $N_{\rm t}$ ($N_{\rm t} \,\gsim\, 8$), except for the high temperature limit where the fixed-scale approach suffers from lattice discretization errors. 

We adopt the fixed-scale approach to calculate the EOS in $2+1$ flavor QCD in Sect.~\ref{sec:EOS}.
We also compute heavy-quark free energies in $2+1$ flavor QCD with the fixed-scale approach in Sect.~\ref{sec:HQFE21}.

\section{Equation of state in $2+1$ flavor QCD with improved Wilson quarks}
\label{sec:EOS}

A systematic study of finite temperature QCD with improved Wilson quarks was made by the CP-PACS Collaboration around the beginning of this century for the case of two-flavor QCD at vanishing chemical potentials using the fixed-$N_{\rm t}$ approach \cite{CPPACS00,CPPACS01}. 
From a series of systematic simulations, they determined the phase structure and LCP's, confirmed the O(4) scaling, and obtained the EOS along several LCP's in the range $m_{\pi}/m_{\rho}$ \gsim\ 0.65 around and above the pseudocritical temperature $T_{\rm pc}$ on $N_{\rm t}=4$ and 6 lattices.

We extend the study to $2+1$ flavor QCD.
By adopting the the fixed-scale approach,
we use zero-temperature configurations generated by the CP-PACS+JLQCD Collaborations \cite{CP-PACS-JLQCD,Ishikawa:2007nn,Aoki:2005et}. 
Our lattice action consists of the RG-improved Iwasaki gauge action\cite{Iwasaki} and the clover-improved Wilson quark action\cite{SW}:
\begin{eqnarray}
S &=& S_g + S_q
\label{eq:gqaction}\\
S_g &=& -\beta \sum_{x} \left\{ 
\sum_{\mu>\nu}c_0W^{1\times 1}_{x,\mu\nu}
+\sum_{\mu,\nu}c_1W^{1\times 2}_{x,\mu\nu}
\right\},
\label{eq:gaction}\\
S_q &=& \sum_{f=u,d,s}\sum_{x,y} \bar{q}_x^f M_{xy}^f q_y^f, 
\label{eq:qaction}\\
M_{xy}^f &=& \delta_{x,y}-\kappa_f
\sum_\mu \left\{ (1-\gamma_\mu)\,U_{x,\mu}\delta_{x+\hat{\mu},y} + 
(1+\gamma_\mu)\,U^\dagger_{x-\hat{\mu},\mu}\delta_{x-\hat{\mu},y}
\right\}
\nonumber\\
&& - \; \delta_{x,y}\, c_{\rm SW}(\beta)\, \kappa_f\sum_{\mu>\nu}\sigma_{\mu\nu}
F_{x,\mu\nu}
\end{eqnarray} 
where $c_1=-0.331$ and $c_0=1-8c_1$.
We set $\kappa_u = \kappa_d \equiv \kappa_{ud}$ and adopt 
the clover coefficient $c_{\rm SW}$ nonperturbatively determined by the Schr\"{o}dinger functional method in \citen{Aoki:2005et}. 
$F_{x,\mu\nu} = (f_{x,\mu\nu}-f^{\dagger}_{x,\mu\nu})/8i$ is the lattice field strength,
with $f_{x,\mu\nu}$ the standard clover-shaped combination of gauge links.
Hadronic properties with this action have been studied down to the physical point by the CP-PACS, JLQCD and PACS-CS Collaborations \cite{CP-PACS-JLQCD, Ishikawa:2007nn, Aoki:2005et, Aoki:2009sf, Aoki:2009pp, Ishikawa:2009su, Aoki:2009ix}.

As the first determination of the EOS with Wilson-type quarks in 2+1 flavor QCD, 
we study at $(\beta,\kappa_{ud},\kappa_{s})=(2.05,0.1356,0.1351)$, 
that corresponds to the smallest lattice spacing and the lightest $u$ and $d$ quark masses ($m_\pi/m_\rho\simeq0.63$) among the zero-temperature configurations generated by the CP-PACS+JLQCD Collaborations\cite{CP-PACS-JLQCD,Ishikawa:2007nn}.
The $s$ quark mass is set around its physical point ($m_{\eta_{ss}}/m_{\phi}\simeq0.74$). 
These $u$ and $d$ quark masses are much larger than their physical values yet.
A study at the physical point \cite{Aoki:2009ix} is reserved for the next step.
The hadronic radius at this simulation point is estimated to be $r_0/a=7.06(3)$ \cite{Maezawa:2009di}.
Setting the lattice scale by $r_0=0.5$ fm, we estimate $1/a\simeq 2.78$ GeV ($a \simeq 0.07$fm).
The lattice size is $28^3 \times 56$ with $N_{\rm s} a \simeq 2$ fm.

\subsection{Beta functions}

   \begin{figure}[b,t]
       \centerline{
    \includegraphics[width=65mm]{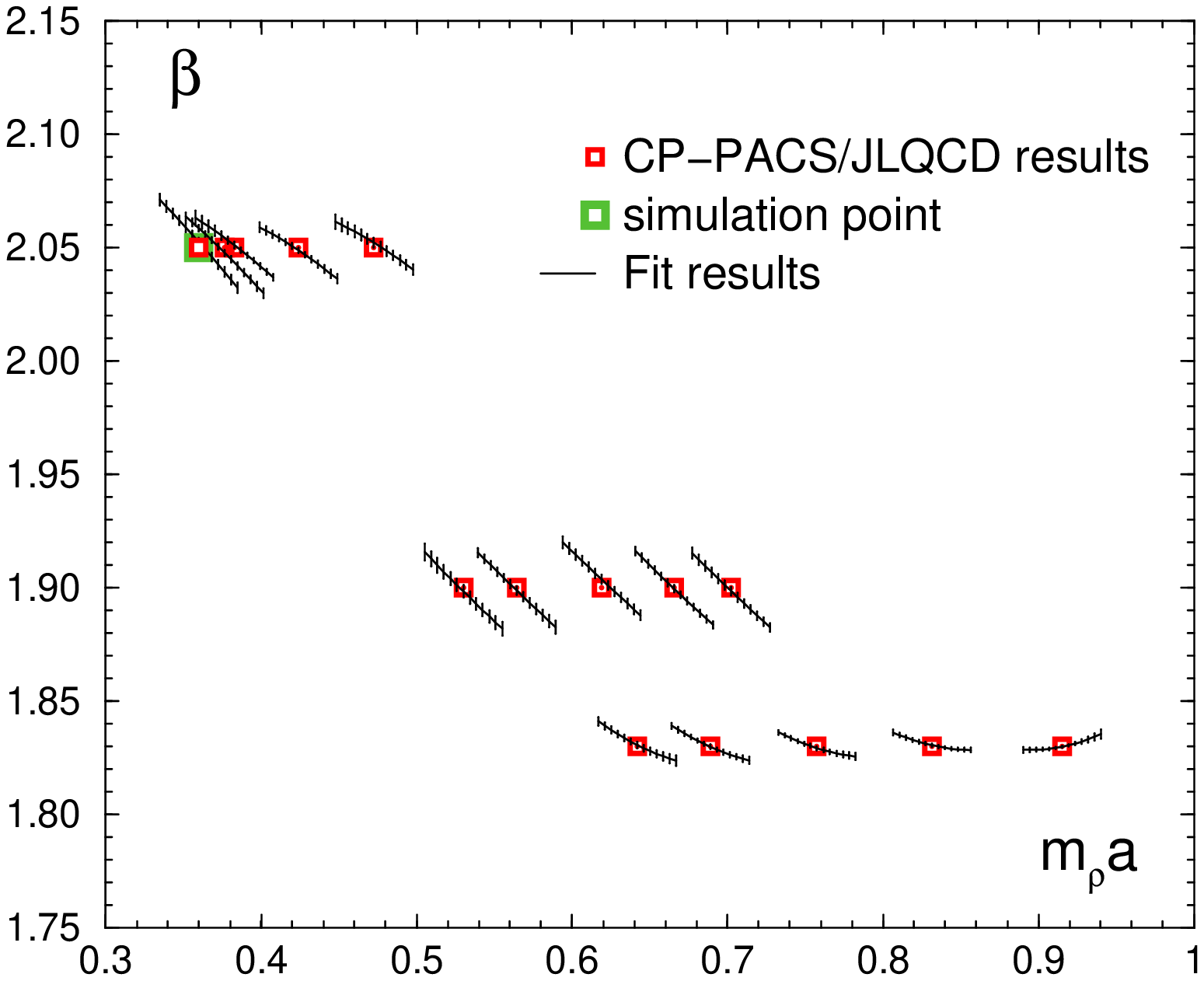}
    \hspace{3mm}
    \includegraphics[width=65mm]{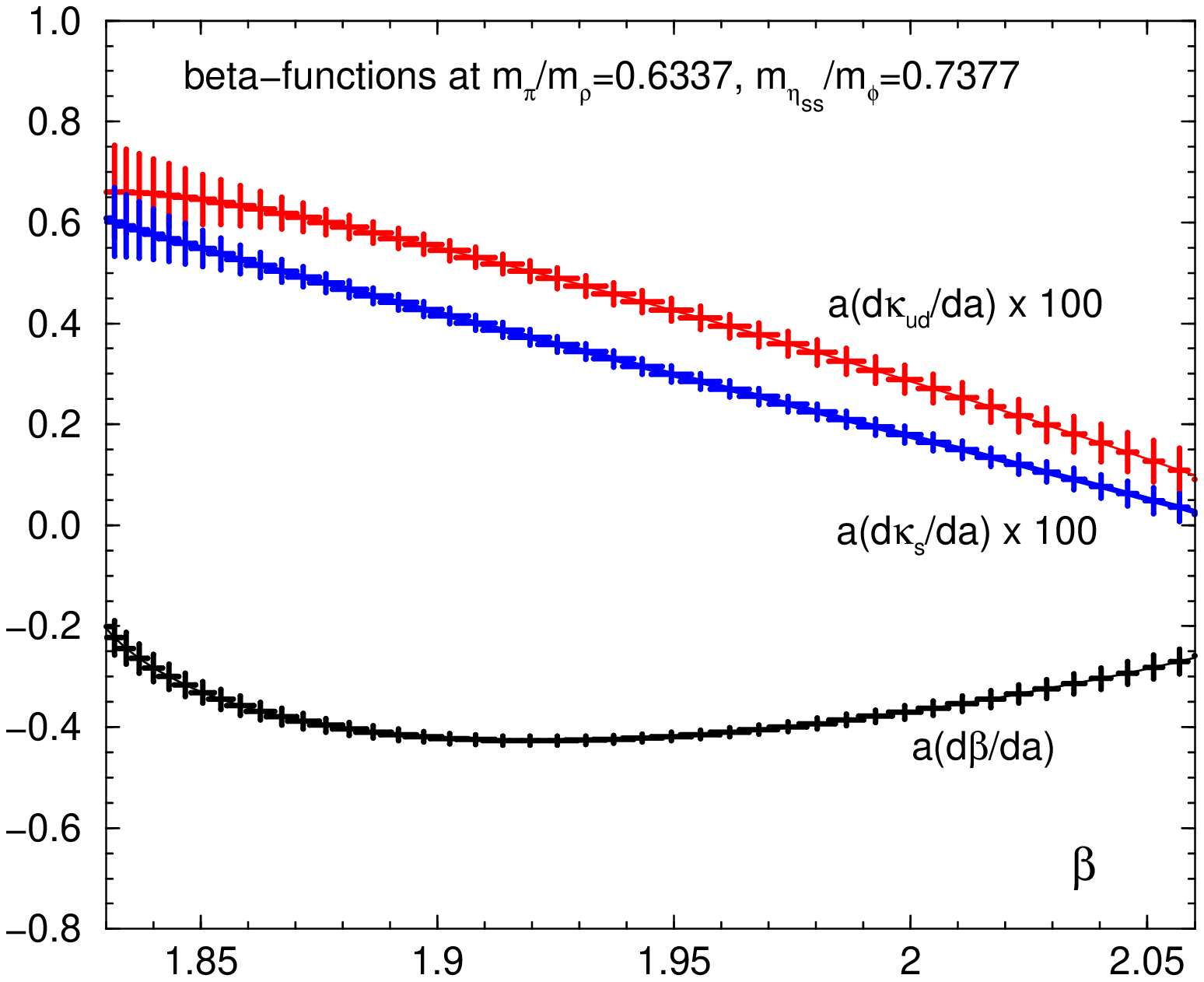}
       }
   \caption{Determination of the beta functions in 2+1 flavor QCD\cite{WEOS12}. 
    {\em Left}: 
    The global fit for $\beta$ as a function of $m_\rho a$ with corresponding 
    $m_\rho/m_\pi$ and $m_{\eta_{ss}}/m_\phi$.
    Square symbols show coupling parameters in the CP-PACS+JLQCD study. 
    To avoid a too busy plot, only half of the data points are shown.
    {\em Right}: 
     Beta functions on our LCP, $m_\pi/m_\rho=0.6337$ and $m_{\eta_{ss}}/m_{\phi}=0.7377$, as functions of $\beta$. The scale setting is made with $am_\rho$. 
    Beta functions for $\kappa_{ud}$ and $\kappa_{s}$ are magnified by factor 100.
   }
   \label{fig:bfunc}
   \end{figure}

Using the same values of the coupling parameters as the zero-temperature simulation, we have generated finite temperature configurations on $32^3\times N_{\rm t}$ lattices with $N_{\rm t}=4$, 6, $\cdots$ 16 \cite{Umeda:2010ye,WEOS12}.
This range of $N_{\rm t}$ corresponds to $T\simeq 170$--700 MeV. 

To evaluate the trace anomaly using (\ref{eqn:e3p}), we need the beta functions $a(d\beta/da)$ and $a(d\kappa_f/da)$ ($f=ud$ and $s$).
These beta functions can be determined nonperturbatively through the coupling parameter dependence of zero-temperature observables.
We use the data of $am_\rho$, $m_\pi/m_\rho$ and $m_{\eta_{ss}}/m_\phi$ at 30 simulation points of the CP-PACS+JLQCD zero-temperature configurations \cite{CP-PACS-JLQCD,Ishikawa:2007nn} to extract the three beta functions.
The first observable $am_\rho$ sets the scale.
A naive method to obtain the beta functions is to fit the data of these observables as functions of the coupling parameters ($\beta$, $\kappa_{ud}$, $\kappa_{s}$), and invert the matrix of the slopes
$ \partial (am_\rho) / \partial\beta $ etc.
However, because $a(d\kappa_f/da)$'s are numerically much smaller than $a(d\beta/da)$, the formers get large relative errors through the matrix inversion procedure, i.e., errors for large components contaminate and dominate the errors for small components.
On the other hand, from the previous experience of two-flavor QCD with improved Wilson quarks in the fixed-$N_{\rm t}$ approach \cite{CPPACS01}, we expect that, although $a(d\kappa_f/da)$'s are small, the quark contribution is important in the trace anomaly.
Therefore, a precise determination of $a(d\kappa_f/da)$'s is required.

To avoid the matrix inversion procedure, we instead fit the coupling parameters $\vec{b} = (\beta, \kappa_{ud}, \kappa_{s})$ as functions of three observables $am_\rho$, $m_\pi/m_\rho \equiv X$ and $m_{\eta_{ss}}/m_\phi \equiv Y$. 
Consulting the overall quality of the fits, we adopt the following third order polynomial function of the observables in this study: 
\begin{eqnarray}
\vec{b} &=& \vec{c}_0 + \vec{c}_1\,(am_\rho) + \vec{c}_2\,(am_\rho)^2
+ \vec{c}_3X
+ \vec{c}_4X^2
+ \vec{c}_5 \,(am_\rho)X
+ \vec{c}_6Y
+ \vec{c}_7Y^2
\nonumber\\
&&+\; \vec{c}_8\,(am_\rho)Y
+ \vec{c}_9X Y
+ \vec{c}_{10}\,(am_\rho)^3
+ \vec{c}_{11}X^3
+ \vec{c}_{12}Y^3
+ \vec{c}_{13} \,(am_\rho)X^2
\nonumber\\
&&+\;  \vec{c}_{14}\,(am_\rho)^2X
+ \vec{c}_{15}\,(am_\rho)Y^2
+ \vec{c}_{16}\,(am_\rho)^2Y 
+ \vec{c}_{17}X Y^2
+ \vec{c}_{18}X^2 Y
\nonumber\\
&&+\; \vec{c}_{19}\,(am_\rho)XY .
\label{eq:beta}
\end{eqnarray}
The global fits for each component of $\vec{b}$ are independent and have dof $=10$.
We find reasonable $\chi^2/$dof of ${\cal O}(1)$\footnote{
We find $\chi^2/{\rm dof}=1.63$, 1.08, and 1.69 for the fit of $\beta$, $\kappa_{ud}$, and $\kappa_s$, respectively.
Here, $\chi^2$ is evaluated using a standard deviation of each coupling parameter estimated by the error propagation rule using the errors of the observables and the partial derivatives of the resulting fitting function (\ref{eq:beta}) with respect to the observables, neglecting the covariance among the observables.}.
The result for $\beta$ is shown in the left panel of Fig.~\ref{fig:bfunc}.

In this study, we define LCP's by $m_\pi/m_\rho$ and $m_{\eta_{ss}}/m_\phi$ at $T=0$.
Therefore, the beta functions are extracted as 
$a\,d \beta/d a = (a m_\rho)\, \partial \beta/\partial (a m_\rho)$ etc., in terms of the coefficients 
$\vec{c}_1$, $\vec{c}_2$, $\vec{c}_5$, $\vec{c}_8$, $\cdots$ in (\ref{eq:beta}).
The resulting beta functions for our LCP  ($m_\pi/m_\rho=0.6337$, $m_{\eta_{ss}}/m_{\phi}=0.7377$) are shown in the right panel of Fig.~\ref{fig:bfunc} as functions of $\beta$.
Their values at $\beta=2.05$ are used to determine the trace anomaly.
As the variable to set the scale, we may alternatively adopt $am_\pi$, $am_{K}$ or $am_{K^*}$ instead of $am_\rho$ in (\ref{eq:beta}).
We use this freedom to estimate a systematic error.
Taking the results from $am_\rho$ as the central value, we obtain 
\begin{eqnarray}
a\frac{d \beta}{d a} = -0.279(24)(^{+40}_{-64})
,\hspace{2mm}
a\frac{d \kappa_{ud}}{d a} = 0.00123(41)(^{+56}_{-68})
,\hspace{2mm}
a\frac{d \kappa_s}{d a} \,=\, 0.00046(26)(^{+42}_{-44})
\nonumber\\
\end{eqnarray}
at our simulation point,
where the first brackets are for statistic errors, and the second brackets are for the systematic error\cite{WEOS12}.

\subsection{Equation of state}

   \begin{figure}[b,t]
       \centerline{
    \includegraphics[width=80mm]{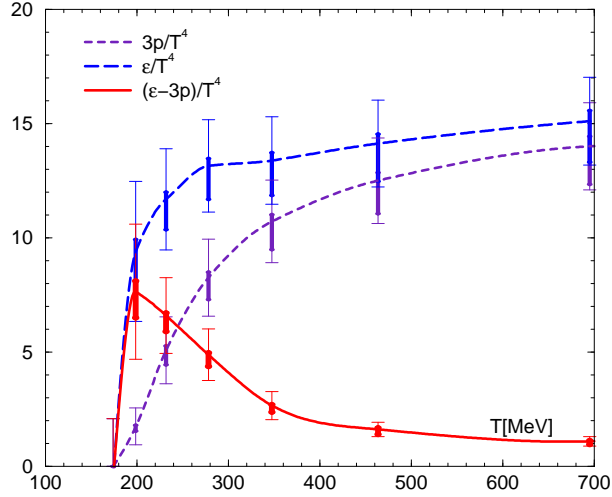}
       }
   \caption{
    Trace anomaly $(\epsilon-3p)/T^4$, energy density $\epsilon/T^4$,
    and pressure $3p/T^4$ in 2+1 flavor QCD\cite{WEOS12}. 
    The thin and thick vertical bars represent statistic and systematic errors, respectively.
    The curves are drawn by the Akima spline interpolation.
    }
   \label{fig:Nf21eos}
   \end{figure}

We now calculate the trace anomaly by (\ref{eqn:e3p}):
\begin{eqnarray}
\frac{\epsilon-3p}{T^4}&=& 
\frac{N_{\rm t}^3}{N_{\rm s}^3}
\left\{
{a\frac{d\beta}{da}}
\left\langle
\frac{\partial S}{\partial\beta}
\right\rangle_{\rm \! sub}
+
{a\frac{d\kappa_{ud}}{da}}
\left\langle
\frac{\partial S}{\partial \kappa_{ud}}
\right\rangle_{\rm \! sub}
+{a\frac{d\kappa_s}{da}}
\left\langle
\frac{\partial S}{\partial \kappa_s}
\right\rangle_{\rm \! sub}
\right\}
\label{eq:tranom}
\end{eqnarray}
with
\begin{eqnarray}
\left\langle 
\frac{\partial S}{\partial \beta} 
\right\rangle_{\rm \! sub} &=&
-\left\langle
\sum_{x,\mu>\nu}c_0W^{1\times 1}_{x,\mu\nu}
+\sum_{x,\mu,\nu}c_1W^{1\times 2}_{x,\mu\nu}
\right\rangle_{\rm \! sub} 
\nonumber\\&&
+\; \frac{d c_{SW}}{d \beta}
\sum_{f=u,d,s}{\kappa_f
\left\langle 
\sum_{x,\mu>\nu}\mbox{tr}^{(c,s)}\sigma_{\mu\nu}F_{x,\mu\nu}\,
(M^{f})^{-1}_{x,x}
\right\rangle_{\rm \! sub} },
\label{eq:dsdb}
\\
\left\langle 
\frac{\partial S}{\partial \kappa_f} 
\right\rangle_{\rm \! sub} &=&
c_f \times
\left(
\left\langle
\sum_{x,\mu}\mbox{tr}^{(c,s)}
\left\{(1-\gamma_\mu)U_{x,\mu}(M^{f})^{-1}_{x+\hat{\mu},x}
\right. \right. \right.
\nonumber\\
&&
\left. \left. \left.
\hspace{32mm} +\;(1+\gamma_\mu)U^\dagger_{x-\hat{\mu},\mu}
(M^{f})^{-1}_{x-\hat{\mu},x}
\right\}
\right\rangle_{\rm \! sub} 
\right.
\nonumber\\&&
\left.
\hspace{12mm}  +\; c_{SW}
\left\langle
\sum_{x,\mu>\nu}\mbox{tr}^{(c,s)}\sigma_{\mu\nu}F_{x,\mu\nu}\,
(M^{f})^{-1}_{x,x}
\right\rangle_{\rm \! sub}
\right),
\label{eq:dsdk}
\end{eqnarray}
where $c_f=2$ for $f=ud$ and 1 for $f=s$.
We evaluate the spatial traces in (\ref{eq:dsdb}) and (\ref{eq:dsdk}) by the random noise method with U(1) random numbers \cite{WHOT10dense}.
The number of noise is 1 for each of the color and spinor indices.

Our result for the trace anomaly $(\epsilon-3p)/T^4$ is shown by a red thick curve in Fig.~\ref{fig:Nf21eos}.
We note that the peak height of about 7 at $T=199$ MeV ($N_{\rm t} = 14$) is roughly consistent with recent results of highly improved staggered quarks obtained with the fixed $N_{\rm t}$ approach at $N_{\rm t} = 6$--12 \cite{Borsanyi:2010cj,Bazavov:2010sb}.
The shape of $(\epsilon-3p)/T^4$ suggests that $T_{\rm pc}$ locates between 174 and 199 MeV.

Carrying out the $T$-integration (\ref{eq:Tintegral}) of the trace anomaly, we obtain the pressure $p$ shown in Fig.~\ref{fig:Nf21eos}.
The energy density $\epsilon$ is calculated from $p$ and $\epsilon -3p$.

Besides the larger errors, our EOS looks roughly consistent with recent results with highly improved staggered quarks near the physical point\cite{Borsanyi:2010cj,Bazavov:2010sb}.
We note that our peak is slightly higher.
This is consistent with the fact that our light quark masses are heavier than their physical values: 
the experience with improved staggered quarks suggests that the peak becomes slightly higher as the light quark masses are increased [see, e.g., Ref.~\citen{Borsanyi:2010cj}].

Our final goal is to extend the study towards the physical point, adopting the on-the-physical-point configurations generated by the PACS-CS Collaboration\cite{Aoki:2009ix}. 
From the present study at $m_{ud}$ heavier than the physical point, we encountered a couple of issues:
Errors of EOS in Fig.~\ref{fig:Nf21eos} are larger than those obtained with the fixed-$N_{\rm t}$ approach\cite{Borsanyi:2010cj,Bazavov:2010sb}.
Besides smaller statistics, this is due to the large statistical error in $(\epsilon-3p)/T^4$ at $T\,\lsim\,200$ MeV, which is caused by the enhancement factor $N_{\rm t}^4$ in (\ref{eq:tranom}) (note that $S$ is proportional to $N_{\rm t}$).
To obtain accurate EOS at low temperatures, we need a large statistics of ${\cal O}(N_{\rm t}^7)$\footnote{A power of $N_{\rm t}$ is reduced by the averaging over the lattice sites.}. 
This is, however, an unavoidable step to suppress discretization errors, and the issue is common with the fixed-$N_{\rm t}$ approach.
Another source of systematic errors in EOS is the limited resolution in $T$ due to the discrete variation of $N_{\rm t}$ in the fixed-scale approach.
Note that the lattice spacings in full QCD studies are usually coarser than those in quenched studies.
Furthermore, in the present study, $N_{\rm t}$ is limited to be even due to the simulation program set we have adopted.
To improve the resolution in $T$, we need simulations at odd values of $N_{\rm t}$ and/or a finer lattice spacing $a$.
An alternative way will be to combine results at different $a$'s, since we can choose $a$'s fine with the fixed-scale approach and thus, after confirming that the discretization effects are sufficiently small in the observables under study, we may combine the results at different $a$'s to more smoothly interpolate in $T$.
We leave these trials to a forthcoming study with lighter quarks.

\section{Finite density QCD on the lattice}
\label{sec:FDLQCD}

Next, let us move on to the issues of finite density QCD.
We consider the action given by (\ref{eq:gqaction})--(\ref{eq:qaction}) with the quark matrix $M_{xy}^f$ replaced by
\begin{eqnarray}
M_{xy}^f &=& \delta_{x,y}-\kappa_f
\sum_{\mu=1}^{3} \left\{ (1-\gamma_\mu)\,U_{x,\mu}\delta_{x+\hat{\mu},y} + 
(1+\gamma_\mu)\,U^\dagger_{x-\hat{\mu},\mu}\delta_{x-\hat{\mu},y}
\right\}
\nonumber\\
&& -\kappa_f \left\{ e^{\mu_f a} (1-\gamma_4)\,U_{x,4}\delta_{x+\hat{4},y} + 
e^{- \mu_f a} (1+\gamma_4)\,U^\dagger_{x-\hat{4},4}\delta_{x-\hat{4},y}
\right\}
\nonumber\\
&& - \; \delta_{x,y}\, c_{\rm SW}(\beta)\, \kappa_f\sum_{\mu>\nu}\sigma_{\mu\nu}
F_{x,\mu\nu}
\label{eq:Mmu}
\end{eqnarray}
Here, the theory at $\mu_f\ne0$ is known to have a serious problem:
In a Monte Carlo simulation, we generate configurations of link variables 
$\{ U_{x,\mu} \}$ with the probability in proportion to the Boltzmann weight  
$(\prod_f \det M^f)\, e^{-S_g}$. 
The expectation value of an operator ${\cal O}[U]$ is then evaluated 
as an average of ${\cal O}$ over the configurations, 
\begin{eqnarray}
\langle {\cal O} \rangle_{(\beta,\vec\kappa,\vec\mu)}
\approx \frac{1}{N_{\rm conf.}} \sum_{ \{ U_{x,\mu} \} } {\cal O}[U],
\end{eqnarray}
where $\vec\kappa = (\kappa_u,\kappa_d,\cdots)$ and $\vec\mu=(\mu_u,\mu_d,\cdots)$.
At $\vec\mu=0$, the quark determinants are real due to the $\gamma_5$-hermiticity of the quark matrices, $(M^f)^\dagger = \gamma_5 \, M^f \,\gamma_5 $.
However, when $\mu_f\ne0$, the $\gamma_5$-hermiticity relation changes to 
\begin{eqnarray}
\left[ M^f(\mu_f) \right]^\dagger = \gamma_5 \, M^f(-\mu_f) \,\gamma_5 ,
\label{eq:gamma5conj}
\end{eqnarray}
and thus $\det M^f$ becomes complex unless $\mu_f=0$.
Because the Boltzmann weight has to be real and positive, 
we cannot perform a Monte Carlo simulation at $\mu_f\ne0$ directly.  

Various methods have been proposed to study finite density QCD avoiding the complex weight problem.
However, at present, all of them are applicable in small $\mu_f/T$ regions only.
In the following subsections, we introduce these methods, mainly focusing on the Taylor expansion and reweighting methods we adopt.

\subsection{Taylor expansion method}
\label{sec:taylor}

The simplest approach to study finite density QCD avoiding the complex weight problem is the Taylor expansion method, in which physical quantities are expended in terms of $\mu_f/T$ around $\vec\mu=0$\cite{BS02,Miya02,GG03}.
For example, the pressure $p = (T/V) \ln Z$ is expanded as
\begin{eqnarray}
\frac{p}{T^4} &=&
\sum_{i,j,k=0}^\infty c_{ijk}(T) \left(\frac{\mu_u}{T}\right)^i
\left(\frac{\mu_d}{T}\right)^j \left(\frac{\mu_s}{T}\right)^k,
\label{eq:p}
\\
c_{ijk} &=& 
\frac{1}{i!j!k!} \frac{1}{VT^3} \left.
\frac{\partial^{i+j+k} \ln Z }{\partial(\mu_u/T)^i
\partial(\mu_d/T)^j \partial(\mu_s/T)^k} \right|_{\vec\mu=0}.
\label{eq:cijk}
\end{eqnarray}
in three-flavor QCD.
The Taylor coefficients $c_{ijk}$ can be evaluated by a conventional Monte Carlo simulation at $\vec\mu=0$ which is free from the complex weight problem.
%
We expect that QCD in the high temperature limit is well described by a free gas of quarks and gluons, in which $p$ consists of terms proportional to $\mu_f^2$ and $\mu_f^4$ only.
Therefore, the expansion will converge well in the high temperature region.

Other observables can also be calculated similarly.
The quark number density $n_f$ is given by
\begin{eqnarray}
\frac{n_f}{T^3} = 
\frac{1}{VT^3} \frac{\partial \ln Z}{\partial (\mu_f/T)} 
= \frac{\partial (p/T^4)}{\partial (\mu_f/T)} ,
\label{eq:qnd}
\end{eqnarray}
where $T$ is fixed in the differentiations.
When we define the light quark number density as $n_q=n_u+n_d$, 
the susceptibilities of the light quark number density $(\chi_q)$ and the strange quark number density $(\chi_s)$ are given by
\begin{eqnarray}
\frac{\chi_q}{T^2} 
= \left( \frac{\partial}{\partial (\mu_u/T)} 
+ \frac{\partial}{\partial (\mu_d/T)} \right) 
\frac{n_u + n_d}{T^3}
, \hspace{5mm} 
\frac{\chi_s}{T^2} 
= \frac{\partial (n_s/T^3)}{\partial (\mu_s/T)} ,
\end{eqnarray}
\label{eq:chiq}
respectively.
The susceptibility of the isospin number can also be given as 
\begin{eqnarray}
\frac{\chi_I}{T^2} 
= \left( \frac{\partial}{\partial (\mu_u/T)} 
- \frac{\partial}{\partial (\mu_d/T)} \right) 
\frac{n_u - n_d}{T^3} .
\label{eq:chii}
\end{eqnarray}
These quantities are expanded around $\vec\mu=0$ in terms of $c_{ijk}$ defined in (\ref{eq:cijk}).
The trace anomaly is given by Eq.~(\ref{eqn:e3p}).
The entropy density is given by 
$ 
s = T^{-1} 
\left(\varepsilon +p - \sum_{f} \mu_f n_f \right).
$ 
The chiral condensate is defined by the derivative of 
$\ln Z$ with respect to the quark mass.
Taylor expansion of them can also be derived.

\subsection{Reweighting method and sign problem}

Another popular approach to finite density QCD is the reweighting method\cite{FK02,Gibbs86,Hase92,Bar97}, adopting the reweighting technique\cite{McDonaldSinger,Ferrenberg:1988yz} to finite density QCD. 
Using the configurations generated at $\vec\mu=0$, expectation values at finite $\vec\mu$ are computed by  correcting the Boltzmann weight with the ``reweighting factor'':
\begin{eqnarray}
\langle {\cal O} \rangle_{(\beta, \vec\kappa, \vec\mu)}
= \frac{\left\langle {\cal O} \times
\prod_f \left[ \det M^f(\mu_f) / \det M^f(0) \right]
\right\rangle_{\! (\beta,\vec\kappa,0)}}{ \left\langle
\prod_f \left[ \det M^f(\mu_f) / \det M^f(0) \right]
\right\rangle_{\! (\beta,\vec\kappa,0)}}. 
\label{eq:murew}
\end{eqnarray} 
The denominator is the ratio of the partition functions at finite $\vec\mu$ and $\vec\mu=0$, 
\begin{eqnarray}
\frac{Z(\beta,\vec\kappa,\vec\mu)}{Z(\beta,\vec\kappa,0)}
=
\left\langle \prod_f \frac{\det M^f(\mu_f)}{\det M^f(0)} \right\rangle_{\! (\beta,\vec\kappa,0)}.
\label{eq:parmu}
\end{eqnarray} 
Here, because $\det M^f(\mu_f)$ is complex, 
the reweighting factor 
\[
\prod_f \frac{\det M^f(\mu_f)}{\det M^f(0) }
\]
has a complex phase $e^{i \hat\theta}$.
When the fluctuation of $\hat\theta$ is larger than ${\cal O}(\pi/2)$,
a reliable calculation of the numerator and denominator in Eq.~(\ref{eq:murew}) is difficult. 
This difficulty is called the ``sign problem (complex phase problem)''.
We actually encounter large fluctuations of $\hat\theta$ at large $\mu_f$ and/or large lattice volume. 

It is worth rewriting the denominator of Eq.~(\ref{eq:murew}) in terms of the distribution function for 
$\hat{P} \equiv -S_g/ (6N_{\rm site} \beta)$  (which is the plaquette if $S_g$ is the standard gauge action) 
and the logarithm of the absolute value of the reweighting factor 
$\hat{F} \equiv \ln \left| \prod_f \left(\det M^f(\mu_f)/ \det M^f(0)\right) \right|$:
\begin{eqnarray}
\left\langle
\prod_f  \frac{\det M^f(\mu_f)}{\det M^f(0)}  \right\rangle_{\!\! (\beta,\vec\kappa,0)} =
\int \left\langle e^{i \hat\theta} \right\rangle_{\! P,F} \, 
e^F w_0(P,F; \beta, \vec\kappa, 0) \, dP dF,
\label{eq:denorew}
\end{eqnarray} 
where $w_0$ is the distribution function for the phase-quenched system,
\begin{eqnarray}
w_0(P,F; \beta, \vec\kappa, \vec\mu) 
= \int \! {\cal D}U\, \delta(P-\hat{P}[U]) \delta({F-\hat{F}[U]}) \left|\prod_f \det M^f(\mu_f)\right| e^{6N_{\rm site} \beta \hat{P}} , 
\nonumber\\
\label{eq:nophaseZ}
\end{eqnarray} 
and 
$\langle e^{i \hat\theta} \rangle_{P,F}$ is the expectation value of the operator $e^{i \hat\theta}$ at $\vec\mu=0$  
with fixing the values of $\hat{P}$ and $\hat{F}$ to $P$ and $F$:
\begin{eqnarray}
\left\langle e^{i \hat\theta} \right\rangle_{\! P,F} = 
\frac{\left\langle \delta(P-\hat{P})\, \delta({F-\hat{F}}) \,
e^{i \hat\theta} \right\rangle_{\!(\beta,\vec\kappa,0)}}{
\left\langle \delta(P-\hat{P}) \,\delta({F-\hat{F}}) 
\right\rangle_{\!(\beta,\vec\kappa,0)}}.
\end{eqnarray} 

By measuring the histogram of $\hat{P}$ and $\hat{F}$ in a phase-quenched simulation, 
we can determine $w_0$ around the peak of the histogram. 
However, in the calculation of Eq.~(\ref{eq:denorew}), precise information of $w_0$ is required around the maximum of the integrant, which may deviate from the peak of $w_0$ due to the factor $\langle e^{i \hat\theta} \rangle_{\! P,F} \,e^F$. 
To calculate $w_0$ in a wider range of $(P,F)$, we combine results of phase-quenched simulations at different points in the coupling parameter space, adopting the reweighting formulae for the phase-quenched theory\cite{WHOT10dense}.
Further demonstration of such calculation will be given in Sect.~\ref{sec:Histogram}.

For Eq.~(\ref{eq:denorew}), we also need to estimate $ \langle e^{i \hat\theta} \rangle_{P,F}$.
Because the total distribution function is real and positive in finite-density QCD due to the charge conjugation symmetry, the imaginary part of $ \langle e^{i \hat\theta} \rangle_{P,F}$ must be averaged out when the statistics is infinite. 
Since the imaginary part is the source of the sign problem, we may remedy or mitigate the problem if we could find a method in which the imaginary part is removed and the real part is reliably estimated.
In the next subsection, we show that 
it is useful to consider a cumulant expansion of $ \langle e^{i \hat\theta} \rangle_{P,F}$ in which $ \ln \langle e^{i \hat\theta} \rangle_{P,F}$ is separated into real and imaginary parts\cite{Ejiri:2007ga,WHOT10dense}.

\subsection{Cumulant expansion and Gaussian approximation}
\label{sec:cumulant}

For simplicity, let us consider the case of $N_{\rm f}$ degenerate quarks in the followings.
A crucial step in handling the fluctuations in the phase $\hat\theta$ is to introduce an appropriate definition of $\hat\theta$ removing the ambiguity of complex phase with modulo $2\pi$.
We uniquely define the complex phase by the Taylor expansion as
\begin{eqnarray}
\hat\theta (\mu) 
& = & N_{\rm f} \, {\rm Im} \,[\ln \det M(\mu)]
\nonumber\\
& \stackrel{\rm def.}{=} & N_{\rm f} \sum_{n=0}^{\infty} \frac{1}{(2n+1)!} {\rm Im}
\left[ \frac{\partial^{2n+1} (\ln \det M(\mu))}{\partial (\mu/T)^{2n+1}} 
\right]_{\mu=0} \left( \frac{\mu}{T} \right)^{2n+1} ,
\label{eq:tatheta}
\end{eqnarray}
where the derivatives of $\ln\det M$ can be unambiguously expressed in terms of $M^{-1}$ and derivatives of $M$.
Note that the expectation value of $\hat\theta$ defined by Eq.~(\ref{eq:tatheta}) is not restricted to the range $(-\pi,\pi)$, and the maximum value of $|\hat\theta|$ is proportional to the volume of the system. 
The conventional complex phase in the range $(-\pi,\pi)$ is recovered by taking the principal value of $\hat\theta$ with modulo $2\pi$.

\begin{figure}[t]
\begin{center}
\includegraphics[width=68mm]{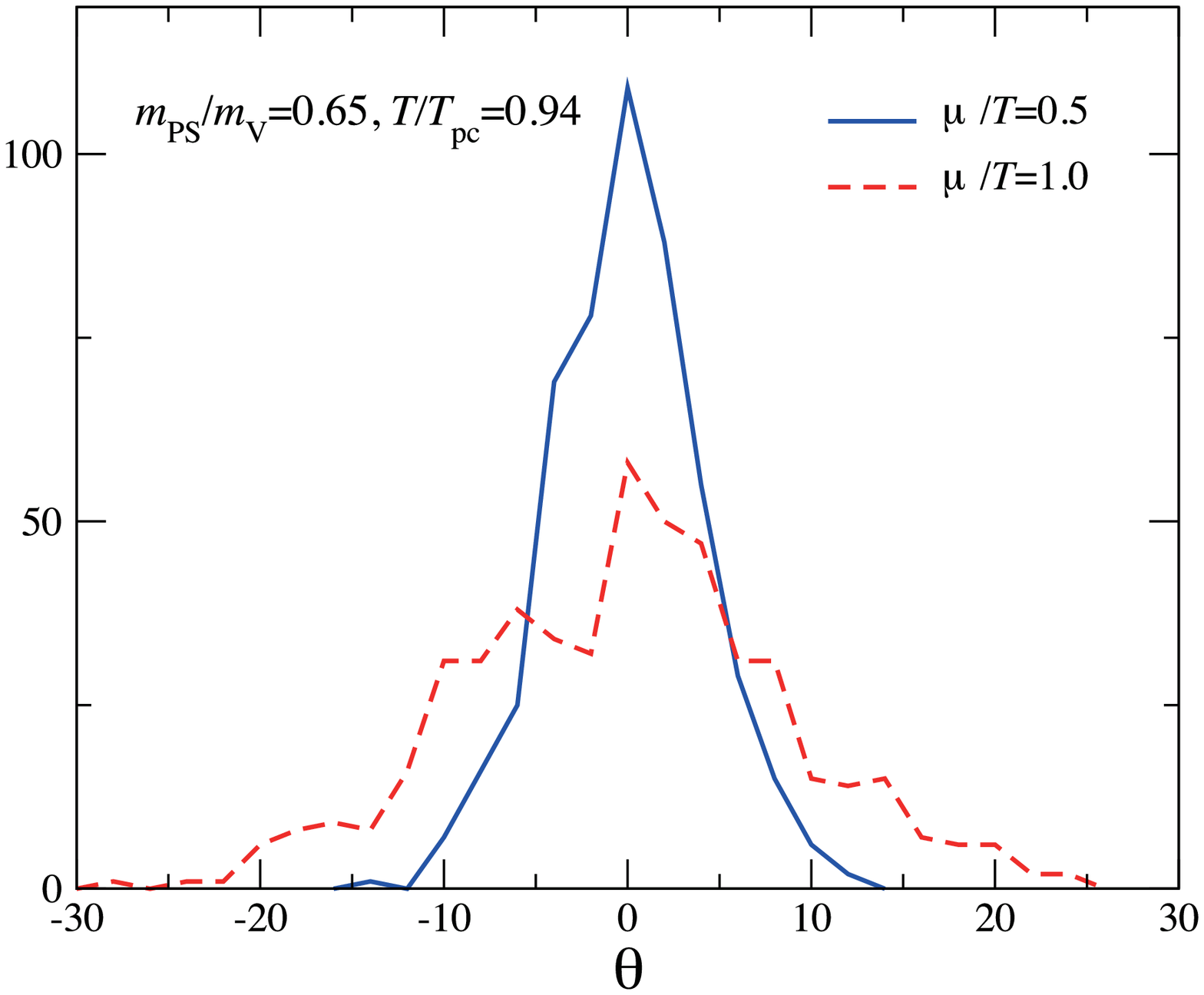}
\hspace{1mm}
\includegraphics[width=68mm]{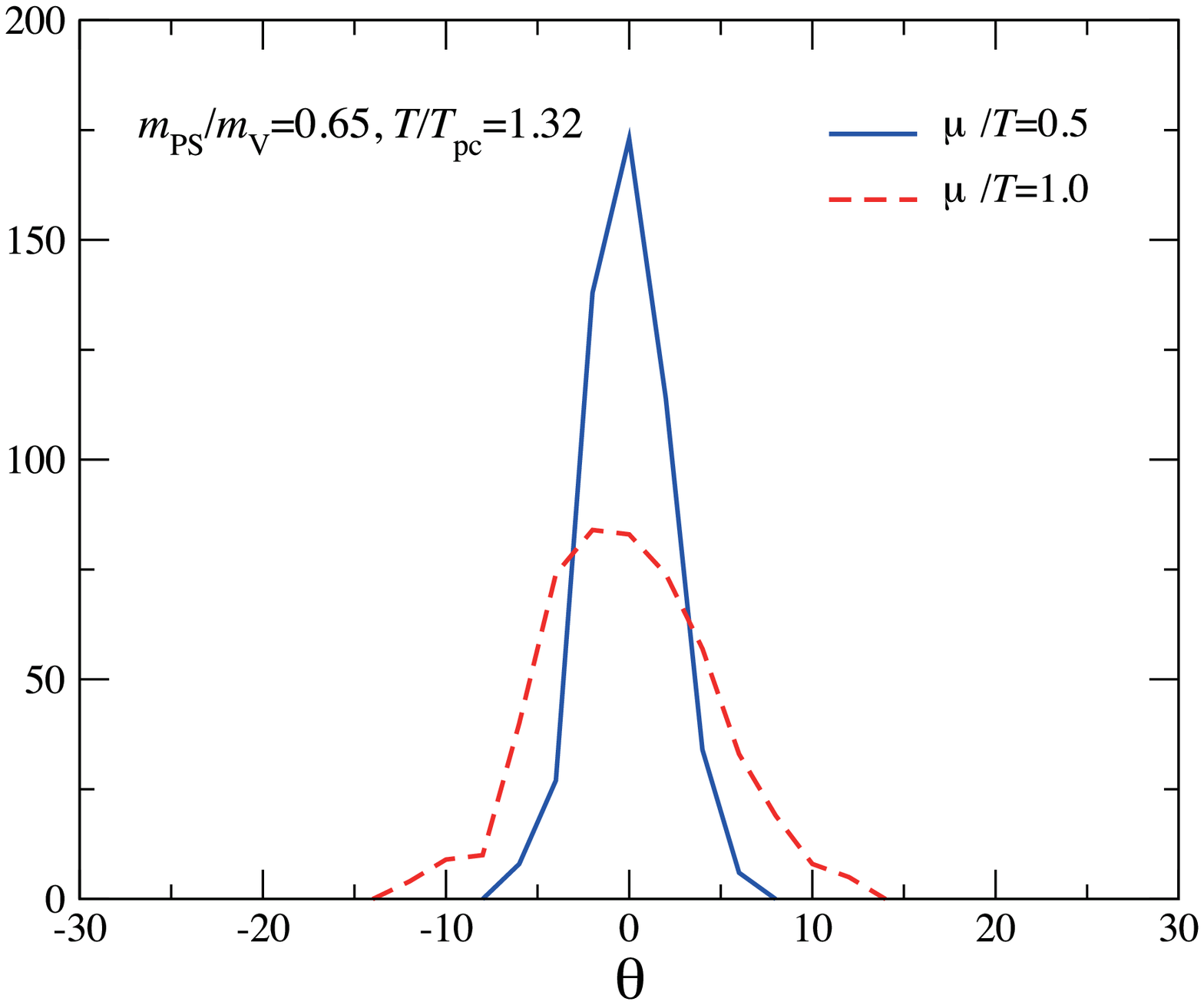}
\caption{The histogram of $\hat\theta$ in $N_{\rm f}=2$ QCD with improved Wilson quarks 
at $(m_{\pi}/m_{\rho}, T/T_{\rm pc})=(0.65, 0.94)$ (left) and 
$(0.65, 1.32)$ (right)\cite{WHOT10dense}.}
\label{fig:phhis}
\end{center}
\end{figure} 

Typical results for the histogram of $\hat\theta$ are shown in Fig.~\ref{fig:phhis} for $N_{\rm f}=2$ QCD with improved Wilson quarks\cite{WHOT10dense}, where the power series in (\ref{eq:tatheta}) is evaluated up to $(\mu/T)^3$ discarding ${\cal O}\left( (\mu/T)^5 \right)$ terms\footnote{In two-flavor QCD with p4 improved staggered quarks\cite{Ejiri:2007ga}, influence of the next order term was shown to be small at $\mu/T \lsim 2.5$. See, e.g., Fig.~9 of Ref.~\citen{Ejiri:2007ga}.}. 
We see that the width of the distribution becomes wider as $\mu/T$ is increased, indicating a severer sign problem.

To mitigate the sign problem, we evaluate $ \langle e^{i \hat\theta} \rangle_{P,F}$ by the cumulant expansion\cite{Ejiri:2007ga}: 
\begin{eqnarray}
\left\langle e^{i \hat\theta} \right\rangle_{\! P,F} \; = \;
\exp \left[i \langle \hat\theta \rangle_c
- \frac{\langle \hat\theta^2 \rangle_c}{2} 
- \frac{i \langle \hat\theta^3 \rangle_c}{3!} 
+ \frac{\langle \hat\theta^4 \rangle_c }{4!} 
+ \frac{i \langle \hat\theta^5 \rangle_c }{5!} 
- \frac{\langle \hat\theta^6 \rangle_c}{6!} + \cdots \right],
\nonumber\\
\label{eq:cum}
\end{eqnarray}
where $\langle \hat\theta^n \rangle_c$ is the $n^{\rm th}$ order cumulant:
$
\langle \hat\theta^2 \rangle_c 
= \langle \hat\theta^2 \rangle_{\! P,F} 
$, 
$
\langle \hat\theta^4 \rangle_c 
= \langle \hat\theta^4 \rangle_{\! P,F}
-3 \langle \hat\theta^2 \rangle_{\! P,F}^2
$, 
$
\langle \hat\theta^6 \rangle_c 
= \langle \hat\theta^6 \rangle_{\! P,F}
-15 \langle \hat\theta^4 \rangle_{\! P,F} 
\langle \hat\theta^2 \rangle_{\! P,F} 
+30 \langle \hat\theta^2 \rangle_{\! P,F}^3 
$, $\cdots$.
A key observation in handling the cumulant expansion is that $\langle \hat\theta^n \rangle_c =0$ for any odd $n$
due to the symmetry under $\hat\theta \rightarrow -\hat\theta$. 
This implies that, provided that the cumulant expansion converges, 
$\langle e^{i \hat\theta} \rangle_{P,F}$ is guaranteed to be real and positive and the sign problem is resolved.
The sign problem is transformed into the convergence problem of the cumulant expansion in this approach.

As shown in Fig.~\ref{fig:phhis}, we find that the distribution of $\hat\theta$ is well described by a Gaussian function up to $\mu/T \sim {\cal O}(1)$.
The Gaussian distribution of $\hat\theta$ is observed also with improved staggered quarks in two-flavor QCD\cite{Ejiri:2007ga}, 
and was discussed in Ref.~\citen{splitt} too.
The Gaussian distribution means that the cumulant expansion (\ref{eq:cum}) is dominated by the lowest non-trivial order of $\langle\hat\theta^2\rangle$, and thus the expansion is well converged. 
Corrections to the Gaussian distribution can be incorporated by taking higher order terms in the cumulant expansion.

Here, we note that $\hat\theta = {\cal O}(\mu/T)$ from Eq.~(\ref{eq:tatheta}).
Therefore, if we take into account the cumulants up to the $n^{\rm th}$ order, the truncation error does not affect the Taylor expansion up to ${\cal O}\left((\mu/T)^n\right)$.
This means that the convergence of the cumulant expansion is closely related to the convergence of the Taylor expansion in $\mu$.
The Gaussian approximation is valid at least at small $\mu$ and the higher order cumulants may become visible at large $\mu$. 
The applicability range of the Gaussian approximation in $\mu$ has to be checked for each cases by calculating higher order cumulants. 

We now argue that the range of applicability does not change with the system volume on sufficiently large lattices when the correlation length of the system is finite\cite{WHOT10dense}:
The expansion coefficients for $\hat\theta$ in Eq.~(\ref{eq:tatheta}) are given by traces of products of $M^{-1}$'s and $\partial^n M/ \partial (\mu/T)^n$'s over the spatial positions.
For example, the first coefficient is given by the trace of 
$N_{\rm f} [M^{-1} (\partial M/ \partial (\mu/T))]$.
We note that the diagonal elements of this matrix are the local quark-number density operators [$\sim \bar{\psi}(x) \gamma_0 \psi(x)$] at $\mu=0$. 
When the correlation length of the local operators is finite and is much shorter than the system size, we may decompose the trace into a summation of independent contributions from spatially separated regions. 
The same discussion is applicable to higher order coefficients too. 
Then, the phase $\hat\theta$ may be written as a sum of local contributions from spatially separated regions, which are statistically independent with each other, i.e.\ $\hat\theta \approx \sum_x \hat\theta_x$,  
where $\hat\theta_x$ is the contribution from a spatial region labeled by $x$.
The average of $\exp(i\hat\theta)$ is thus 
\begin{eqnarray}
\left\langle e^{i\hat\theta} \right\rangle 
\approx \prod_x \left\langle e^{i\hat\theta_x} \right\rangle 
= \exp \left( \sum_x \sum_n \frac{i^n}{n!} 
\left\langle \hat\theta_x^n \right\rangle_c \right).
\end{eqnarray}
This suggests that all cumulants $\langle \hat\theta^n \rangle_c 
\approx \sum_x \langle \hat\theta_x^n \rangle_c$ 
increase in proportion to the volume as the volume increases, 
that is, $\langle \hat\theta^n \rangle_c \propto$ volume for any $n$, in contrast with a na\"ive expectation that $\hat\theta^n$ may be proportional to (volume)$^n$ since $\hat\theta$ is proportional to the volume.\footnote{
This property of the higher order cumulants can be understood also from the effective potential $V_{\rm eff} (P,F) = - \ln w_0(P,F) - \ln \langle e^{i\hat\theta} \rangle_{(P,F)}$, which will be studied in Sect.~\ref{sec:Histogram}.
Because $V_{\rm eff}$ and $\ln w_0$ are proportional to the system volume, each term in the expansion of $ \ln \langle e^{i\hat\theta} \rangle_{(P,F)}= \sum_n (i^n/n!) \langle \hat\theta^n \rangle_c $ should not increase faster than (volume)$^1$ for any $n$.
Otherwise, $V_{\rm eff}$ becomes singular in the thermodynamic limit. 
}
Therefore, while the width of the distribution, i.e.\ the phase fluctuation, increases in proportion to the volume, 
the ratios of the cumulants are independent of the volume
--- the higher order terms in the cumulant expansion are well under control in the large volume limit.

Thus, the range where the cumulant expansion is applicable is independent of the volume.
This is a good news in attacking the sign problem, which is known to become severer with increasing the lattice volume.
Furthermore, we find that, when the system size is sufficiently larger than the correlation length, the distribution of $\hat\theta/$volume tends to a Gaussian distribution, since the distribution function of an average over many independent variables $\hat\theta_x$ tends to a Gaussian function by the central limit theorem.

\subsection{Other approaches}

Besides the Taylor expansion method and the reweighting method, as well as combinations of these two methods, various methods have been proposed as alternative approaches to study QCD at finite densities.
An approach is to perform analytic continuation from simulations at imaginary chemical potentials\cite{dFP,dEL}. 
For a complex $\mu_f$, Eq.~(\ref{eq:gamma5conj}) is generalized to 
$[M^f(\mu_f)]^{\dagger} = \gamma_5 M^f(-\mu_f^*) \gamma_5$.
Therefore, when $\mu_f$ is purely imaginary, the Boltzmann weight is real, and we can simulate the system without the sign problem.
Using results of simulations performed at imaginary $\mu$'s, information at  small real $\mu$ can be obtained by an analytic continuation.
The analytic continuation is usually based on a Taylor expansion in terms of $\mu$ around $\mu=0$,
i.e.\ we fit observables obtained at imaginary $\mu$'s with the Taylor expansion ansatz and extrapolate the resulting fitting function to a small real $\mu$.
Improvements of the analytic continuation procedure have also been discussed in Refs~\citen{Cea07,Sakai09} to obtain results in a wider range of real $\mu$.
Systematic high precision simulations in a wide range of the imaginary $\mu$ are required for a reliable analytic continuation.

Another approach is to construct the canonical 
partition function $Z_{\rm C} (T,N)$ by fixing the total 
quark number $N$ or the quark number density\cite{Gibbs86,Hase92,Bar97,Alex05,Krat05,EjiriC}. 
Using the canonical partition function, we can also discuss the effective potential as a function of the quark number.
The relation between the grand canonical partition function $Z(T,\mu)$ and the canonical partition function $Z_{\rm C} (T,N)$ is given by 
\begin{eqnarray}
Z(T,\mu) 
= \int \! {\cal D}U \left[ \det M(\mu) \right]^{N_{\rm f}} e^{-S_g}
= \sum_{N} \, Z_{\rm C}(T,N) \, e^{\, N \mu/T} 
\label{eq:cpartition} 
\end{eqnarray}
for the degenerate $N_{\rm f}$-flavor case. 
Because this is a Laplace transformation, $Z_{\rm C}$ is obtained from $Z$ by an inverse Laplace transformation.
To find $N$ that gives the largest contribution to $Z$, it is worth introducing an effective potential as a function of $N$, 
\begin{eqnarray}
V_{\rm eff}(N,T,\mu) 
\equiv - \ln Z_{\rm C}(T,N) -N \,\frac{\mu}{T} 
= \frac{f(T,N)}{T} -N \,\frac{\mu}{T} , 
\label{eq:effp} 
\end{eqnarray}
where $f$ is the Helmholtz free energy. 
$V_{\rm eff}$ is useful to study the nature of phase transitions\cite{EjiriC}.

\section{Two flavors of improved Wilson quarks at finite density}
\label{sec:Nf2Mu}

Adopting the Taylor expansion method and the Gaussian method discussed in the previous section,
we made the first study of finite-density QCD with Wilson-type quarks\cite{WHOT10dense}.
We study two-flavor QCD with RG-improved gauge action and the clover-improved Wilson quark action.
%
A systematic study of finite-temperature QCD with this action has been done by the CP-PACS Collaboration\cite{CPPACS00,CPPACS01}.
The phase diagram at $\mu=0$ as well as LCP's determined with $m_\pi/m_\rho$ are given in Refs.~\citen{CPPACS00,CPPACS01,WHOT10dense}.
We extend the study to $\mu\ne0$.

\subsection{Taylor expansion for EOS}

We study the pressure and quark number densities defined by Eqs.~(\ref{eq:p}) and (\ref{eq:qnd}) for the case of two-flavor QCD.
Defining the quark chemical potential $\mu_q = (\mu_u + \mu_d)/2$ and the isospin chemical potential $\mu_I = (\mu_u - \mu_d)/2$, 
the quark number and isospin susceptibilities are given by
\begin{eqnarray}
\frac{\chi_q}{T^2}= \frac{\partial^2 (p/T^4)}{\partial (\mu_q/T)^2},
\hspace{5mm} 
\frac{\chi_I}{T^2}= \frac{\partial^2 (p/T^4)}{\partial (\mu_I/T)^2},
\end{eqnarray}
which measure the fluctuations in baryon and isospin numbers in the medium\cite{Gottlieb}.

We study the isosymmetric case $\mu_u=\mu_d=\mu$, $\mu_I=0$.
The pressure is given by
\begin{equation}
\frac{p}{T^4} =
\sum_{n=0}^\infty c_n(T) \left(\frac{\mu}{T}\right)^n,
\hspace{5mm}
c_n (T)= 
\frac{1}{n!} \, \frac{N_{\rm t}^{3}}{N_{\rm s}^3} \left.
\frac{\partial^n \ln Z}{\partial(\mu/T)^n} \right|_{\vec\mu=0}.
\label{eq:cn}
\end{equation}
Here, $c_0(T)$ is the pressure at $\vec\mu=0$ and has been 
computed by the CP-PACS Collaboration.\cite{CPPACS00,CPPACS01} 
%
The susceptibilities are then expanded as 
\begin{eqnarray}
\frac{\chi_q(T,\mu)}{T^2}
= 2c_2+12c_4\left(\frac{\mu}{T}\right)^2+\cdots ,
\;\;\;
\frac{\chi_I(T,\mu)}{T^2}
= 2c^I_2+12c^I_4\left(\frac{\mu}{T}\right)^2+\cdots ,
\;\;\;\;
\end{eqnarray}
where
\begin{equation}
c^I_n= \left. \frac{1}{n!} \, \frac{N_{\rm t}^{3}}{N_{\rm s}^3} \,
\frac{\partial^n \ln Z(T,\mu+\mu_I,\mu-\mu_I)}
{\partial(\mu_I/T)^2 \, \partial(\mu/T)^{n-2}}
\right|_{\vec\mu=0} .
\label{eq:cn_I}
\end{equation}

In this study, we compute the Taylor expansion coefficients for the second and fourth derivatives in (\ref{eq:cn}).
This enables us to compute $\chi_q$ and $\chi_I$ to the lowest non-trivial order in $\mu$. 

\subsection{Random noise method}
\label{sec:Noise}

To evaluate the Taylor coefficients (\ref{eq:cn}) and (\ref{eq:cn_I}), 
we calculate
\begin{eqnarray}
{\cal D}_n = N_{\rm f} \left. \frac{\partial^n \ln \det M}{\partial (\mu a) ^n} \right|_{\vec\mu=0} . 
\label{eq:calDn}
\end{eqnarray}
up to $n=4$\cite{WHOT10dense}. We thus study
\begin{eqnarray}
{\cal D}_1
&=& N_{\rm f} \, {\rm tr} \left( M^{-1} \frac{\partial M}{\partial (\mu a)} \right)_{\! \vec\mu=0} ,
\nonumber\\
{\cal D}_2
& = & N_{\rm f} \left[ {\rm tr} \left( M^{-1} \frac{\partial^2 M}{\partial (\mu a)^2} \right)
 - {\rm tr} \left( M^{-1} \frac{\partial M}{\partial (\mu a)}
                   M^{-1} \frac{\partial M}{\partial (\mu a)} \right) \right]_{\vec\mu=0} , 
\label{eq:dermu}
\end{eqnarray}
etc., where 
\begin{eqnarray}
\left. \left(\frac{\partial^{n} M}{\partial (\mu a)^{n}} \right)_{\! x,y} \right|_{\vec\mu=0} 
= - \kappa \left[
 (1-\gamma_4) U_{x,4} \delta_{x+\hat{4},y} 
  +(-1)^n (1+\gamma_4) U_{x-\hat{4},4}^\dagger \delta_{x-\hat{4},y} \right].
\nonumber\\
\end{eqnarray}
In terms of ${\cal D}_n$, we have 
$c_2 = (N_{\rm t}/2N_{\rm s}^3)\{ \left\langle {\cal D}_2 \right\rangle +\left\langle {\cal D}_1^2 \right\rangle \}$, 
$c_2^I =(N_{\rm t}/2N_{\rm s}^3) \left\langle {\cal D}_2 \right\rangle $, $\cdots$.
Note that ${\cal D}_n$ is real for even $n$ and purely imaginary for odd $n$ \cite{BS02}. 

These traces, say ${\rm tr} A$, can be evaluated by ``the random noise method''.
In this method, instead of calculating the diagonal elements $A_{ii}$ individually for all $i$, we calculate $\eta^\dag \!A \eta$ for several random noise vectors $\eta$.
The contributions of the off-diagonal elements $A_{ij}$ ($i\ne j$) in this quantities are removed by averaging over the random noises, $\overline{\eta_i^* \eta_j} = \delta_{ij}$.

As is clear from the construction, the random noise method is effective when the contaminations from off-diagonal elements are small.
Because the propagator $(M^{-1})_{x,y}$ decreases rapidly with increasing $|x-y|$, the random noise method will work well to suppress small contaminations of spatially off-diagonal elements. 
On the other hand, the off-diagonal elements in the color and spinor indices are from the same spatial point and thus are not suppressed by $|x-y|$ --- they are suppressed only by $1/\sqrt{N_{\rm noise}}$, where $N_{\rm noise}$ is the number of noise vectors.
This motivates us to apply the random noise method for the spatial coordinates only. 
For the trace over the color and spinor indices, we just repeat the calculation generating the random noise vectors for each color-spin index\footnote{
Because a staggered-type quark does not have the spinor index at a spatial point, the number of off-diagonal elements is only 6 in the color $3 \times 3$ matrix, the contamination of off-diagonal elements is less serious. 
This is a reason that the random noise method is adopted more naively with the staggered-type quarks.
However, with Wilson-type quarks, 
the number of the off-diagonal elements with similar magnitude in the quark matrix is 11 times larger than the diagonal one.
Therefore, the color-spinor index should be treated more carefully with Wilson-type quarks. 
}.

For a product of traces, the random noise vectors for each trace must be independent.
We compute such product by subtracting the contribution of the same noise vector from the naive product of two noise averages for each trace.
This effectively increases the number of noises to $N_{\rm noise}\cdot (N_{\rm noise} - 1)$ for the products and thus suppresses their errors due to the noise method. 

In practice, $N_{\rm noise}$ can be small when the error due to the smallness of $N_{\rm noise}$ is smaller than the statistical errors from the averaging over the configurations.
The required number of noise vectors depends on each operator.
In Ref.~\citen{WHOT10dense}, we have tested the random noise method in the evaluation of ${\cal D}_n$ (with $n=1$--4). 
We find that ${\cal D}_1$ has larger fluctuations under a variation of the noise vectors than ${\cal D}_2$ etc., 
and the error in ${\cal D}_1$ dominates in the errors of $c_4$ and $c_4^I$ when we  adopt the same $N_{\rm noise}$.
From this test, we choose $N_{\rm noise}=100$--400 for ${\rm tr}[(\partial^n M/\partial (\mu a)^n) M^{-1}]$ ($n=1$--4), and $N_{\rm noise}=10$ for other operators.


\subsection{Quark number densities and susceptibilities at $\mu\ne0$}
\label{sec:Nf2MuResults}

Simulations are done on a $16^3\times 4$ lattice in the fixed $N_{\rm t}$ approach in the range $\beta=1.50$--2.40 ($T/T_{\rm pc} \approx 0.8$--3 or 4 on LCP's corresponding to $m_\pi/m_\rho = 0.65$ and 0.80, 
where $T_{\rm pc}$ is the pseudo-critical temperature for each LCP.
See Ref.~\citen{WHOT10dense} for details.

\begin{figure}[tb]
       \centerline{
       \includegraphics[width=6.6 cm]{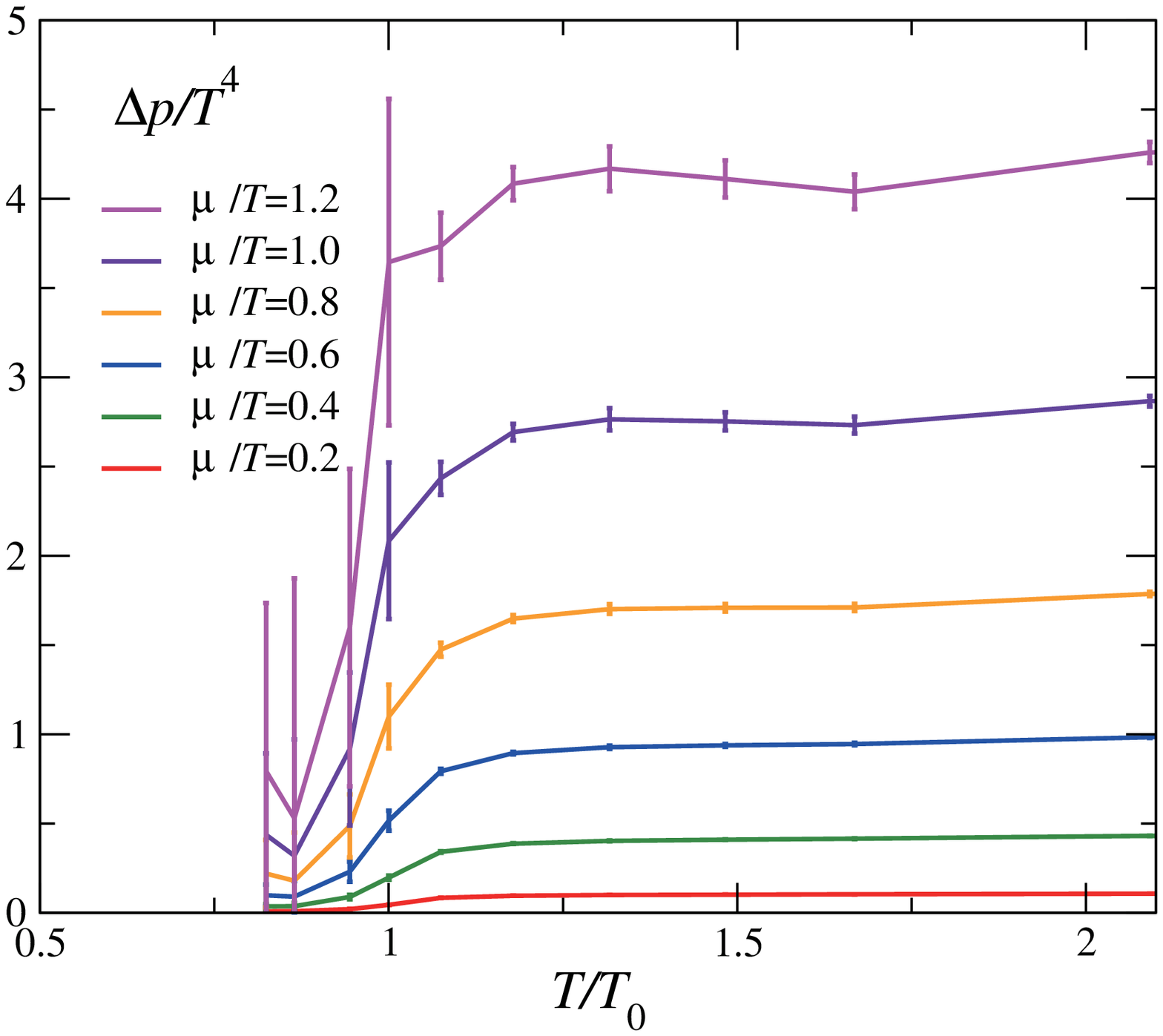}
       \hspace{2mm}
       \includegraphics[width=6.7 cm]{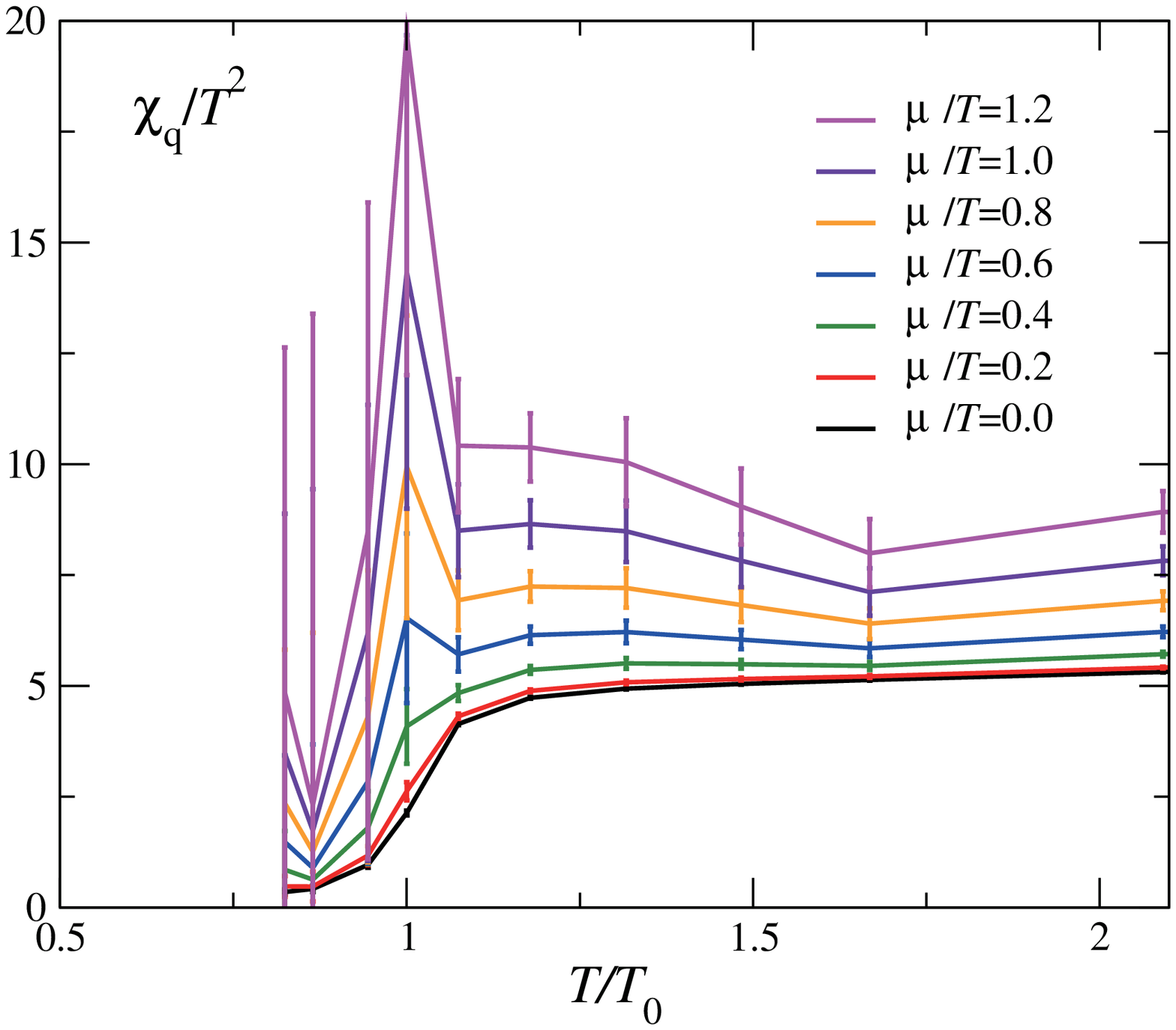}
       }
   \caption{Results of two-flavor QCD at finite density by the standard Taylor expansion method.\cite{WHOT10dense}
  Results are obtained at $m_\pi/m_\rho=0.65$. The truncation error of the Taylor expansion is ${\cal O}(\mu^6)$. 
   $T_0$ is the pseudo-critical temperature at $\mu=0$.
   {\em Left}: $\mu$-dependent contribution to the pressure, $\Delta p/T^4 \equiv p(\mu)/T^4 - p(0)/T^4$.
  {\em Right}: Quark number susceptibility $\chi_q$. 
  }
   \label{fig:wdense1}
   \end{figure}

Our results with the standard Taylor expansion method are shown in Fig.~\ref{fig:wdense1} for LCP at $m_\pi/m_\rho=0.65$. 
The left panel is for the finite density correction $\Delta p/T^4 \equiv p(\mu)/T^4 - p(0)/T^4$ of the pressure at finite $\mu$. 
$T_{0}$ is $T_{pc}$ at $\mu=0$.
Recall that we have a crossover at $T_{\rm pc}$ at $\mu=0$.
The pressure changes more sharply as $\mu$ is increased.
When we increase $\mu$, $\Delta p/T^4$ becomes the same size as 
$p/T^4$ at $\mu=0$ around $\mu/T \sim O(1)$.
In the right panel of Fig.~\ref{fig:wdense1}, we show the results of the quark number susceptibility $\chi_q(\mu)$ at $\mu\ne0$.
We see that a peak seems to be formed when we increase $\mu$.
However, in spite of various improvements in random noise estimators etc.\ as discussed in the previous subsection, the statistical errors due to the complex phase fluctuation of the quark determinant are still a bit too large to draw a definite conclusion about the peak.

   \begin{figure}[tb]
       \centerline{
       \includegraphics[width=6.6 cm]{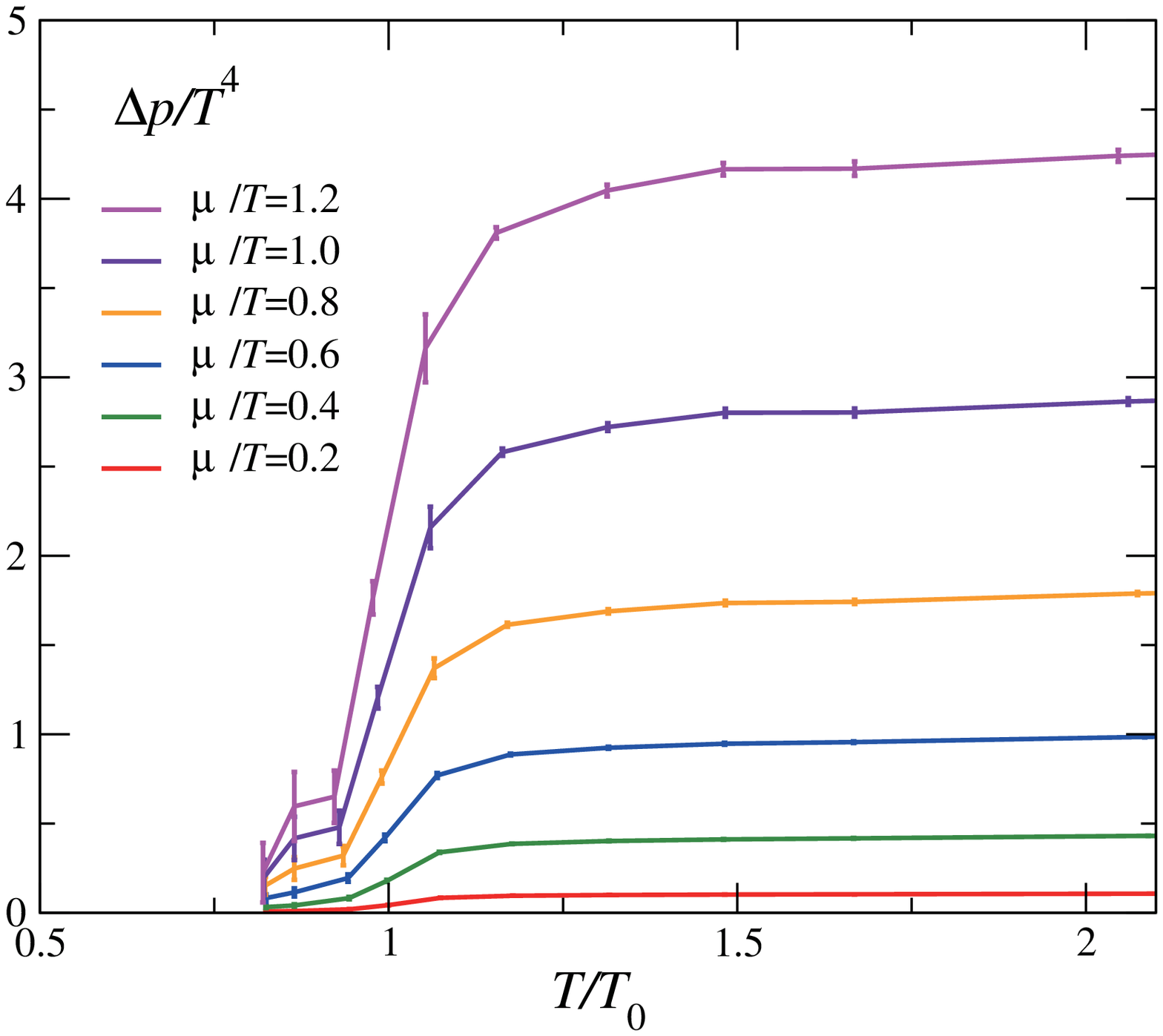}
       \hspace{2mm}
       \includegraphics[width=6.7 cm]{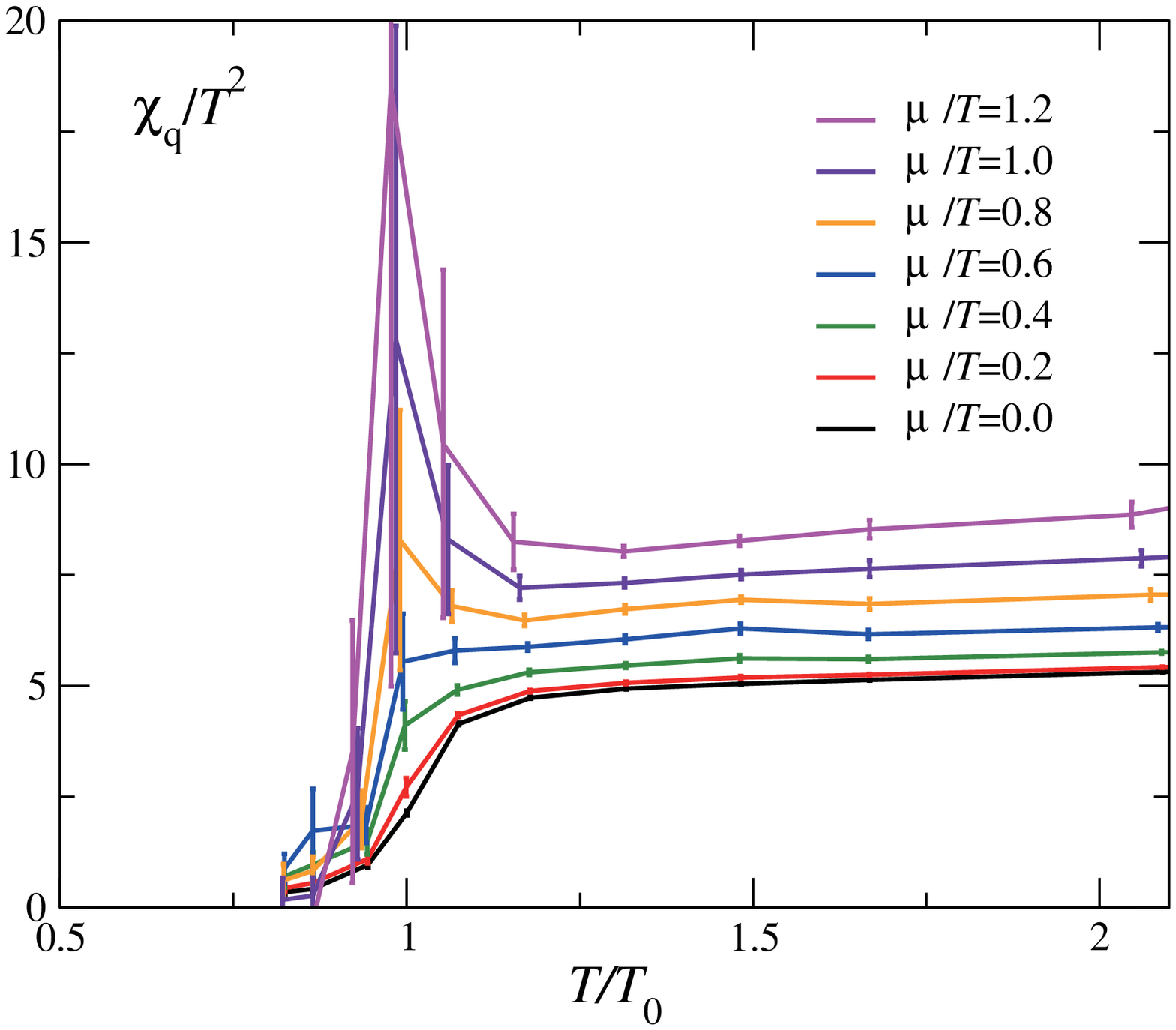}
       }
   \caption{The same as Fig.~\ref{fig:wdense1} but with the combined hybrid and Gaussian methods.\cite{WHOT10dense}
   {\em Left}: $\Delta p/T^4 \equiv p(\mu)/T^4 - p(0)/T^4$.
   {\em Right}: Quark number susceptibility $\chi_q$. 
   }
   \label{fig:wdense2}
   \end{figure}

   \begin{figure}[tb]
       \centerline{
       \includegraphics[width=6.6 cm]{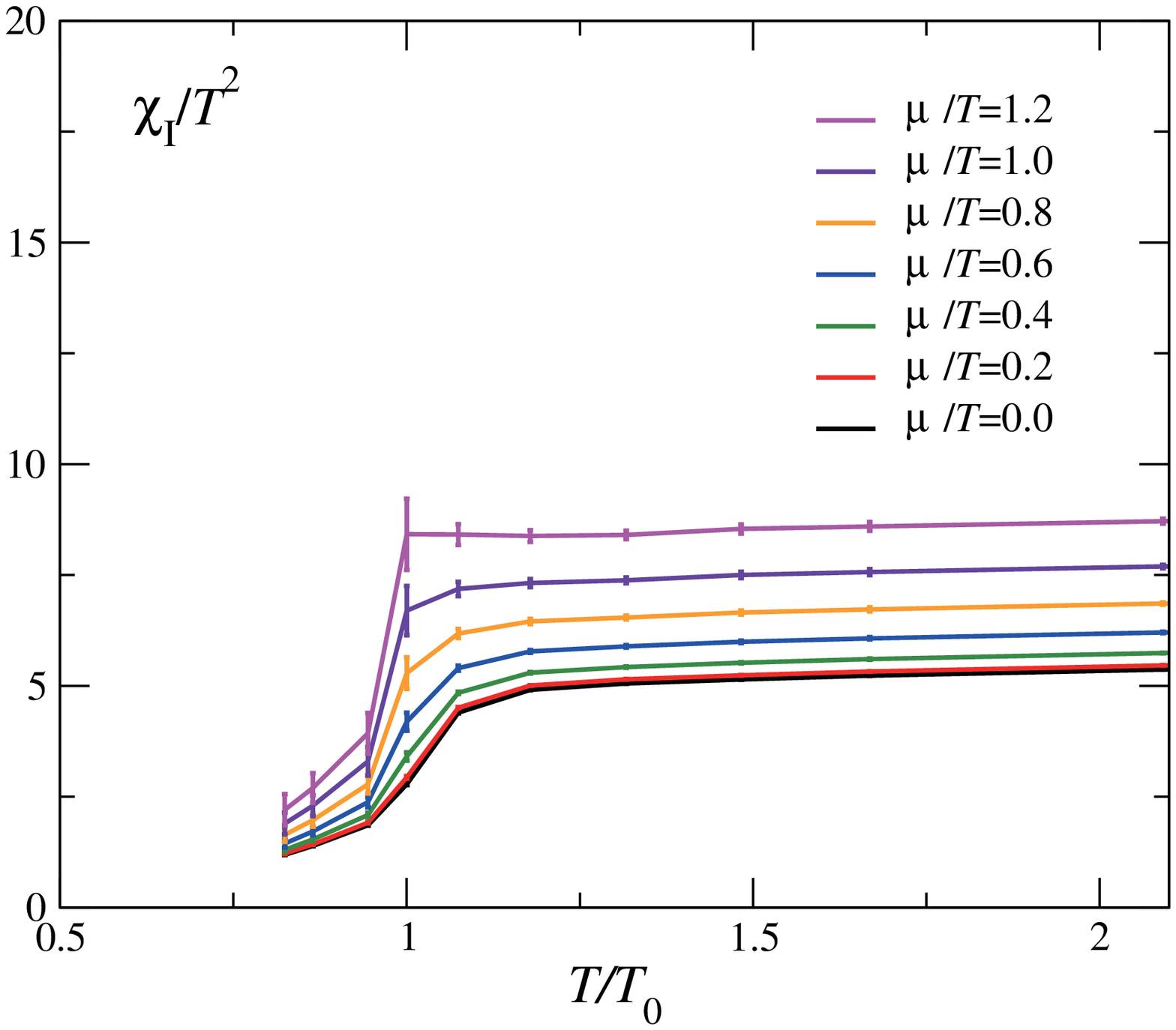}
       \hspace{2mm}
       \includegraphics[width=6.6 cm]{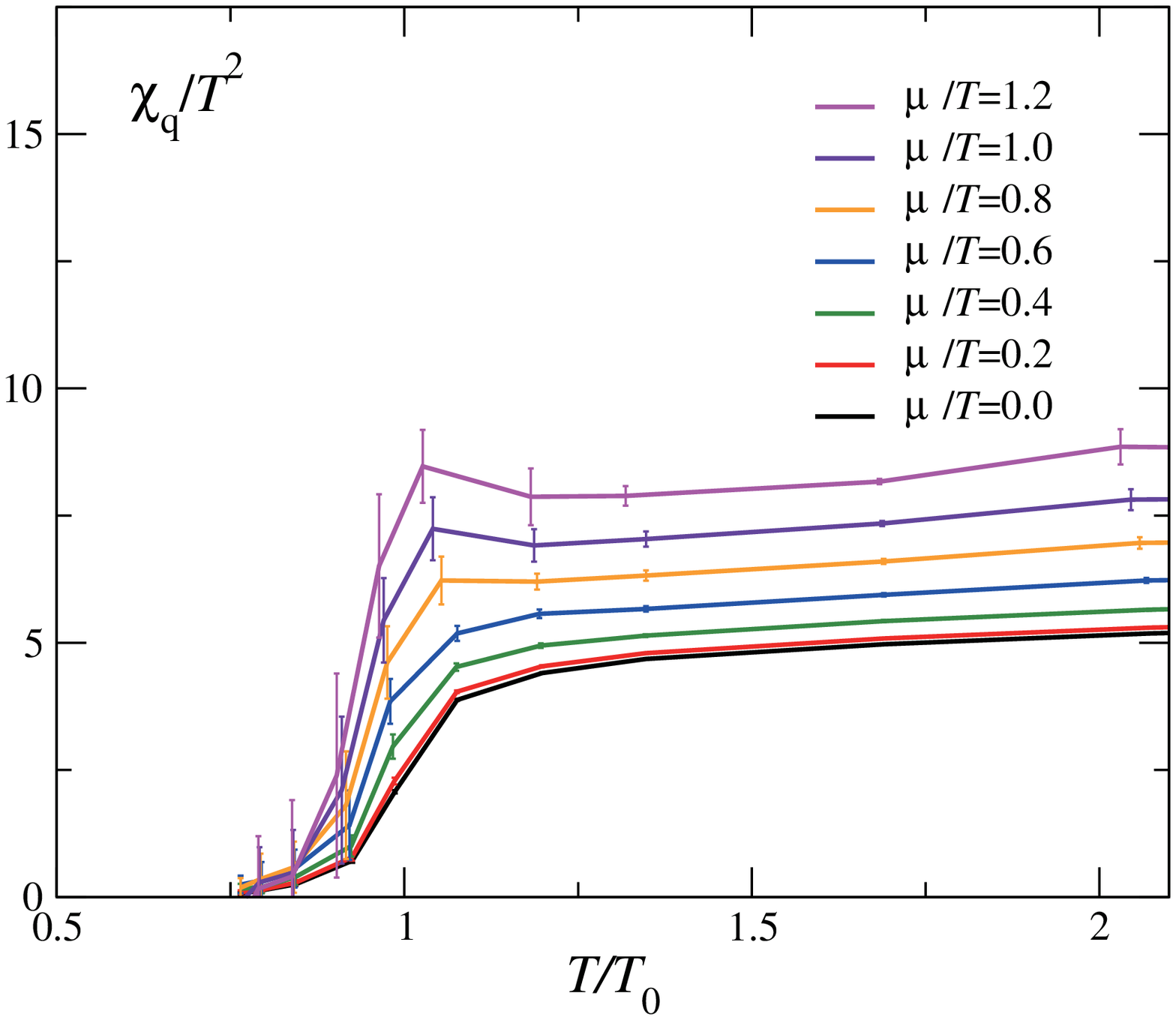}
       }
   \caption{Results of two-flavor QCD at finite density.\cite{WHOT10dense}
   {\em Left}: Isospin susceptibility $\chi_I$ at $m_\pi/m_\rho=0.65$ by the standard Taylor expansion method. 
   {\em Right}: Quark number susceptibility $\chi_q$ at $m_\pi/m_\rho=0.80$ by the combined hybrid and Gaussian methods. 
   }
   \label{fig:wdense3}
   \end{figure}

In order to suppress the errors from the phase fluctuation, we apply the cumulant expansion method discussed in Sect.~\ref{sec:cumulant}. 
We compute the quark determinant ratio in Eq.~(\ref{eq:denorew}) by the Taylor expansion up to $O(\mu^4)$,
\begin{eqnarray}
\hat{F} (\mu) 
& \equiv & N_{\rm f}\,  {\rm Re} \left[ \ln \left( \frac{\det M(\mu)}{\det M(0)} \right) \right]
\;\approx\; \sum_{n=1}^{N_{\rm max}} \frac{1}{(2n)!} \, {\rm Re} {\cal D}_{2n} \, (\mu a)^{2n} ,
\label{eq:teabs} \\
\hat\theta (\mu) 
& = & N_{\rm f}\, {\rm Im} [\ln \det M(\mu)]
\;\approx\; \sum_{n=1}^{N_{\rm max}} \frac{1}{(2n-1)!} \, {\rm Im} {\cal D}_{2n-1} \, (\mu a)^{2n-1} ,
\label{eq:tatheta2}
\end{eqnarray}
with $N_{\rm max}=2$, 
and estimate the phase factor by the second order cumulant assuming the Gaussian distribution of $\theta$,
$\langle e^{i\hat\theta} \rangle \approx \exp[-\langle \hat\theta^2 \rangle /2]$.
We moreover shift $\beta$ from the simulation point $\beta_0$ such that the statistical error due to the $F$-integration is minimized in Eq.~(\ref{eq:denorew}). 
We perform a fit of the resulting pressure ($=$ free energy) in terms of $\mu$ to obtain $\chi_q$.
The results for $\Delta p/T^4$ and $\chi_q/T^2$ are shown in Fig.~\ref{fig:wdense2}.
We find that the statistical errors are appreciably suppressed by these methods. 
We also note that, although the simulations at different $T$ are independent, the $T$-dependences of $\Delta p$ and $\chi_q$ are smooth and natural.
We can now clearly identify a sharp peak in $\chi_q/T^2$ that appears around $T_{\rm pc}$ when $\mu/T \ge {\cal O}(1)$. 
The peak becomes higher as $\mu$ increases.
These are consistent with the observations with staggered-type quarks and suggest a critical point at finite $\mu$.

In contrast with the peak in $\chi_q$, the isospin susceptibility shows no sharp peak, as shown in the left panel of Fig.~\ref{fig:wdense3}.
This is in accordance with the expectation that $\chi_I$ is analytic at the critical point since the iso-triplet mesons remain massive. 

Results at $m_\pi/m_\rho=0.80$ are similar, but with milder peaks in $\chi_q$ than those in Fig.~\ref{fig:wdense2}, as shown in the right panel of Fig.~\ref{fig:wdense3}.
This may be explained in part by the expectation that the critical point locates at larger $\mu$ because the quark mass is larger than that for $m_\pi/m_\rho= 0.65$.
See Ref.~\citen{WHOT10dense} for more discussions.

\section{Histogram method and QCD phase structure at zero and finite densities}
\label{sec:Histogram}

In a study of the QCD phase structure shown in Figs.~\ref{fig:QCDpd}, and \ref{fig:QCDpd2}, identification of first-order transition region is quite important.
Among several methods to study the nature of phase transitions, the  probability distribution of physical observables provides us with the most intuitive way:  
The probability distribution function is defined as the generation rate of configurations with fixed expectation values of physical observables.
Double or multiple peaks in the probability distribution function for observables which are sensitive to the phase, such as the energy density, chiral order parameter, etc., give a signal of first order phase transition.
A problem is that, in order to trace the variation of the shape of a probability distribution function, we need a statistically reliable data of the distribution function in a wide range of the expectation values. 
In the identification of a first order transition, a correct evaluation of the double-peak distribution requires a quite long simulation with sufficiently many flip-flops among different phases [see, e.g., Ref.~\citen{QCDPAXsu3}].
This is computationally quite demanding with dynamical quarks.

Here, we note that the calculation of the probability distribution function is required also for the reweighting method for the components of the action \cite{McDonaldSinger,Ferrenberg:1988yz}.
Therefore, when we adopt observables which are the components of the action, the computation of the distribution function at different points in the coupling parameter space is straightforward by the reweighting method\cite{Ejiri:2007ga}.
This helps us to obtain the probability distribution function in a wide range of expectation values.
Therefore, the method is quite powerful to study the location of first order transitions.

We apply the method to explore the phase structure of QCD.
In this section, we introduce the method, which may be viewed as a variant of the histogram method or the density of state method\cite{DOS}, and test it in the heavy-quark region of QCD\cite{Saito1,Saito2}.
We then present our on-going project to study finite density QCD with light dynamical quarks by combining the histogram method with phase-quenched simulations\cite{Nakagawa}.

\subsection{Histogram method}
\label{subsec:Histogram}

As a demonstration, let us consider the simplest lattice QCD: the combination of the plaquette gauge action with unimproved Wilson quarks.
\begin{eqnarray}
S &=& S_g + S_q ,
\nonumber\\
S_g &=& -6 N_{\rm site} \beta \, \hat{P}, \\
S_q &=& \sum_{f=1}^{N_{\rm f}} \sum_x 
\left\{ \bar{q}_x^f q_x^f
-\kappa_f  \sum_{\mu=1}^3 \bar{q}_x^f \left[ (1-\gamma_{\mu})U_{x,\mu} q_{x+\hat{\mu}}^f + (1+\gamma_{\mu})U_{x-\hat{\mu},\mu}^{\dagger} q_{x-\hat{\mu}}^f \right] \right.
\nonumber\\
&&
\left.
-\kappa_f  \bar{q}_x^f \left[ e^{\mu_f a} (1-\gamma_{4})U_{x,4} q_{x+\hat{4}}^f + e^{-\mu_f a} (1+\gamma_{4})U_{x-\hat{4},4}^{\dagger} q_{x-\hat{4}}^f \right] \right\}
\nonumber\\
& \stackrel{\rm def.}{=}  & 
\sum_{f=1}^{N_{\rm f}} \sum_{x,y} \bar{q}_x^f \, M_{xy} (\kappa_f,\mu_f) \, q_y^f ,
\end{eqnarray} 
where $N_{\rm site}=N_{\rm s}^3 \times N_{\rm t}$ is the lattice volume and
\begin{eqnarray}
\hat{P}= \frac{1}{6 N_{\rm site}} \displaystyle \sum_{x,\, \mu < \nu} 
\frac{1}{3}\, {\rm Re \ tr} \left[ U_{x,\mu} U_{x+\hat{\mu},\nu}
U^{\dagger}_{x+\hat{\nu},\mu} U^{\dagger}_{x,\nu} \right]
\label{eq:plsaq}
\end{eqnarray}
is the plaquette.
Note that $M$ does not depend on $\beta$.\footnote{
When we consider improved gauge actions such as (\ref{eq:gaction}), we replace $\hat{P}$ with the operator appearing in the gauge action: $\hat{P} \stackrel{\rm def.}{=} - S_g / (6N_{\rm site}\beta)$.
On the other hand, when $M$ depends on $\beta$ as in the case of improved quark actions, more careful treatments are required. See discussions in Sect.~\ref{sec:PQ}.
}

Denoting the values of the operators ($\hat{P}$, $\cdots$) as ($P$, $\cdots$), 
the probability distribution function for ($P$, $\cdots$) is defined by
\begin{eqnarray}
  w(P,\cdots; \beta, \vec\kappa, \vec\mu) 
 &=& \int \! {\cal D} U\, \delta\left(\hat{P}[U]-P\right) \cdots \prod_f \det M(\kappa_f, \mu_f) \, e^{6N_{\rm site} \beta \hat{P}}
  \nonumber\\
 &=& e^{6N_{\rm site} \beta P} \int\! {\cal D} U\, \delta\left(\hat{P}[U]-P\right) \cdots \prod_f \det M(\kappa_f, \mu_f),
\end{eqnarray}
where ``$\cdots$'' in the r.h.s.\ means the product of delta functions for other operators.
We now define the effective potential as
\begin{equation}
V_{\rm eff}(P, \cdots ; \beta, \vec\kappa, \vec\mu) = -\ln   w(P, \cdots ; \beta, \vec\kappa, \vec\mu) .
\end{equation}
Note that $P$ represents the freedom of gauge internal energy, and thus should be sensitive to the phase structure of the system.

A useful property of the plaquette distribution function and the effective potential is 
\begin{eqnarray}
w(P, \cdots ; \beta, \vec\kappa, \vec\mu)
&=& w(P, \cdots ; \beta_0, \vec\kappa, \vec\mu) \, e^{6(\beta-\beta_0)N_{\rm site}P} ,
\\
V_{\rm eff}(P, \cdots ; \beta, \vec\kappa, \vec\mu) 
&=& V_{\rm eff}(P, \cdots ; \beta_0, \vec\kappa, \vec\mu) - 6(\beta - \beta_0)N_{\rm site}P.
\label{eq:Veff_beta+shift}
\end{eqnarray}
We thus find that 
\begin{eqnarray}
\frac{d V_{\rm eff}}{dP} (P, \cdots ; \beta, \vec\kappa, \vec\mu)
= \frac{d V_{\rm eff}}{dP} (P, \cdots ; \beta_0, \vec\kappa, \vec\mu) 
-6 (\beta - \beta_0) N_{\rm site},
\label{eq:derrewbeta}
\end{eqnarray}
and $d^2 V_{\rm eff}/dP^2$ is independent of $\beta$.

The $\vec\kappa$ and $\vec\mu$-dependences of $V_{\rm eff}$ can also be computed by the reweighting method as follows:
\begin{equation}
V_{\rm eff}(P, \cdots ; \beta, \vec\kappa, \vec\mu) 
= V_{\rm eff} (P, \cdots ; \beta, \vec\kappa_0, \vec\mu_0) 
- \ln R(P, \cdots ; \vec\kappa, \vec\mu, \vec\kappa_0, \vec\mu_0),
\end{equation}
where the reweighting factor $R$ is evaluated as
\begin{eqnarray}
R(P, \cdots ; \vec\kappa, \vec\mu, \vec\kappa_0, \vec\mu_0) 
& \stackrel{\rm def.}{=} & 
\frac{w(P, \cdots ; \beta, \vec\kappa, \vec\mu)}
{ w(P, \cdots ; \beta, \vec\kappa_0, \vec\mu_0)} 
\nonumber\\
&=& 
\frac{\int \!{\cal D} U \delta(\hat{P}-P) \cdots 
\prod_f \det M(\kappa_f,\mu_f)\, e^{6\beta N_{\rm site} \hat{P} }}
{\int \!{\cal D} U \delta(\hat{P}-P) \cdots  
\prod_f \det M(\kappa_f^0,\mu_f^0)\, e^{6\beta N_{\rm site} \hat{P} }} 
\nonumber\\
&=& 
\frac{\int \!{\cal D} U \delta(\hat{P}-P) \cdots \prod_f \det M(\kappa_f,\mu_f)}
{\int \!{\cal D} U \delta(\hat{P}-P) \cdots \prod_f \det M(\kappa_f^0,\mu_f^0)} 
\nonumber\\
&=& 
\frac{ \left\langle \delta(\hat{P}-P) \cdots \prod_f
\frac{ \det M(\kappa_f,\mu_f)}{ \det M(\kappa_f^0,\mu_f^0)} \right\rangle_{\!\!(\vec\kappa_0, \vec\mu_0)} }
{ \left\langle \delta(\hat{P}-P) \cdots \right\rangle_{\!(\vec\kappa_0, \vec\mu_0)} }
\nonumber \\
& = &
\left\langle \prod_f
\frac{\det M(\kappa_f,\mu_f)}{\det M(\kappa_f^0,\mu_f^0)} \right\rangle_{\!\!(\vec\kappa_0, \vec\mu_0); P,\cdots} .
\label{eq:rkdef}
\end{eqnarray}
Note that $R$ is independent of $\beta$ and thus can be evaluated at any $\beta$.
By adjusting and combining $\beta$, we can obtain precise values of $R$ in a wide range of $P,\cdots$.

\subsection{QCD in the heavy-quark region}

We first test the method in the heavy-quark region.
As shown in Fig.~\ref{fig:QCDpd2}, we have the first order deconfinement transition of the SU(3) Yang-Mills theory in the limit of infinite quark masses.
The transition is expected to turn into a crossover when we decrease the quark masses.
We study the boundary of the first order region by the histogram method.
To take advantage of light computational costs in quenched simulations, 
we choose $\kappa_f^0 = \mu_f^0 = 0$ $(f=1,\cdots,N_{\rm f})$.

For simplicity, let us consider the case of degenerate quarks: $\kappa_f=\kappa$, $\mu_f=\mu$ $(f=1,\cdots,N_{\rm f})$.
Generalization to non-degenerate cases is easy.

To the lowest order of the hopping parameter expansion, 
the quark determinant ratio appearing in the r.h.s.\ of (\ref{eq:rkdef}) is evaluated as
\begin{equation}
   \frac{\det M(\kappa, \mu)}{\det M(0,0)} 
  =  \exp \left[ 288N_{\rm site}\kappa^4\hat{P}+3N_{\rm s}^32^{N_{\rm t}+2}\kappa^{N_{\rm t}}\left\{\cosh \left(\frac{\mu}{T}\right) \hat\Omega_{\rm R}+i\sinh \left( \frac{\mu}{T} \right) \hat\Omega_{\rm I}\right\} \right],
  \label{eq:detM}
\end{equation}
where 
\begin{equation}
\hat{\Omega} = \frac{1}{3N_{\rm s}^3} \sum_{\mathbf{x}} {\rm Tr} \left[ 
U_{\mathbf{x},4} U_{\mathbf{x}+\hat{4},4} 
\cdots U_{\mathbf{x}+(N_{\rm t} -1)\hat{4},4} \right]
\end{equation}
is the Polyakov loop, and 
$\hat\Omega_{\rm R} = {\rm Re} \hat\Omega$ and $\hat\Omega_{\rm I} = {\rm Im} \hat\Omega$ 
are its real and imaginary parts. 
We note that the contribution of $\hat{P}$ in the r.h.s.\ can be absorbed by a shift of the gauge coupling  $\beta  \rightarrow \beta + 48N_{\rm f}\kappa^4$.

The term proportional to $\hat\Omega_{\rm I}$ induces the complex phase at $\mu\ne0$, which is the origin of the sign problem at large $\mu$.

   \begin{figure}[tb]
       \centerline{
       \includegraphics[width=68mm]{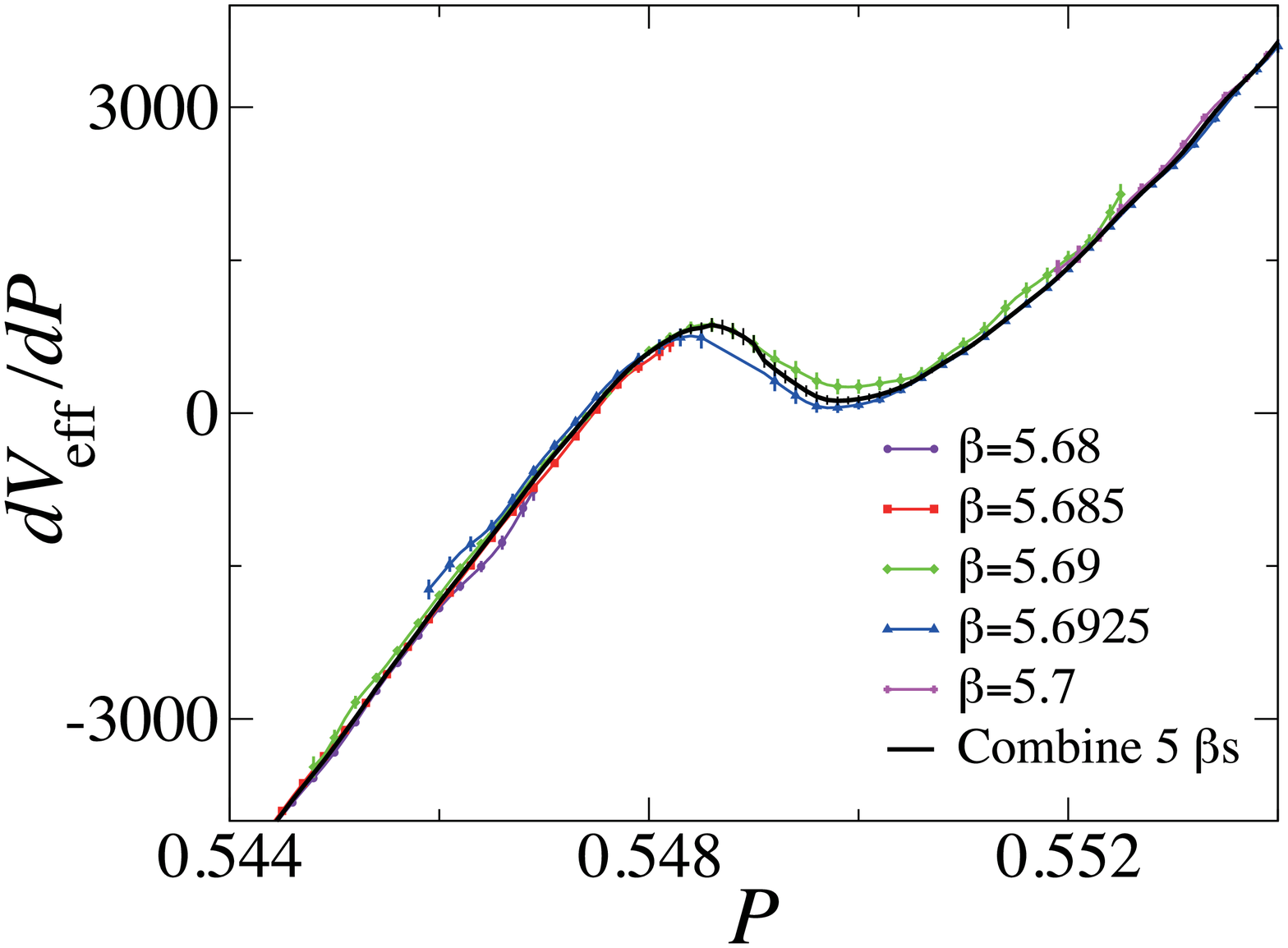}
       \hspace{2mm}
       \includegraphics[width=66mm]{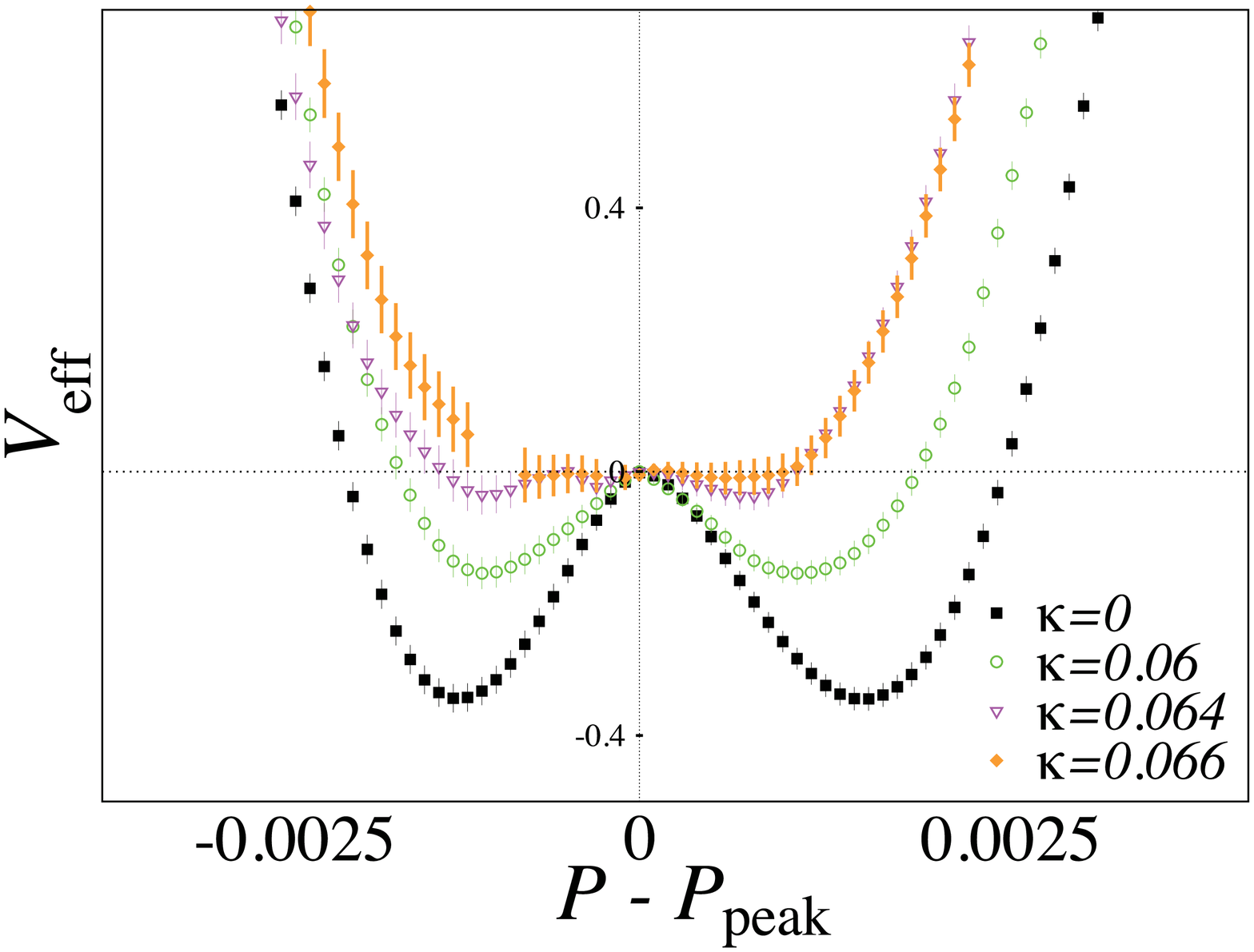}
       }
\caption{Effective potential at $\mu=0$ as a function of $P$ in the heavy-quark region\cite{Saito1}.
 {\em Left}: 
$dV_{\rm eff}(P;\beta,0)/dP$ in the heavy quark limit.
Data obtained at five different values of $\beta$ in the range 5.68--5.70 are converted to $\beta=5.69$ by (\ref{eq:derrewbeta}). 
 {\em Right}: 
$V_{\rm eff}(P;\beta,\kappa)$ for $\kappa > 0$, where $\beta$ is adjusted such that the two minima of $V_{\rm eff}$ have the same depth and the constant term of $V_{\rm eff}$ is adjusted such that $V_{\rm eff}=0$ at the central peak point $P_{\rm peak}$.
}
   \label{fig:HQVeff}
   \end{figure}

\subsection{Results at $\mu=0$ in the heavy-quark region}

We are now ready to calculate the effective potential.
Let us first study the case $\mu=0$\cite{Saito1,Saito2}.
In this case, the complex phase term is absent  in (\ref{eq:detM}).

The double-well nature of the effective potential is clearly seen when we consider the effective potential for $P$ only:
\[
V_{\rm eff}(P;\beta,\kappa) \stackrel{\rm def.}{=}  - \ln \int \! w(P,\Omega_{\rm R};\beta,\kappa) \, d\Omega_{\rm R} .
\]
Our results for $dV_{\rm eff}(P;\beta,0)/dP$ at $\kappa=0$ are shown in the left panel of Fig.~\ref{fig:HQVeff}.
Using (\ref{eq:derrewbeta}), we shift the results obtained at five different $\beta$'s to $\beta=5.69$.
With different $\beta$, the range of $P$ in which we can reliably obtain $V_{\rm eff}$ is different.
We find that the results of $dV_{\rm eff}(P;\beta,0)/dP$ at different $\beta$'s are smoothly connected with each other by (\ref{eq:derrewbeta}).
We can thus obtain accurate values of $dV_{\rm eff}(P;\beta,0)/dP$ in a wide range of $P$.

Similarly, we calculate $dV_{\rm eff}(P;\beta,\kappa)/dP$ at $\kappa>0$ by using the reweighting factor $R$ and (\ref{eq:detM}).
We then integrate $dV_{\rm eff}(P;\beta,\kappa)/dP$ in $P$ to get $V_{\rm eff}(P;\beta,\kappa)$ as shown in the right panel of Fig.~\ref{fig:HQVeff}.
We find that the double-well structure of $V_{\rm eff}(P)$ becomes weaker with increasing $\kappa$, and eventually disappears at finite $\kappa$, say $\kappa_{\rm cp}$.
Examining the shape of $V_{\rm eff}$ more closely, we obtain 
$\kappa_{\rm cp} = 0.0658(3)(^{+4}_{-11})$ for two-flavor QCD in the lowest order of the hopping parameter expansion on an $N_{\rm t}=4$ lattice\cite{Saito1}.

   \begin{figure}[tb]
       \centerline{
       \includegraphics[width=80mm]{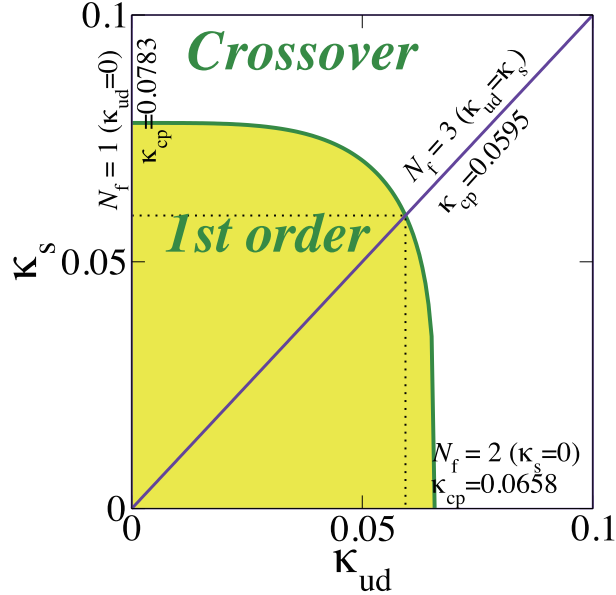}
       }
\caption{The phase boundary separating the first order transition region and crossover region in the $(\kappa_{\rm ud}, \kappa_{\rm s})$ plane at $\mu=0$ in the heavy-quark region\cite{Saito1}. 
}
   \label{fig:HQ_PD}
   \end{figure}

The argument can be easily generalized to the case of non-degenerate quark masses.
Our results for the critical point $\kappa_{\rm cp}$ in $2+1$ flavor QCD is shown in Fig.~\ref{fig:HQ_PD}\cite{Saito1}. 
The top-right corner of Fig.~\ref{fig:QCDpd2} corresponds to this plot rotated by $180^\circ$. 
Our results are consistent with that obtained in an effective model\cite{Alexandrou99} and with a recent study using the hopping parameter expansion\cite{Fromm12}.

The expression (\ref{eq:detM}) suggests us to adopt $\hat\Omega_{\rm R}$ as an additional operator for the effective potential $V_{\rm eff}$.
Then, the $\hat\Omega_{\rm R}$ in the r.h.s.\ of (\ref{eq:detM}) is simply replaced by its expectation value $\Omega_{\rm R}$, and the reweighting factor $R$ is just a given function of $P$ and $\Omega_{\rm R}$ in this case.
We have  
\begin{eqnarray}
  \frac{\partial V_{\rm eff}}{\partial P}(P, \Omega_{\rm R}; \beta, \kappa)
  &=& \frac{\partial V_0}{\partial P}(P, \Omega_{\rm R}; \beta_0)  - 6N_{\rm site} \left(\beta + 48N_{\rm f}\kappa^4 -\beta_0 \right) ,
  \label{eq:dV0dP}
\\
  \frac{\partial V_{\rm eff}}{\partial \Omega_{\rm R}}(P, \Omega_{\rm R};\kappa) 
  &=& \frac{\partial V_0}{\partial \Omega_{\rm R}}(P, \Omega_{\rm R})  - 3 N_{\rm s}^3 2^{N_{\rm t}+2} N_{\rm f}  \kappa^{N_{\rm t}} ,
  \label{eq:dV0dOmega}
\end{eqnarray}
where $V_0$ is the effecvtive potential in the heavy quark limit (SU(3) pure gauge theory).
The argument $\beta$ in $\partial V_{\rm eff}/\partial \Omega_{\rm R}$ and $\partial V_0/\partial \Omega_{\rm R}$ is omitted in (\ref{eq:dV0dOmega}) since they are independent of $\beta$ due to (\ref{eq:Veff_beta+shift}).
Note that, besides known overall constants [the last terms in (\ref{eq:dV0dP}) and (\ref{eq:dV0dOmega})], 
the dependences of $\partial V_{\rm eff}/\partial P$ and $\partial V_{\rm eff}/\partial \Omega_{\rm R}$ on $P$ and $\Omega_{\rm R}$ are independent of $\beta$ and $\kappa$.

\begin{figure}[tb]
       \centerline{
       \includegraphics[width=100mm]{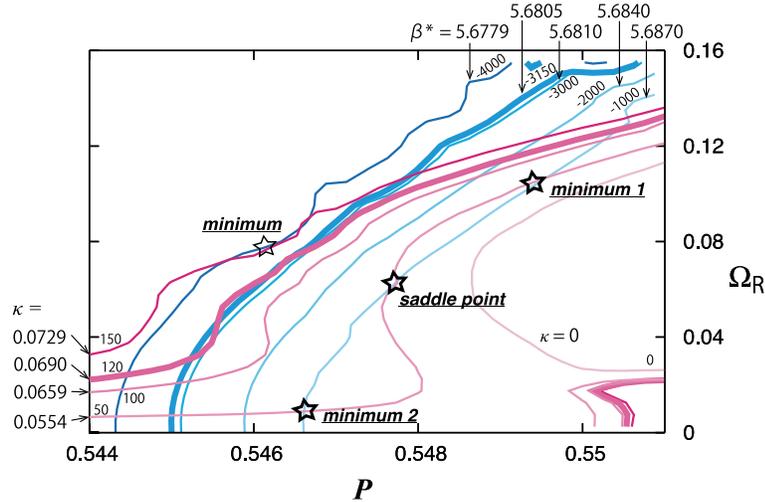}
       }
\caption{Contour plot of $\partial V_0/\partial P$ (blue curves) and $\partial V_0/\partial \Omega_{\rm R}$ (red curves) at $\mu=0$ in the heavy-quark region\cite{Saito2}. Values of $\beta^* = \beta + 48N_{\rm f}\kappa^4$ and $\kappa$ for the corresponding curves of $\partial V_{\rm eff}/\partial P = 0$ and $\partial V_{\rm eff}/\partial \Omega_{\rm R} = 0$ are also given.
}
   \label{fig:HQ_PDm}
   \end{figure}

Because $\Omega_{\rm R}$ represents the freedom of heavy-quark free energy, we expect it be sensitive to the phase structure of the system. 
Around the first order transition point, we will have a double-well structure of $V_{\rm eff}$ in the two-dimensional plane $(P,\Omega_{\rm R})$.
To study the phase structure, it is useful to examine the curves $\partial V_{\rm eff}/\partial P = 0$ and $\partial V_{\rm eff}/\partial \Omega_{\rm R} = 0$.
From (\ref{eq:dV0dP}) and (\ref{eq:dV0dOmega}), these curves at different $(\beta,\kappa)$ corresponds to different contour curves of $\partial V_0/\partial P$ and $\partial V_0/\partial \Omega_{\rm R}$.
A contour plot for $\partial V_0/\partial P$ and $\partial V_0/\partial \Omega_{\rm R}$ is given in Fig.~\ref{fig:HQ_PDm}.
When the curves $\partial V_{\rm eff}/\partial P = 0$ and $\partial V_{\rm eff}/\partial \Omega_{\rm R} = 0$ cross at only one point, we have just one minimum of $V_{\rm eff}$.
In this case, we have no first-order transitions around this $(\beta,\kappa)$.
On the other hand, when we have three intersection points, we have two minima and one saddle point, implying the existence of the first order transition.
In particular, from the merger of three intersection points, we can determine the critical point where the first order transition line terminates.
From Fig.~\ref{fig:HQ_PDm}, we find that the S-shaped curve of $\partial V_{\rm eff}/\partial \Omega_{\rm R}=0$ leads to three intersection points at small $\kappa$, 
and that the S-shape becomes weaker with increasing $\kappa$. 
A preliminary estimate for the critical point is $\kappa_{\rm cp}\approx 0.0690(7)$\cite{Saito2}, shown by thick contour curves in Fig.~\ref{fig:HQ_PDm}.
We are currently testing a refinement of the method to extract smoother contour curves.

To examine the quality of the hopping parameter expansion, we have studied the effect of the next-leading order terms in the evaluation of the critical point $\kappa_{\rm cp}$\cite{WHOTprep,Okuno12}.
To this order, we need to incorporate  $\kappa^6$-loops with length six and generalized Polyakov loops with length $N_t+2$ in the quark determinant ratio. 
The effects of $\kappa^6$-loops may be absorbed by a shift of $\beta$.
Examining the effects of generalized Polyakov loops, we find  that $\kappa_{\rm cp}$ shifts only by about 3\% on the $N_t=4$ lattice due to the next-leading order terms. 
We also find that the contribution of the next-leading order terms becomes comparable to that of the leading order terms at $\kappa \sim 0.18$. 
Because $\kappa_{\rm cp}$ is much smaller than this, we conclude that the hopping parameter expansion is well valid up to $\kappa \sim \kappa_{\rm cp}$.
Accordingly, an estimation of the pseudo-scalar meson mass around $\kappa_{\rm cp}$ leads to $m_\pi \approx T/0.023$ for $N_f=2$\cite{Saito1}, i.e., $m_\pi \sim 7$--9 GeV with $T\sim T_{\rm pc} \sim 160$--200 MeV.
Thus, $\kappa_{\rm cp}$ locates well in the heavy-quark region.

\subsection{Results at $\mu\ne0$ in the heavy-quark region}

In order to extend the study to finite densities, we have to calculate the reweighting factor due to the complex phase,
\begin{equation}
\left\langle e^{ i \hat\theta} \right\rangle_{\! P, \Omega_{\rm R}} \quad
\mbox{with} \hspace{3mm}
\hat\theta=3 N_{\rm s}^3 2^{N_{\rm t}+2} N_{\rm f} \lambda\hat\Omega_{\rm I}
,\hspace{3mm} 
\lambda = \kappa^{N_{\rm t}}\sinh\left( \mu/T \right).
\label{eq:thetaOmegaI}
\end{equation}
where $\langle\cdots\rangle_{\! P, \Omega_{\rm R}}$ is the expectation value at fixed $P$ and $\Omega_{\rm R}$ in quenched QCD.
When $\hat\theta$ fluctuates a lot at large $\mu$, a reliable estimation of $ \langle e^{ i\hat\theta} \rangle_{\! P, \Omega_{\rm R}} $ becomes difficult (the sign problem).\footnote{
If we could treat $\hat\Omega_{\rm I}$ as a variable for $V_{\rm eff}$ too, the reweighting factor $\langle e^{ i \hat\theta} \rangle$ is just a given function of $\Omega_{\rm I}$ (or equivalently $\theta$).
However, when the fluctuation in $\hat\theta$ is large, then it is difficult to determine a reliable $V_{\rm eff}$ unless quite a high statistics is accumulated. 
This is a rephrasing of the sign problem.
}.

Before evaluating $ \langle e^{ i\hat\theta} \rangle_{\! P, \Omega_{\rm R}} $, let us consider the case of phase-quenched finite density QCD, in which the complex phase term is removed in the quark determinant. 
In two-flavor QCD, this corresponds to the case of the isospin chemical potential, $\mu_u=-\mu_d \equiv \mu_I$.
From (\ref{eq:detM}), we find that, after shifting $\beta \rightarrow \beta + 48N_{\rm f}\kappa^4$,  the effects of $\mu_I$ are just to further shift $\kappa \rightarrow  \kappa\, \cosh^{1/N_{\rm t}} (\mu_I/T)$.
Therefore, we have 
\begin{equation}
\kappa_{\rm cp}^I(\mu_I)=\kappa_{\rm cp}(0) / \cosh^{1/N_{\rm t}}(\mu_I/T)
\end{equation}
for the critical point in the phase-quenched QCD to the lowest order of the hopping parameter expansion, where $\kappa_{\rm cp}(0)$ is the critical point at $\mu_u=\mu_d=0$.
Note that, with increasing $\mu_I$, the critical point approaches towards $\kappa=0$ where the hopping parameter expansion is reliable.

   \begin{figure}[tb]
       \centerline{
       \includegraphics[width=70mm]{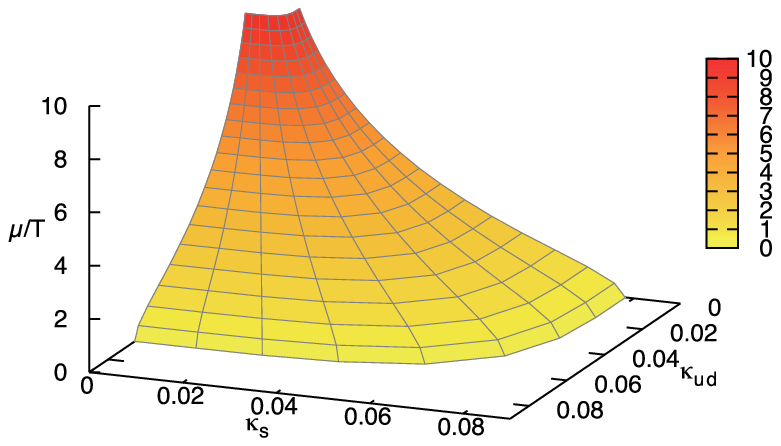}
       \includegraphics[width=70mm]{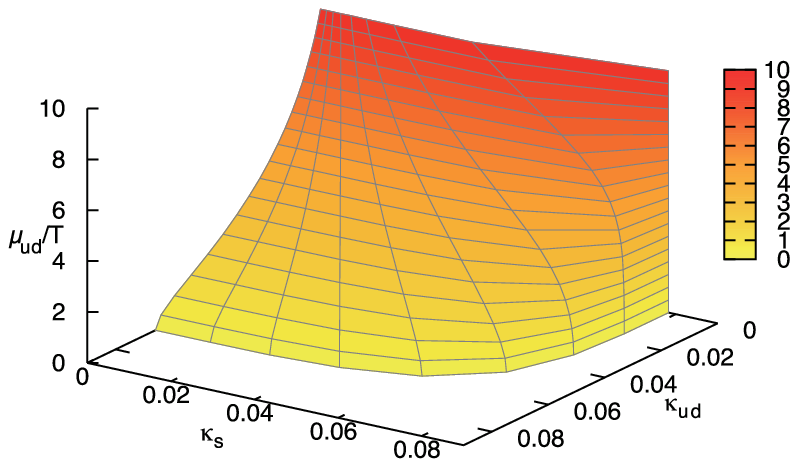}
       }
\caption{Critical surface separating the first order transition and crossover regions in the heavy-quark region.
 {\em Left}: 
The case $\mu_{u} = \mu_{d} = \mu_{s} \equiv \mu$.
 {\em Right}: 
The case that may be realized in heavy ion collisions: $\mu_{u} = \mu_{d} \equiv \mu_{ud}$ and $\mu_{s} = 0$.
}
   \label{fig:HQ_PD2}
   \end{figure}

We now compute the effect of the phase. 
To evaluate $ \langle e^{ i\hat\theta} \rangle_{\! P, \Omega_{\rm R}} $, we adopt the Gaussian approximation with the cumulant expansion \cite{Ejiri:2007ga,WHOT10dense} discussed in Sect.~\ref{sec:cumulant}.
In the heavy-quark region, we study $\langle \hat\theta^{2n} \rangle_c
 = \left( 3N_{\rm s}^3  2^{N_{\rm t}+2}N_{\rm f} \lambda \right)^{2n}  \langle \hat\Omega_{\rm I}^{2n} \rangle_c$.
We find that $\langle \hat\theta^2\rangle_c \gg \langle \hat\theta^4 \rangle_c$ around the critical point\cite{Saito2}.
This confirms the validity of the Gaussian approximation.
We thus have 
\begin{eqnarray}
   \frac{\partial V_{\rm eff} }{\partial P} 
   &=& \frac{\partial V_0}{\partial P}-6N_{\rm site} \left(\beta + 48N_{\rm f}\kappa^4 - \beta_0\right)
          +\frac{(3N_{\rm s}^3 2^{N_{\rm t}+2}N_{\rm f} \lambda)^2}{2} \frac{\partial \langle \hat\Omega_{\rm I} ^2\rangle_c}{\partial P},
   \label{eq:Veff_mu_P}
\\
   \frac{\partial V_{\rm eff}}{\partial \Omega_{\rm R}} 
   &=& \frac{\partial V_0}{\partial \Omega_{\rm R}}-3N_{\rm s}^3 2^{N_{\rm t}+2}N_{\rm f}\kappa^{N_{\rm t}}\cosh \left(\frac{\mu}{T}\right)
           +\frac{(3N_{\rm s}^3 2^{N_{\rm t}+2}N_{\rm f} \lambda)^2}{2} \frac{\partial \langle \hat\Omega_{\rm I} ^2\rangle_c}{\partial \Omega_{\rm R}}.
   \label{eq:Veff_mu_Omega}
\end{eqnarray}   
When the last term in (\ref{eq:Veff_mu_Omega}) modifies the S-shape of the curve $\partial V_{\rm eff} / \partial \Omega_{\rm R} = 0$ shown in Fig.~\ref{fig:HQ_PDm}, $\kappa_{\rm cp}(\mu)$ deviates from $\kappa_{\rm cp}^I(\mu)$.
By evaluating these quantities, we however find that the contribution of the last term in (\ref{eq:Veff_mu_Omega}) is at most about 3\% of the second term around $\kappa_{\rm cp}^I(\mu)$ even in the large $\mu$ limit.
Therefore, $\kappa_{\rm cp}(\mu)$ is indistinguishable from $\kappa_{\rm cp}^I(\mu)$ within the current statistical errors\cite{Saito2}:
\begin{equation}
\kappa_{\rm cp}(\mu) \;\approx\; \kappa_{\rm cp}(0) / \cosh^{1/N_{\rm t}}(\mu/T)
\end{equation}
Generalization of this result to non-degenerate cases such as the $2+1$ flavor QCD ($\mu_{u} = \mu_{d} \equiv \mu_{ud}$) is straightforward:
\begin{equation}
2\, [\kappa_{ud}(\vec\mu)]^{N_{\rm t}} \cosh(\mu_{ud}/T) 
+ [\kappa_{s}(\vec\mu)]^{N_{\rm t}} \cosh(\mu_s/T) 
\;\approx\; 2\, [\kappa_{\rm cp}^{N_{\rm f}=2}(0) ]^{N_{\rm t}} .
\end{equation}
Critical surfaces for the symmetric case $\mu_{u} = \mu_{d} = \mu_{s} \equiv \mu$ and a more realistic case of $\mu_{u} = \mu_{d} \equiv \mu_{ud}$ and $\mu_{s} = 0$, that may be realized in heavy ion collisions, are shown in Fig.~\ref{fig:HQ_PD2}.

\subsection{Phase-quenched simulations towards the physical point}
\label{sec:PQ}

We are challenging to apply the histogram method to explore the phase structure of finite-density QCD in the light quark region.
For the sake of notational simplicity, we consider the degenerate case in this section too.
Generalization to non-degenerate cases is straightforward.

When quarks are light, the Polyakov loop $\hat\Omega$ plays no more a decisive role in the dynamics of the system.
We thus consider $\det M$ itself as the additional operator for the effective potential. 
$\det M$ represents the freedom of quark internal energy, and thus should be sensitive to the phase structure of the system.
We denote the absolute value and the complex phase of the quark determinant as follows:
\begin{eqnarray}
\hat{F}(\beta,\kappa,\mu) & = & 
N_{\rm f} \ln \left|\det M(\beta,\kappa,\mu)/\det M(\beta,\kappa,0)\right|
\nonumber\\ & = &  
N_{\rm f} \int_0^{\mu} \! {\rm Re}
\left[ \frac{\partial \,\ln\det M(\beta,\kappa,\mu')}{\partial \mu'} \right] d\mu' ,
\\
\hat\theta(\beta,\kappa,\mu) & = & 
N_{\rm f}\, {\rm Im} \left[ \ln\det M(\beta,\kappa,\mu) \right] 
\nonumber\\ & = &  
N_{\rm f} \int_0^{\mu} \! {\rm Im}
\left[ \frac{\partial \,\ln\det M(\beta,\kappa,\mu')}{\partial \mu'} \right] d\mu' ,
\label{eq:thetaPQ}
\end{eqnarray}
where $ \partial(\ln\det M)/\partial\mu' = {\rm tr} [M^{-1} (\partial M/\partial \mu') ]  $ can be evaluated by the random noise method discussed in Sect.~\ref{sec:Noise}. 
Note that $\hat\theta$ is uniquely defined in the range $(-\infty,\infty)$, as discussed in Sect.~\ref{sec:cumulant}.
%
The distribution function and the effective potential for $P$ and $F$ are now given by 
\begin{eqnarray}
w(P,F;\beta,\kappa,\mu)
&=& \int \!{\cal D}U \delta(\hat{P}-P) \delta(\hat{F}-F)\, e^{i\hat\theta} \,
\left|\det M(\beta,\kappa,\mu)\right|^{N_{\rm f}} e^{6\beta N_{\rm site} \hat{P}}
\nonumber\\
&=& \left\langle e^{i\hat\theta} \right\rangle_{(0:\beta,\kappa,\mu);P,F}
w_0(P,F;\beta,\kappa,\mu),
\label{eq:w_from_pq}\\
V_{\rm eff}(P,F;\beta,\kappa,\mu) 
&=& V_0(P,F;\beta,\kappa,\mu) - \ln \left\langle e^{i\hat\theta} \right\rangle_{(0:\beta,\kappa,\mu);P,F} ,
\label{eq:part_func}
\end{eqnarray}
where $w_0$ and $V_0 = - \ln w_0$ are the distribution function and the effective potential for the phase-quenched system,
and the expectation value $\langle e^{i\hat\theta} \rangle_{(0:\beta,\kappa,\mu);P,F}$ is evaluated with fixed $P$ and $F$ in the phase-quenched simulation. 

In the following, let us consider a simpler case where the quark matrix $M$ can be treated as independent of $\beta$. 
E.g., in a study of the phase structure at a value of $\beta$, we may treat $\beta$ in $M$ as fixed to that value.
Physical properties such as the phase structure will not be affected by this procedure\footnote{
When we calculate observables as functions of $\beta$, we need to incorporate the effects from the $\beta$-dependence in $M$. See \citen{CswBeta} as a trial to incorporate the $\beta$ dependence of $c_{\rm SW}$ in the reweighting factor.
}.
Then, $\langle e^{i\hat\theta} \rangle_{(0:\beta,\kappa,\mu);P,F}$ becomes independent of $\beta$ and can be evaluated at any $\beta$ to cover a wider range of $P$ and $F$.

   \begin{figure}[tb]
       \centerline{
       \includegraphics[width=68mm]{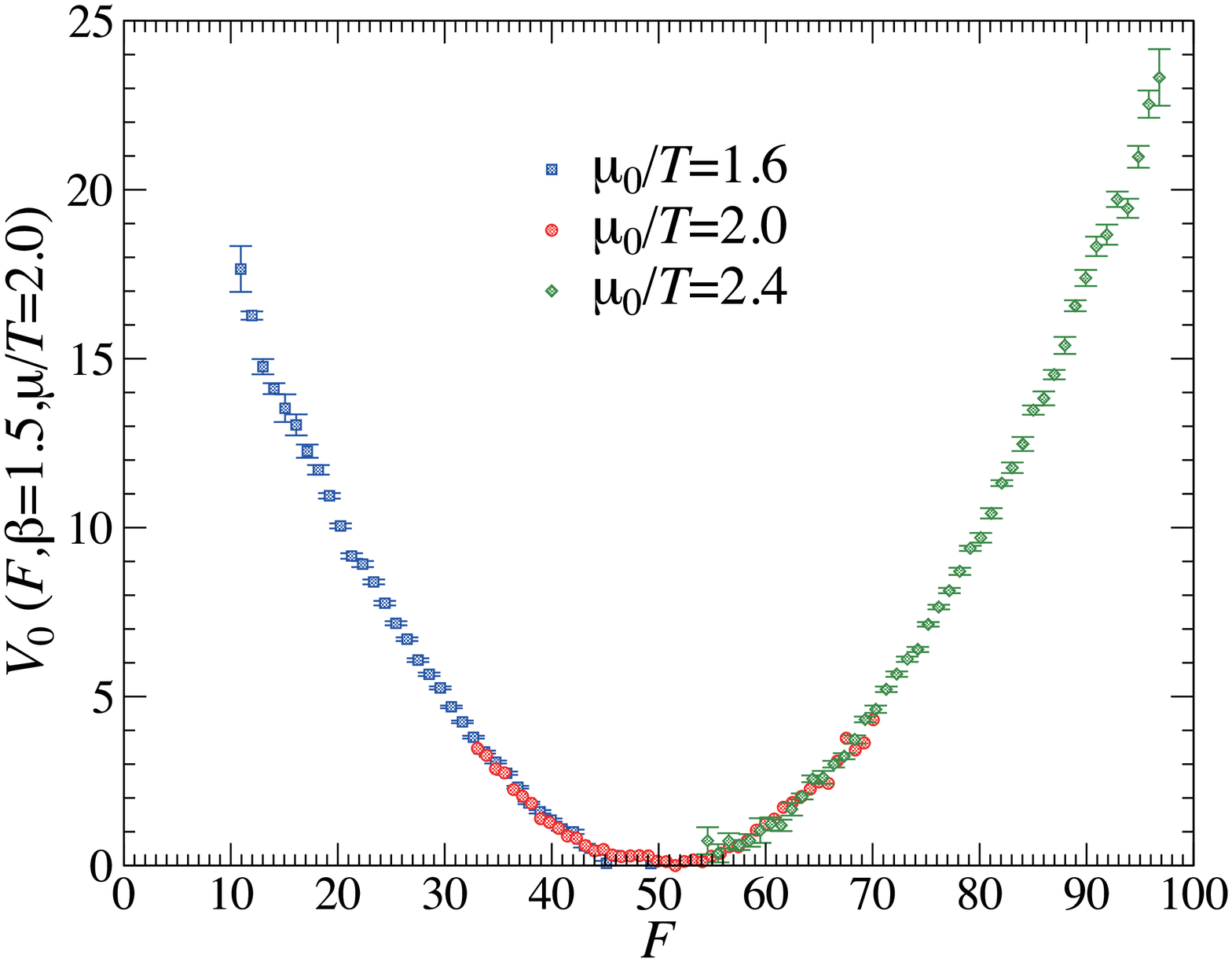}
       \hspace{0.2mm}
       \includegraphics[width=67mm]{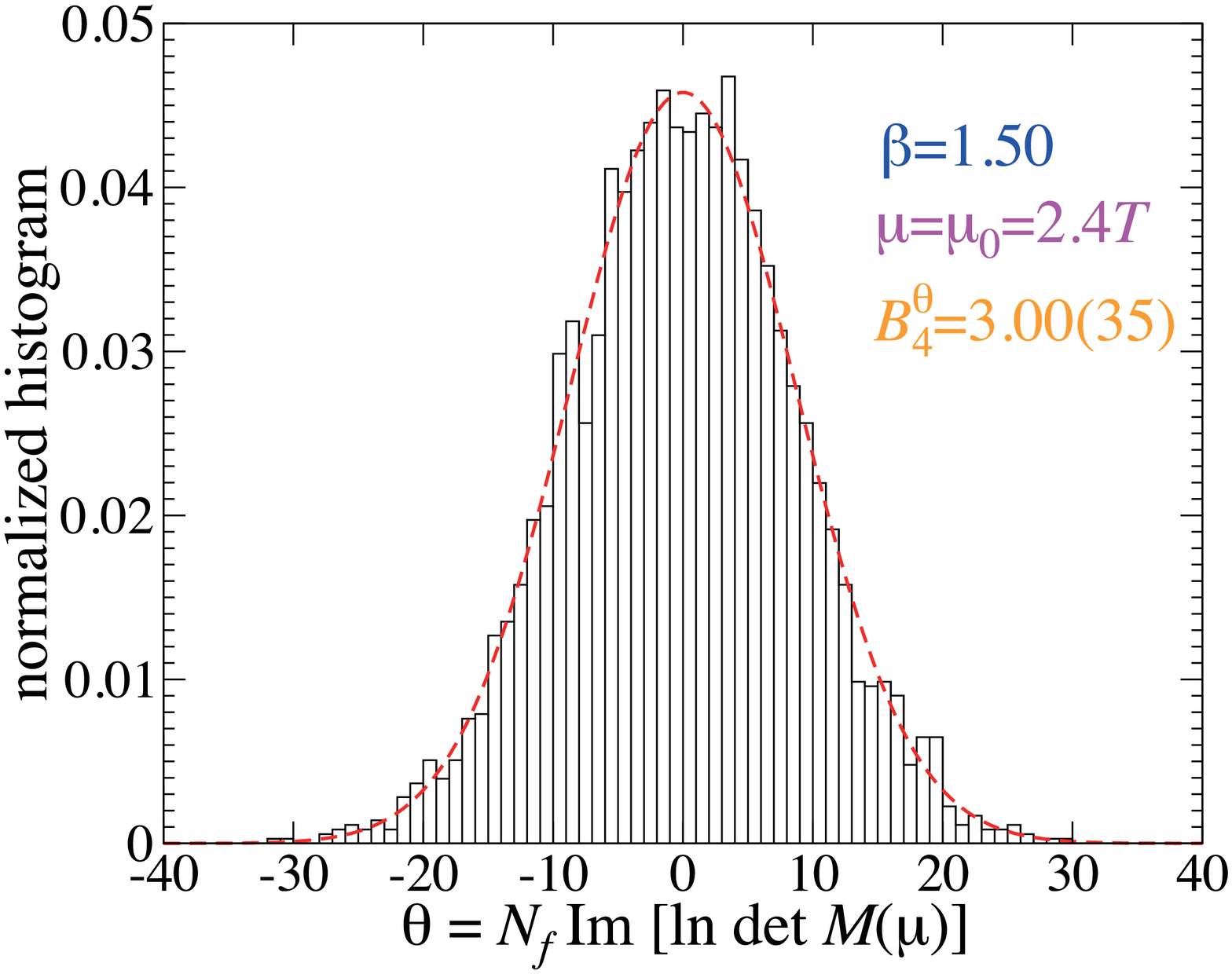}
       }
\caption{Results of phase-quenched simulations in two-flavor QCD with RG-improved gauge action at $\beta=1.5$\cite{Nakagawa}.
 {\em Left}: The effective potential $V_0(F)$ at $\mu/T = 2.0$ evaluated at three different simulation points.
 {\em Right}: The distribution of the phase of the quark determinant at $\mu/T = 2.4$. The dashed curves are the fitted results with the Gaussian function. $B^\theta_4$ is the fourth-order Binder cumulant normalized such that $B^\theta_4 = 3$ for the Gaussian function.
}
   \label{fig:PQ}
   \end{figure}

Because the phase-quenched simulations in two-flavor QCD correspond to the case of isospin chemical potentials, a comment is in order about the influence of the pion-condensed phase which exist at large isospin chemical potentials\cite{PionCondensation}.
In the pion-condensed phase, $\langle e^{i\hat\theta} \rangle$ is expected to vanish by model studies\cite{PionCondensation2,PionCondensation3}. 
According to (\ref{eq:w_from_pq}), this means that the configurations in the pion-condensed phase have no contributions to the physics of phase-unquenched QCD --- $w$ and $V_{\rm eff}$ are dominated by phase-quenched configurations out of the pion-condensed phase, and we only need to generate configurations outside the pion-condensed phase.

We test the method in two-flavor QCD adopting the RG-improved Iwasaki gauge action (\ref{eq:gaction}) and the clover-improved Wilson quark action (\ref{eq:qaction}).
According to the footnote in Sect.~\ref{subsec:Histogram}, $\hat{P}$ is identified as
\begin{equation}
\hat{P} = \frac{1}{6 N_{\rm site}} \sum_{x} \left\{ 
\sum_{\mu>\nu}c_0W^{1\times 1}_{x,\mu\nu}
+\sum_{\mu,\nu}c_1W^{1\times 2}_{x,\mu\nu}
\right\}.
\end{equation}
in the present case.
We perform simulations at $m_\pi/m_\rho \approx 0.8$ on an $8^3\times4$ lattice
with a non-perturbatively estimated $c_{\rm SW}$.
In the present study of the phase structure at each $\beta$, we treat $c_{\rm SW}$ as a constant independent of $\beta$.

In the left panel of Fig.~\ref{fig:PQ}, we show our results of $V_0$ as a function of $F$ (after integrating out $P$). 
The results of simulations at three values of $\mu_0/T$ are translated to $\mu/T=2.0$ using a reweighting formula for $V_0$ in the $\mu$-direction: 
\begin{eqnarray}
V_0(P,F;\beta,\kappa,\mu) &=& V_0(P,F;\beta,\kappa,\mu_0)- \ln R_0(P,F;\kappa,\mu,\mu_0)
\label{eq:V0reweightingMu}
\\
R_0(P,F;\kappa,\mu,\mu_0) &\stackrel{\rm def.}{=}& 
\frac{w_0(P, F ; \beta, \kappa, \mu)}
{ w_0(P, F ; \beta, \kappa, \mu_0)} 
\nonumber\\
&=& 
\frac{\int \!{\cal D} U \delta(\hat{P}-P) \delta(\hat{F}-F) 
\, |\det M(\kappa,\mu)|^{N_{\rm f}} \, e^{6\beta N_{\rm site} \hat{P}} }
{\int \!{\cal D} U \delta(\hat{P}-P) \delta(\hat{F}-F) 
\, |\det M(\kappa,\mu_0)|^{N_{\rm f}} \, e^{6\beta N_{\rm site} \hat{P}} }
\nonumber \\
&=& 
\frac{\int \!{\cal D} U \delta(\hat{P}-P) \delta(\hat{F}-F) 
\, |\det M(\kappa,\mu)|^{N_{\rm f}} }
{\int \!{\cal D} U \delta(\hat{P}-P) \delta(\hat{F}-F) 
\, |\det M(\kappa,\mu_0)|^{N_{\rm f}} }
\nonumber \\
& = &
\left\langle \left|
\frac{\det M(\kappa,\mu)}{\det M(\kappa,\mu_0)} \right|^{N_{\rm f}} \right\rangle_{\!(0: \kappa, \mu_0); P,F} ,
\label{eq:rkdefpq}
\end{eqnarray}
where $R_0$ does not dependent on $\beta$.
From this figure, we confirm that the data obtained at different simulation points form a smooth $V_0$.
We can thus obtain precise $V_0$ in a wide range of $F$.

To calculate $\langle e^{i\hat\theta} \rangle_{(0:\kappa,\mu);P,F}$ , 
we adopt the cumulant expansion method.
Typical result for the distribution of $\theta$ is shown in the right panel of Fig.~\ref{fig:PQ}.
We find that the distribution is well approximated by a Gaussian function.
We also note that the second-order cumulant increases with increasing $\mu$, while the forth-order cumulant is consistent with zero within the statistical error, though the statistical error increases with $\mu$\cite{Nakagawa}.
Therefore, the cumulant expansion is well controlled by the leading term and we may reliably evaluate the complex phase factor even at these relatively high values of $\mu$.
A project towards clarification of the phase structure at the physical point is under way in this direction.

\section{Heavy-quark free energy}
\label{sec:HQFE}

Finally, we study the heavy-quark free energies and screening masses in QGP.
The free energies for static quark-antiquark and two static quarks characterize inter-quark interactions in QGP,  and their Debye screening masses describe the thermal fluctuation of quarks and gluons in QGP.
In a phenomenological model, they are relevant to the fate of heavy-quark bound states such as $J/\psi$ and $\Upsilon$ in QGP created in the relativistic heavy-ion collisions at RHIC and LHC\cite{Matsui-Satz,Satz:2008zc}.
On the lattice, studies in quenched QCD \cite{Kaczmarek:1999mm,NakamuraSaito,Kaczmarek:2004gv,Rothkopf:2011db}
and in full QCD with staggered-type quark actions \cite{Kaczmarek:2005ui,Doring,Fodor:2005qy,Doring:2005ih} 
or with Wilson-type quark actions \cite{Bornyakov:2004ii,WHOT07,WHOT10dense,WHOT10screening,Maezawa2011} have been made. 
Comparisons with analytic studies \cite{Laine:2006ns,Burnier:2009bk} have also been attempted.

Heavy-quark free energies on the lattice are defined through the correlation functions of the local Polyakov line operator
$\hat\Omega ( {\bf x} )  \stackrel{\rm def.}{=} \prod_{ \tau = 1}^{N_{\rm t}} U_{({\bf x},\tau),4}$.
Note that the trace over the color indices is not taken in $\hat\Omega ( {\bf x} )$.
With a gauge-fixing, we define the free energy $F^R$ in various color channels $R$\cite{Nadkarni}:
\begin{eqnarray}
e^{-F^{\bf 1}(r,T,\mu)/T}
 &=&
  \frac{1}{3} \langle {\rm Tr} 
\hat\Omega^\dagger({\bf x}) \hat\Omega ({\bf y})
\rangle
, 
\label{eq:singlet}
\\
e^{-F^{\bf 8}(r,T,\mu)/T}
 &=& 
\frac{1}{8} \langle {\rm Tr} \hat\Omega^\dagger({\bf x})
{\rm Tr} \hat\Omega ({\bf y})
\rangle
-
\frac{1}{24} \langle {\rm Tr} \hat\Omega^\dagger({\bf x})
 \hat\Omega ({\bf y}) \rangle
, \\
e^{-F^{\bf 6}(r,T,\mu)/T}
 &=& 
\frac{1}{12} \langle {\rm Tr} \hat\Omega({\bf x})
{\rm Tr} \hat\Omega ({\bf y})
\rangle
+
\frac{1}{12} \langle {\rm Tr} \hat\Omega({\bf x})
 \hat\Omega ({\bf y}) \rangle
, \\
e^{-F^{{\bf 3}^*}(r,T,\mu)/T}
 &=& 
\frac{1}{6} \langle {\rm Tr} \hat\Omega({\bf x})
{\rm Tr} \hat\Omega ({\bf y})
\rangle
-
\frac{1}{6} \langle {\rm Tr} \hat\Omega({\bf x})
 \hat\Omega ({\bf y}) \rangle
,
\label{eq:sextet}
\end{eqnarray}
where $r = |{\bf x} - {\bf y}|$, and 
$R={\bf 1}$ for the color singlet $Q\overline{Q}$ channel,
${\bf 8}$ for the color octet $Q\overline{Q}$ channel,
${\bf 3}^*$ for the color anti-triplet $QQ$ channel,
and ${\bf 6}$ for the color sextet $QQ$ channel, respectively.
To preserve the free energy interpretation of $F^R$ by the transfer matrix theory, the gauge-fixing procedure should not include the temporal links. 
We thus adopt the Coulomb gauge.

Above $T_{\rm pc}$, we also introduce normalized free energies $V^R$
whose constant terms are adjusted such that they vanish at large distances $r \rightarrow\infty$. 
This is equivalent to defining the free energies by dividing the r.h.s.\ of Eqs.~(\ref{eq:singlet})--(\ref{eq:sextet}) by $\langle {\rm Tr} \hat\Omega({\bf x}) \rangle^2$.

\subsection{Heavy-quark free energies in two-flavor QCD}

We first compute the free energies (\ref{eq:singlet})--(\ref{eq:sextet}) in two-flavor QCD at zero and finite densities.\cite{WHOT07}
We consider the case $\mu_u=\mu_d=\mu$.
We use the gauge configurations generated for the studies discussed in Sect.~\ref{sec:Nf2Mu}. 
As mentioned in Sect.~\ref{sec:Nf2MuResults}, the configurations were generated on a $16^3\times 4$ lattice on LCP's corresponding to $m_\pi/m_\rho = 0.65$ and 0.80. 
We thus adopt the fixed-$N_{\rm t}$ approach for this study.

\subsubsection{Heavy-quark free energies at $\mu=0$ in two-flavor QCD}
 
   \begin{figure}[tb]
       \centerline{
       \includegraphics[width=69mm]{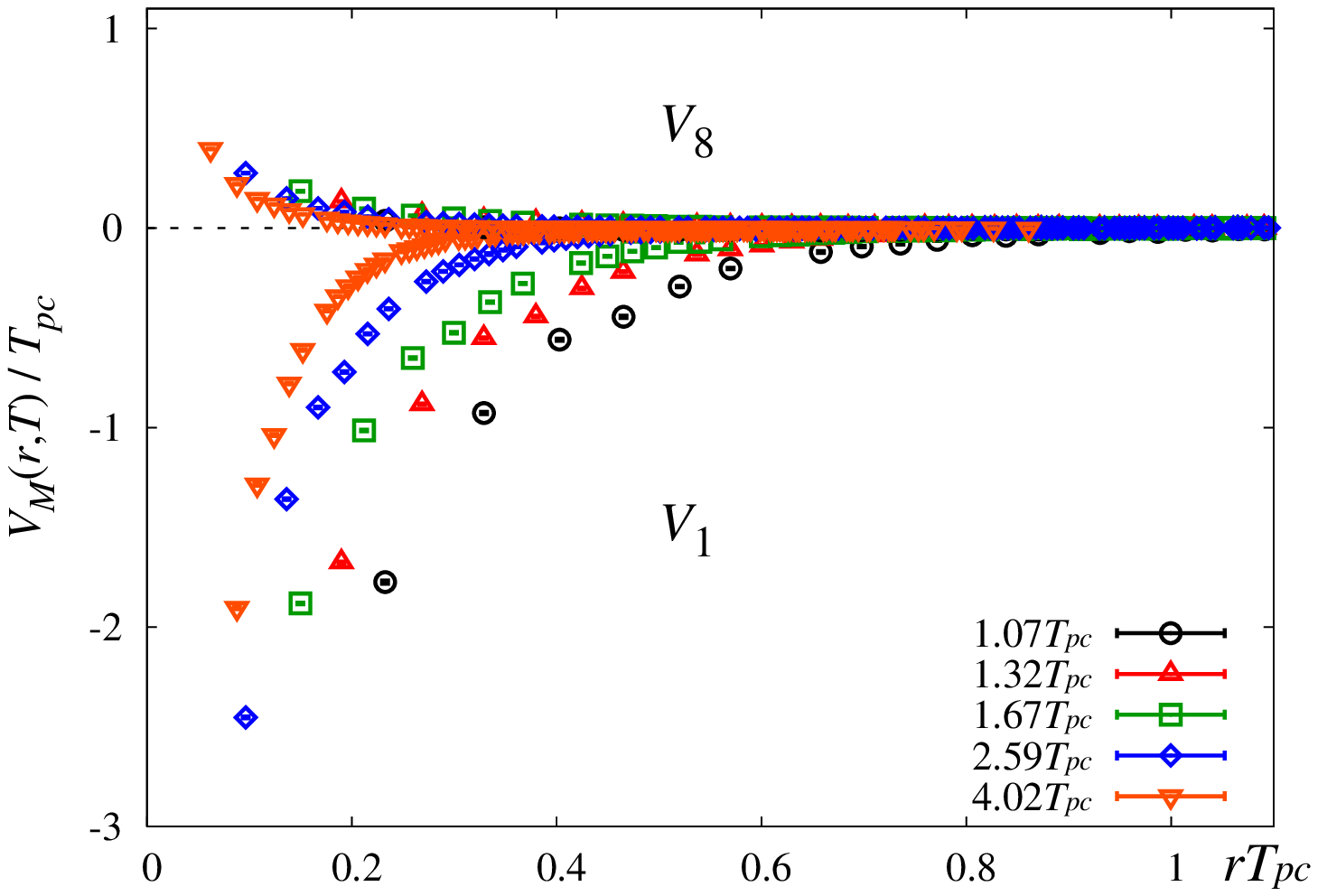}
       \hspace{1mm}
       \includegraphics[width=69mm]{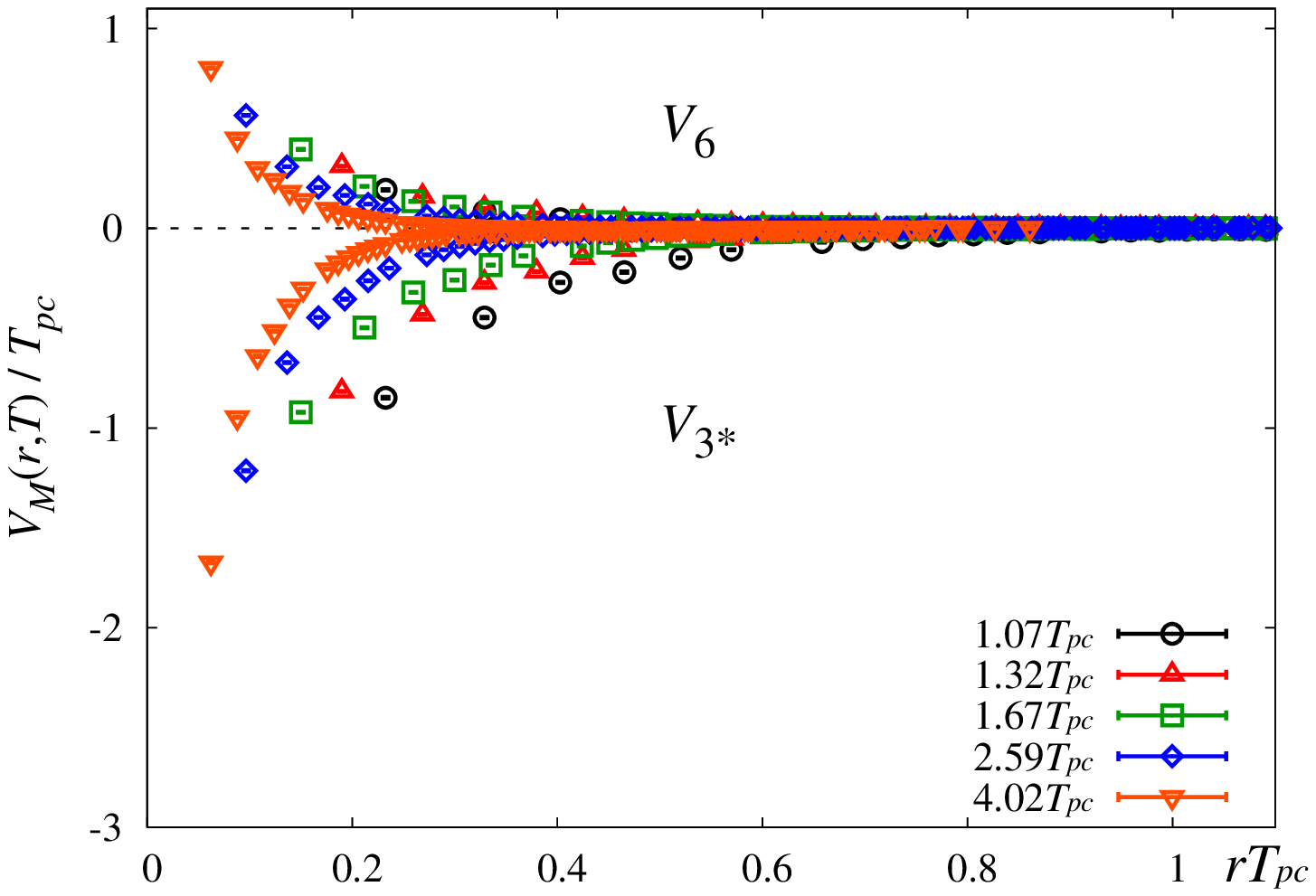}
       }
\caption{Heavy-quark free energies in two-flavor QCD
    for color singlet and octet $Q\overline{Q}$ channels (left)
    and color anti-triplet and sextet $QQ$ channels (right)
    obtained at $m_\pi/m_\rho = 0.65$ on a $16^3\times 4$ lattice\cite{WHOT07}. 
    The free energies are normalized such that they vanish at large distances.
}
   \label{fig:Nf2HQFE}
   \end{figure}

   \begin{figure}[tb]
       \centerline{
       \includegraphics[width=69mm]{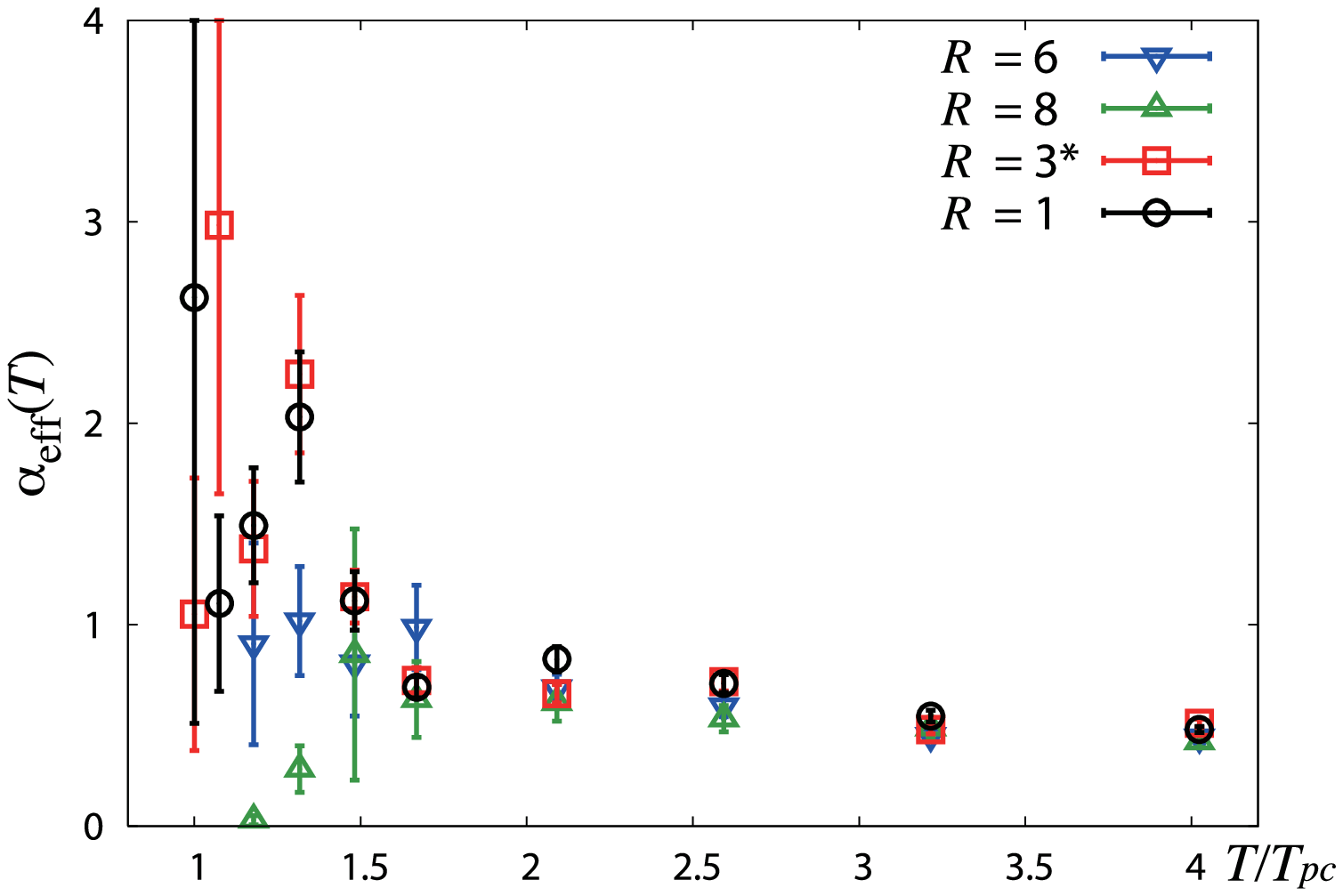}
       \hspace{1mm}
       \includegraphics[width=69mm]{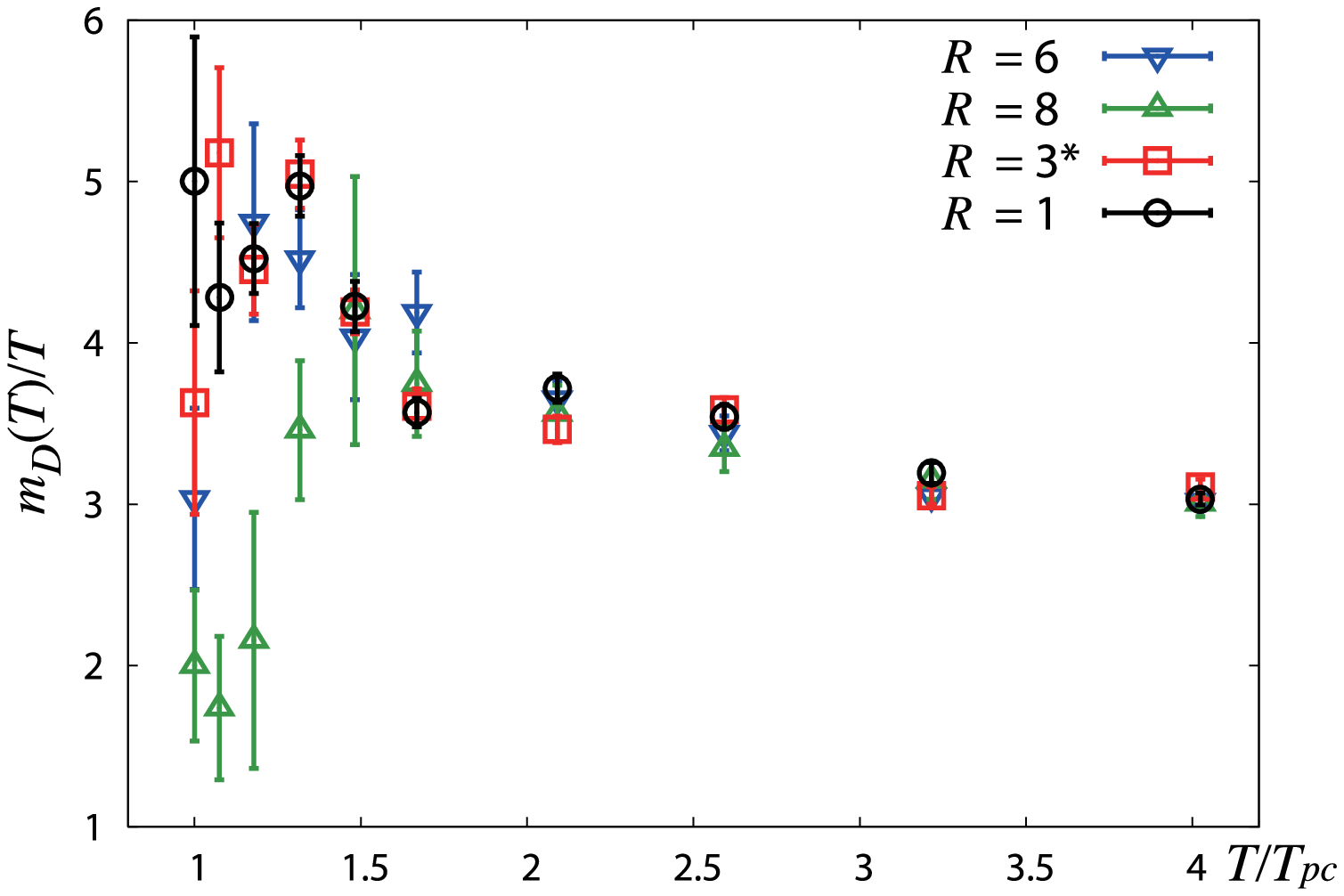}
       }
\caption{The effective running coupling $\alpha_{\rm eff}(T)$ (left) and
    Debye screening mass $m_D(T)$ (right) in two-flavor QCD, obtained at $m_\pi/m_\rho = 0.65$ on a $16^3\times 4$ lattice\cite{WHOT07}.
}
   \label{fig:Nf2HQFE2}
   \end{figure}

The heavy-quark free energies at $\mu=0$ are shown in Fig.~\ref{fig:Nf2HQFE} for $m_\pi/m_\rho= 0.65$ and $T \ge T_{\rm pc}$.
Results for $m_\pi/m_\rho= 0.80$ are similar.
We find that the inter-quark interaction is ``attractive'' in the color-singlet and antitriplet channels and is ``repulsive'' in the color octet and sextet channels. 
We also see that, irrespective of the channels, the inter-quark interaction becomes rapidly weak at long distances as $T$ increases, as expected from the Debye screening at high temperatures.

We find that the heavy-quark  free energies in the high temperature phase are well described by the screened Coulomb form, 
\begin{equation}
V^R(r,T) \;=\; C_{\! R} \, \frac{\alpha_{\rm eff}(T)}{r} \, e^{-m_D(T)\, r} ,
\label{eq:SCP}
\end{equation}
where $\alpha_{\rm eff}(T)$ and $m_D(T)$ are the effective running coupling and Debye screening mass, respectively.
From the fits, we note that the color-channel dependence of the free energies can be absorbed by the kinematical Casimir factor inspired by the lowest-order perturbation theory:
\begin{eqnarray}
C_{\bf 1}     = -\frac{4}{3}, \ \ \
C_{\bf 8}     =  \frac{1}{6}, \ \ \
C_{\bf 6}     =  \frac{1}{3}, \ \ \
C_{{\bf 3}^*}   = -\frac{2}{3},
\label{eq:Casimir}
\end{eqnarray}
This Casimir dominance at high temperatures was reported also in quenched studies using the Lorentz gauge \cite{NakamuraSaito}. 
With the Casimir factors, $\alpha_{\rm eff}(T)$ and $m_D(T)$ are well universal to all color channels, as shown in Fig.~\ref{fig:Nf2HQFE2}.
The magnitude and the $T$-dependence of the Debye mass are also consistent with the next-to-leading-order thermal perturbation theory.\cite{WHOT07}

\subsubsection{Heavy-quark free energies at $\mu\ne0$ in two-flavor QCD}
 
    \begin{figure}[tb]
       \centerline{
       \includegraphics[width=69mm]{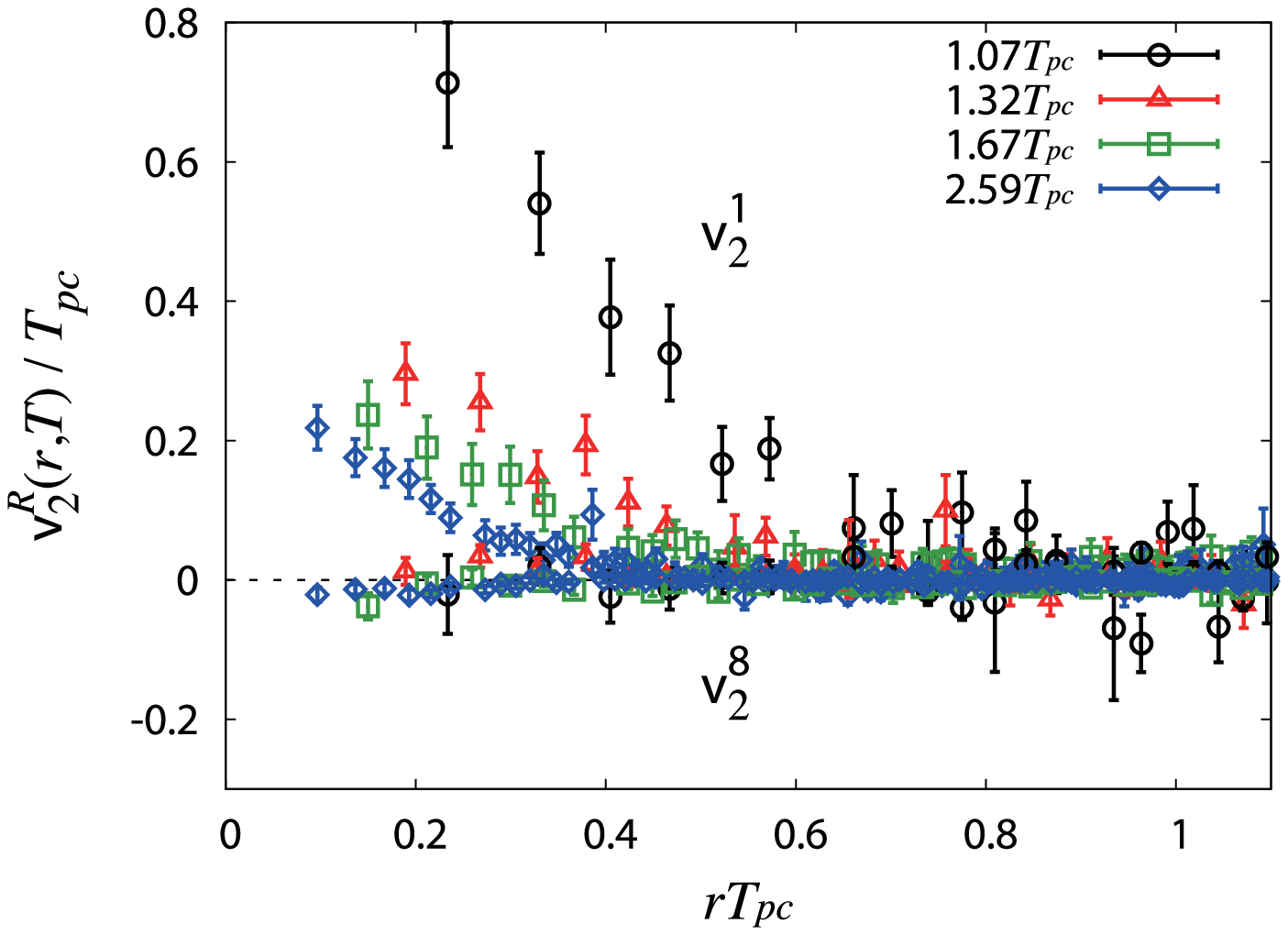}
       \hspace{1mm}
       \includegraphics[width=69mm]{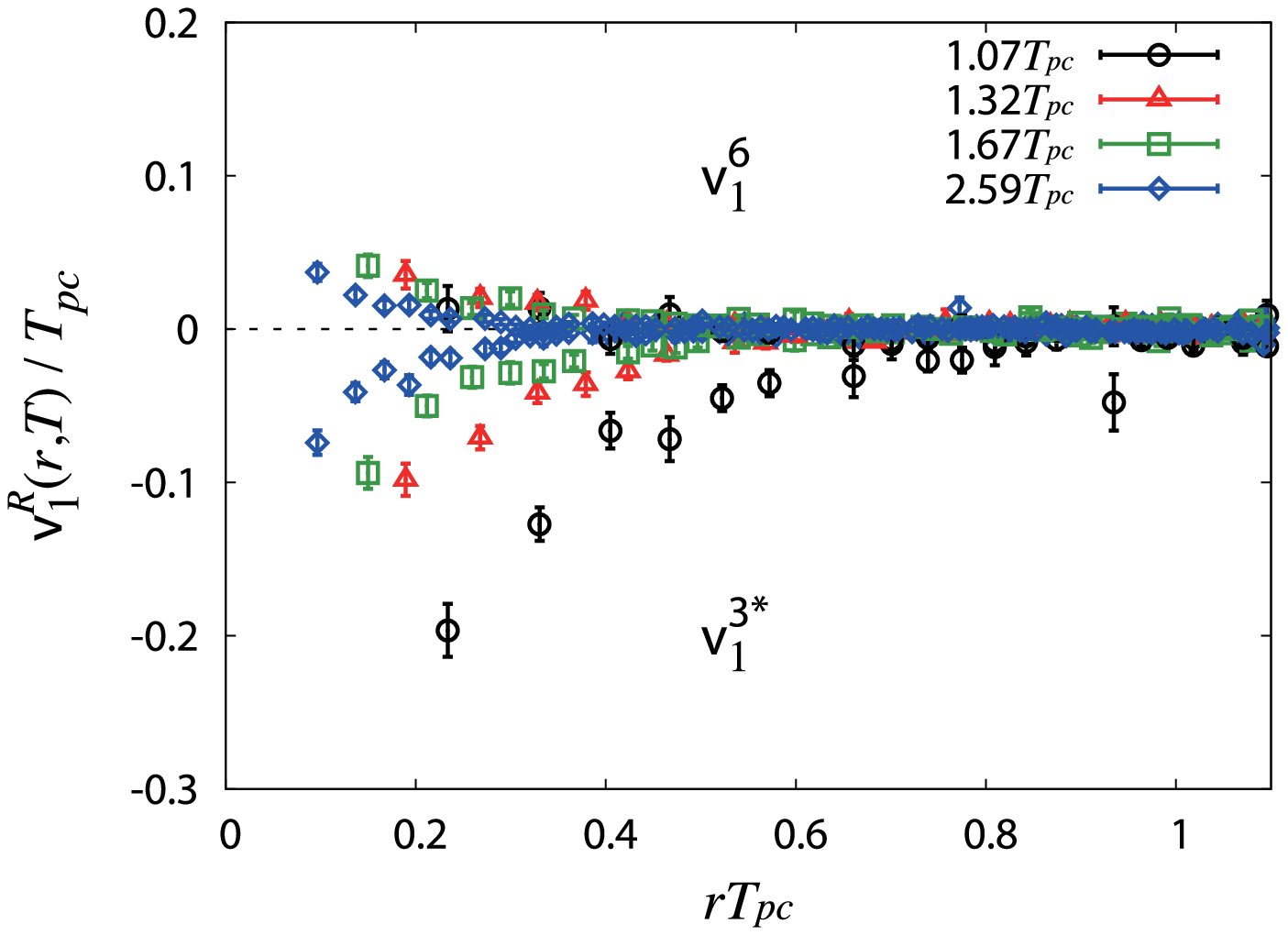}
       }
\caption{Taylor coefficients for heavy-quark free energies in two-flavor QCD,
    obtained at $m_\pi/m_\rho = 0.65$ on a $16^3\times 4$ lattice\cite{WHOT10dense}. 
 {\em Left}:  $v_2^R$ for color singlet and octet $Q\overline{Q}$ channels.
 {\em Right}: $v_1^R$ for color anti-triplet and sextet $QQ$ channels. 
 }
   \label{fig:Nf2HQFEmu}
   \end{figure}

The Taylor expansion of $V^R$ with respect to $\mu_{q}/T$ is given by
\begin{eqnarray}
V^R(r,T,\mu_{q}) = 
v^R_0(r,T) +
v^R_1(r,T) \left( \frac{\mu_{q}}{T} \right) +
v^R_2(r,T) \left( \frac{\mu_{q}}{T} \right)^2 +
O(\mu^3)
,
\label{eq:ENFE}
\end{eqnarray}
where concrete forms of the Taylor coefficients $v^R_n$ are given in Ref.~\citen{WHOT10dense}.

The color singlet and octet channels do not have the odd orders in the Taylor expansion since the $Q\bar{Q}$ free energies are invariant under the charge conjugation.
$v_0^R$ is the normalized free energy at $\mu=0$ shown in Fig.~\ref{fig:Nf2HQFE}.
Results for the lowest non-trivial order are shown in Fig.~\ref{fig:Nf2HQFEmu}.
See Ref.~\citen{WHOT10dense} for higher orders as well as those at $m_\pi/m_\rho = 0.80$.
From these figures, we find that, both around $T_{\rm pc}$ and at higher temperatures, 
the sign of $v_1^R$ is the same with that of $v_0^R$,
whereas the sign of a $v_2^R$ is the opposite of that of $v_0^R$:
\begin{eqnarray}
  v_1^R \cdot v_0^R > 0  \ \  \text{(only for $QQ$ free energies)} ,
\hspace{5mm}
  v_2^R \cdot v_0^R < 0 .
\end{eqnarray}
Because $v_1^R$ is absent for $Q\bar{Q}$ free energies, this means that, in the leading-order of $\mu_{q}$,
the inter-quark interaction between $Q$ and $\bar{Q}$ becomes weak at finite $\mu_{q}$, 
while that between $Q$ and $Q$ becomes strong. 
In other words, $Q \bar{Q}$ ($QQ$) free energies are screened (anti-screened)
by the internal quarks induced at finite $\mu_{q}$.

Taylor expansion coefficients for $\alpha_{\rm eff}(T,\mu)$ and $m_D(T,\mu)$ can be computed similarly.
We find that the leading correction in $m_D(T,\mu)$ due to finite $\mu$ is larger than a prediction of the leading-order thermal perturbation theory.\cite{WHOT10dense}

\subsection{Heavy-quark free energies in $2+1$ flavor QCD}
\label{sec:HQFE21}

We now extend the study to the more realistic $2+1$ flavor QCD\cite{Maezawa2011,Maezawa:2009di}.
As discussed in Sect.~\ref{sec:EOS}, we adopt the {\em fixed-scale approach} for $2+1$ flavor QCD.
We thus vary $T$ by varying $N_{\rm t}$ with fixed coupling parameters.
We use the finite-temperature configurations generated in Sect.~\ref{sec:EOS} to compute the heavy-quark free energies at $m_\pi/m_\rho\simeq0.63$ and $m_{\eta_{ss}}/m_{\phi}\simeq0.74$.
In this study of $2+1$ flavor QCD, we restrict ourselves to the case $\mu=0$.

A good feature of the fixed-scale approach is that the renormalization factors, which are determined on a zero-temperature lattice depending on the coupling parameters, are common to all $T$'s in the fixed-scale approach, because the coupling parameters are fixed for all $T$'s.
This is so also for the heavy-quark free energies.
Therefore, we can directly compare the bare free energies at different $T$'s in the fixed-scale approach.

\begin{figure}[tbp]
  \begin{center}
    \includegraphics[width=70mm]{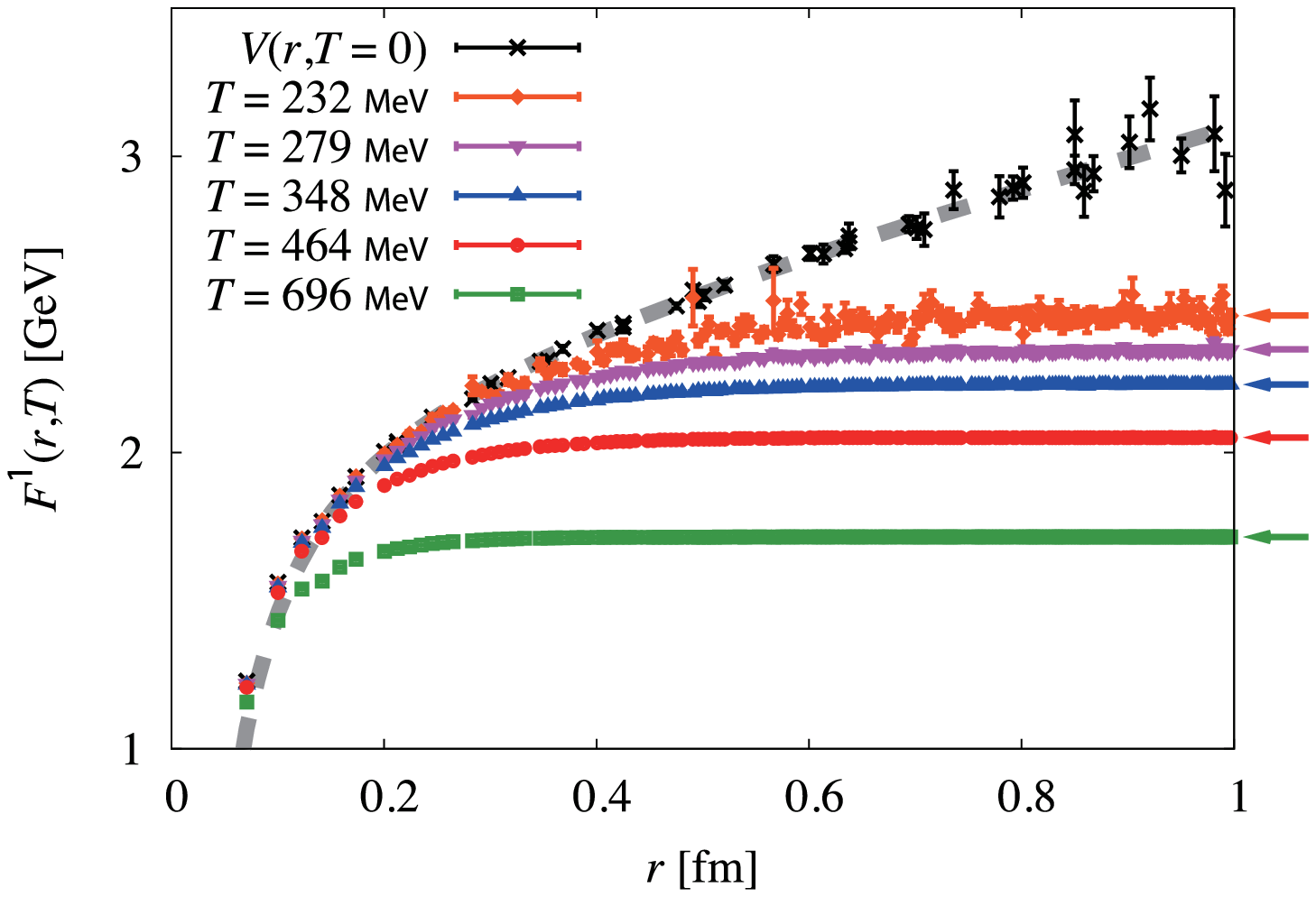}
    \hspace{0.5mm}
    \includegraphics[width=67mm]{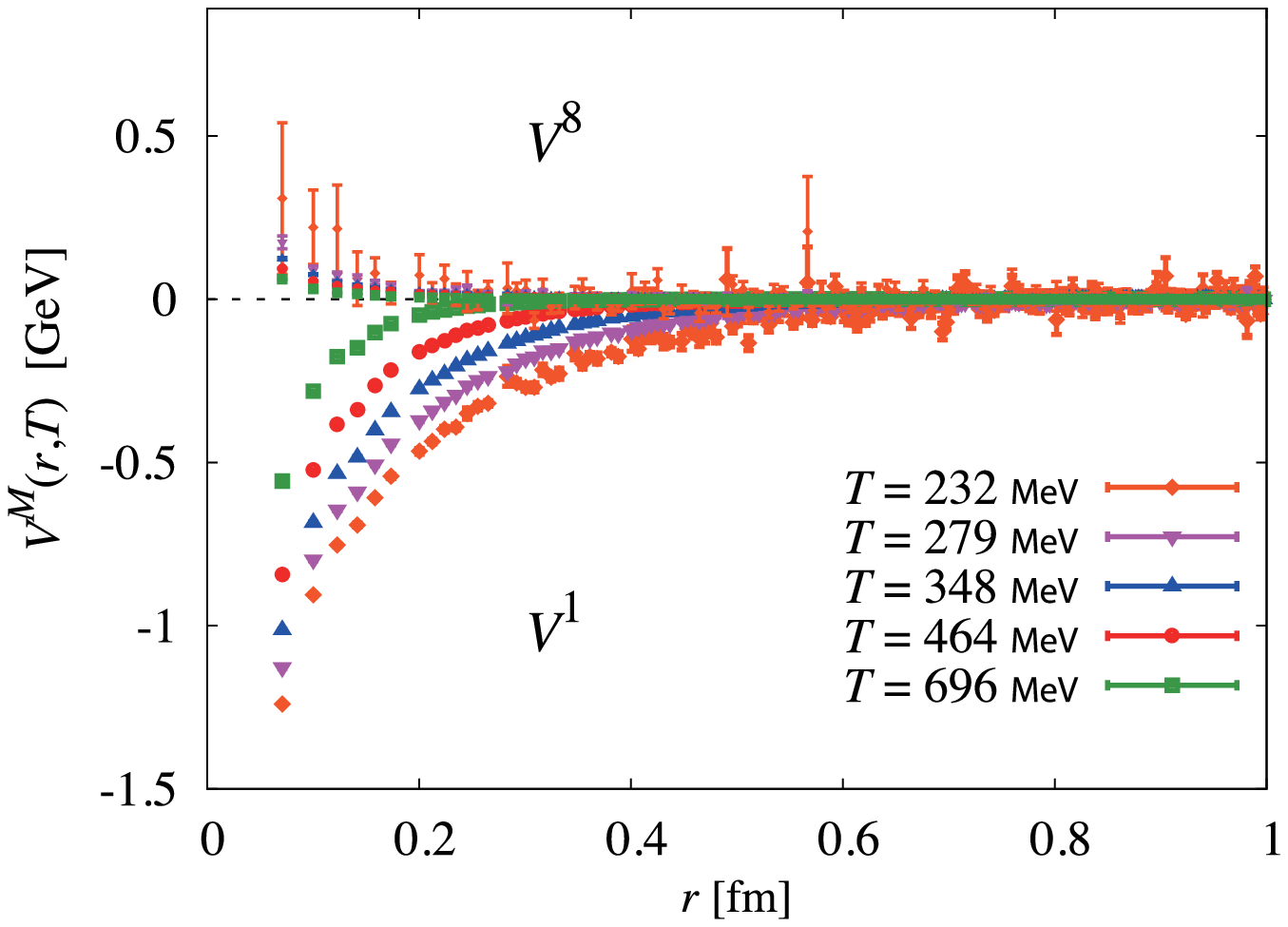} 
    \caption{Free energies of static quarks in 2+1 flavor QCD at finite temperatures with the fixed-scale approach.\cite{Maezawa2011}
The scale was set by the Sommer scale $r_0 = 0.5$ fm.
 {\em Left}: Bare free energy in the color-singlet channel. 
    The static quark-antiquark potential $V(r)$ at $T=0$ has been calculated
     by the CP-PACS and JLQCD Collaborations \cite{Ishikawa:2007nn}.
    The fit result of $V(r)$ by the Coulomb $+$ linear form is shown by the dashed gray curve.
    The arrows on the right side denote twice the single-quark free energy $2 F_Q$.
 {\em Right}: Normalized free energies for color singlet and octet $Q\bar{Q}$ channels.
     }
    \label{fig:HQFE21}
  \end{center}
\end{figure}

Our results for the bare free energies are shown in the left panel of Fig.~\ref{fig:HQFE21}.
Plotted are the data for the color singlet $Q\bar{Q}$ channel at various $T$'s in the high temperature phase, 
together with the zero-temperature static quark-antiquark potential $V(r)$ defined by the Wilson loop operator:
\begin{equation}
V(r) = - \lim_{\ell \rightarrow \infty} \left[  \ell^{-1} \ln \left\langle W_{i4}^{r \times \ell} \right\rangle \right].
\end{equation}
We find that singlet free energy $F^1(r, T)$ at any $T$ converges to $V(r)$ at short distances.
This is in accordance with the expectation that the short distance physics is insensitive to temperature. 
With the fixed-scale approach, we directly confirm that the theoretical expectation is actually satisfied on the lattice.\footnote{
In the case of the conventional fixed-$N_{\rm t}$ approach, different renormalization factor is required at each $T$.
In early studies, the insensitivity at short distances was just assumed and used to adjust the constant term of $F^1(r, T)$ at different $T$\cite{Kaczmarek:2004gv,Kaczmarek:2005ui}.
In more recent studies, the renormalization factors are computed by a series of zero-temperature simulations at the same coupling parameters as the finite temperature simulations.
See, e.g., Ref.~\citen{RBCB2008}.
}

At large distances, $F^1(r, T)$ departs from $V(r)$ and eventually becomes flat due to Debye screening.
On the right edge of the left panel of Fig.~\ref{fig:HQFE21}, we show twice the single-quark free energy defined by $ 2 F_Q = - 2 T \ln \langle {\rm Tr} \Omega({\bf x}) \rangle $ at each $T$ by the arrows with the same color.
We confirm that $F^1(r, T)$ converges to $2 F_Q(T)$  quite accurately.

By subtracting $2F_Q$, we obtain normalized free energies shown in the right panel of  Fig.~\ref{fig:HQFE21} for the $Q\bar{Q}$ channels. 
See \citen{Maezawa2011} for the results in the $QQ$ channels.
Performing fits with the screened Coulomb form (\ref{eq:SCP}), we confirm the Casimir dominance (\ref{eq:Casimir}) as in the case of two-flavor QCD.

\subsection{Gauge-independent screening masses}

The Debye screening masses and the effective couplings computed in the previous subsections are dependent on the choice of the gauge. 
Therefore, their physical meanings are not quite clear. 
In Ref.~\citen{WHOT10screening}, we have proposed a gauge-independent definition of screening masses for electric and magnetic channels.

Under the Euclidean time-reflection $\R$ and the charge conjugation $\Ca$, the gluon fields transform as,
\begin{eqnarray}
 {A}_i(\tau,{\bf x}) \xrightarrow{\R} {A}_i(-\tau,{\bf x}) ,\hspace{1mm}
 A_4 (\tau,{\bf x}) \xrightarrow{\R}  - A_4(-\tau,{\bf x}) ,\hspace{1mm}
 A_\mu(\tau,{\bf x}) \xrightarrow{\Ca}  - A^*_\mu(\tau,{\bf x}) .
 \end{eqnarray}
We call an operator magnetic (electric) if it is even (odd) under $\R$. 
It is natural to extract magnetic and electric properties by decomposing observables using these symmetries\cite{Arnold:1995bh}: 
Under these transformations, the Polyakov line operator $\hat\Omega({\bf x})$ transforms as
\begin{eqnarray}
\hat\Omega({\bf x}) \xrightarrow{\R} \Omega^\dag({\bf x}),
\quad
\hat\Omega({\bf x}) \xrightarrow{\Ca} \Omega^\ast({\bf x})
\end{eqnarray}
Magnetic (electric) Polyakov line operator can be defined as $\R$-even ($\R$-odd) part of $\hat\Omega({\bf x})$,
\begin{eqnarray}
\Om({\bf x}) \equiv \frac{1}{2} \left(\hat\Omega({\bf x}) + \hat\Omega^\dagger({\bf x}) \right),
\quad
\Oe({\bf x}) \equiv \frac{1}{2} \left(\hat\Omega({\bf x}) - \hat\Omega^\dagger({\bf x}) \right),
\end{eqnarray}
which can be further decomposed into  $\Ca$-even and odd parts as
\begin{eqnarray}
\hat\Omega_{{\rm M} \pm}({\bf x}) \equiv \frac{1}{2} \left( \Om({\bf x}) \pm \Om^*({\bf x}) \right),
\quad
\hat\Omega_{{\rm E} \pm}({\bf x}) \equiv \frac{1}{2} \left( \Oe({\bf x}) \pm \Oe^*({\bf x}) \right),
\label{eq:Omega-E} 
\end{eqnarray}
where  $\pm$ stands for even or odd under $\Ca$. 
Note that ${\rm Tr} \Omo = {\rm Tr} \Oee = 0$ and ${\rm Tr} \Ome$ (${\rm Tr} \Oeo $) is nothing but the real (imaginary) part of ${\rm Tr} \hat\Omega$. 

Then the magnetic (electric) screening mass is extracted from the long-distance behavior of generalized gauge-invariant Polyakov loop correlation functions,
\begin{eqnarray}
C_{\rm M +} (r,T) &\equiv&
 \left\langle {\rm Tr} \Ome ({\bf x})\, {\rm Tr} \Ome ({\bf y}) \right\rangle 
 - \left|\langle {\rm Tr}\hat\Omega \rangle \right|^2
 \xrightarrow[r \rightarrow \infty]{ }
 A \, \frac{e^{-m_{\rm M+}(T)\, r} }{rT}
 , 
\nonumber \\
C_{\rm E-} (r,T) &\equiv&
 \left\langle {\rm Tr} \Oeo ({\bf x})\, {\rm Tr} \Oeo ({\bf y}) \right\rangle
\xrightarrow[r \rightarrow \infty]{ }
 B \, \frac{e^{-m_{\rm E-}(T)\, r} }{rT}
 , \label{eq:Cmm}
\end{eqnarray}
where $r = |{\bf x} - {\bf y}|$ and $\langle {\rm Tr} \hat\Omega \rangle$ is real due to the $\Ca$ symmetry.
Note that the conventional gauge-invariant Polyakov loop correlation function is given by 
\begin{eqnarray}
C_{\Omega} (r,T) \equiv 
\left\langle {\rm Tr} \hat\Omega^\dagger ({\bf x})\, {\rm Tr} \hat\Omega ({\bf y}) \right\rangle 
- \left|\langle {\rm Tr}\hat\Omega \rangle\right|^2
= C_{\rm M+} (r,T)  - C_{\rm E-} (r,T) .
\end{eqnarray}
 
\begin{figure}[t]
  \begin{center}
    \includegraphics[width=75mm]{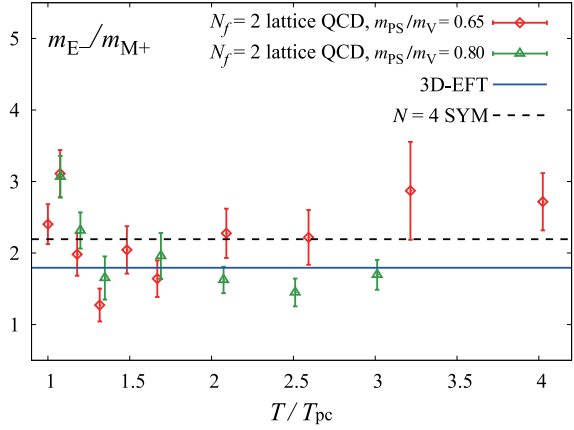} 
    \caption{Comparison\cite{WHOT10screening} of the screening ratio, $\Meoi/\Mmei$,
    with predictions in the dimensionally-reduced effective field theory (3D-EFT)\cite{Hart:2000ha}
    and ${\cal N}=4$ supersymmetric Yang-Mills theory\cite{Bak:2007fk}. }
    \label{fig:SMratio}
  \end{center}
  \vspace{-5mm}
\end{figure}

Using the configurations generated for \citen{WHOT07}, we computed these screening masses in two-flavor QCD in the high temperature phase.
We find that (i) $C_{\rm M +}$ and $C_{\rm E-}$ have opposite sign, 
and (ii) $C_{\rm M +}$ has larger magnitude and longer range than $C_{\rm E-}$ at long distances. 
The latter implies that the the conventional $C_{\Omega}$ is dominated by the magnetic sector at long distances, and thus $m_{_\Omega}(T) \simeq m_{\rm M+}(T)$.
We also find $m_{\rm M+}(T) < m_{\rm E-}(T)$.
A comparison with a dimensionally-reduced effective field theory\cite{Hart:2000ha} and an ${\cal N}=4$ supersymmetric Yang-Mills theory with AdS/CFT correspondence\cite{Bak:2007fk} 
lead to good agreements of $\Meoi/\Mmei$ for $1.5 < T/T_{\rm pc} < 3$, as shown in Fig.~\ref{fig:SMratio}\cite{WHOT10screening}. 
Further study is needed to clarify the meanings and implications of these results.

\section{Summary}
\label{sec:conclusions}

The WHOT-QCD Collaboration is pushing forward a series of projects to clarify the phase structure and thermodynamic properties of the QCD matter at finite temperatures and densities, mainly adopting improved Wilson quarks.
Wilson-type quarks do not have the theoretical unclearness of the staggered-type quarks, but require more computational resources. 
Thus, development and application of various computational techniques are mandatory.
We have developed the $T$-integration method to calculate the equation of state in the fixed-scale approach, the Gaussian method based on the cumulant expansion to tame the sign problem, and the histogram method combined with the reweighting technique to explore the phase structure of QCD.
By adopting them, we have made a series of studies in QCD at finite temperatures and densities with improved Wilson quarks.
In particular, we have carried out the first study of finite-density QCD with two flavors of Wilson quarks and the first calculation of the equation of state with $2+1$ flavors of Wilson quarks.
We are extending the studies towards our final objective of $2+1$ flavor QCD directly at the physical point.

\section*{Acknowledgements}
We are grateful to the members of the WHOT-QCD Collaboration.
We also thank the members of the CP-PACS and JLQCD Collaborations for 
providing us with their high-statistics 2+1 flavor QCD configurations with improved Wilson quarks at $T=0$.
This work is in part supported 
by Grants-in-Aid of the Japanese Ministry
of Education, Culture, Sports, Science and Technology, 
(Nos.\ 
22740168, 
21340049 , 
23540295),  
by the Large Scale Simulation Program of High Energy Accelerator Research Organization (KEK), 
and by the Large-Scale Numerical Simulation Projects of CCS/ACCC, Univ.\ of Tsukuba.
Parts of the simulations were performed also on supercomputers at RCNP, Osaka Univ.\ and YITP, Kyoto Univ.
SE is supported by the Grant-in-Aid for Scientific Research on Innovative Areas
No.\ 2004:2310576.

%

\end{document}